\documentclass[pre,twocolumn,showpacs,superscriptaddress,floatfix]{revtex4}
\pdfoutput=1 

\usepackage{graphicx}
\usepackage{amssymb}
\usepackage{subfigure}

\renewcommand{\vec}[1]{{\bf #1}}

\begin{document}

\title{Effects of Friction and Disorder on the Quasi-Static\\ Response of
  Granular Solids to a Localized Force}

\author{C. Goldenberg}
\email{chayg@pmmh.espci.fr}
\affiliation{Laboratoire de Physique et M\'ecanique des Milieux
  H\'et\'erog\`enes (CNRS UMR 7636), ESPCI, 10 rue Vauquelin, 75231 Paris Cedex
  05, France.}
\author{I. Goldhirsch}
\email{isaac@eng.tau.ac.il}
\affiliation{Department of Fluid Mechanics and Heat Transfer,
Faculty of Engineering, Tel Aviv University,
Ramat-Aviv, Tel Aviv 69978, Israel.}

\date{\today}

\begin{abstract}
  The response to a localized force provides a sensitive test for different
  models of stress transmission in granular solids. The elasto-plastic models
  traditionally used by engineers have been challenged by theoretical and
  experimental results which suggest a wave-like (hyperbolic) propagation of
  the stress, as opposed to the elliptic equations of static elasticity.
  Numerical simulations of two-dimensional granular systems subject to a
  localized external force are employed to examine the nature of stress
  transmission in these systems as a function of the magnitude of the applied
  force, the frictional parameters and the disorder (polydispersity). The
  results indicate that in large systems (typically considered by engineers),
  the response is close to that predicted by isotropic elasticity whereas the
  response of small systems (or when sufficiently large forces are applied) is
  strongly anisotropic. In the latter case the applied force induces changes in
  the contact network accompanied by frictional sliding.  The larger the
  coefficient of static friction, the more extended is the range of forces for
  which the response is elastic and the smaller the anisotropy. Increasing the
  degree of polydispersity (for the range studied, up to 25\%) decreases the 
  range of elastic  response. This article is an extension of a previously
  published letter~\cite{Goldenberg05}. 
\end{abstract}

\maketitle
\section{Introduction}
\label{sec:introduction}
The description of the static phase of  granular matter, much like 
its dynamical phases, is of
significant current interest.  Some studies focus on  constitutive relations
in order to describe the stress distribution in these
materials whereas other investigations probe their microscopic characteristics,
such as the nature of the  force distributions and  correlations.

Static and quasi-static properties of granular materials are commonly modeled
by engineers using elasto-plastic ~\cite{Nedderman92,Savage98b} and
hypoplastic~\cite{Wu96,Gudehus04} models.  Other types of models have been put
forward (mostly) by physicists.  Some of these models comprise hyperbolic
partial differential equations for the stress transmission through a granular
material ~\cite{Wittmer97,Tkachenko99,Head01,Edwards01,Ball02,Blumenfeld04}.
This is in contrast with the elliptic, non-propagating nature of the classical
equations of static elasticity.  It seems that this dichotomy in the modeling
of granular statics started~\cite{Savage98b,Bouchaud98} in the context of the
interpretation of experiments performed on conical
piles~\cite{Smid81,Brockbank97}, where a pressure dip below the apex of the
pile has been observed (the presence or absence of such a dip in conical piles
and in wedges was found to depend on the construction method~\cite{Vanel99b}).

In order to obtain insights into this problem it is convenient to consider the
simpler geometry of a granular rectangular layer (or slab) resting on a solid
support~\cite{DeGennes99}, not the least since this class of systems is well
researched experimentally
~\cite{Silva00,Geng01,Reydellet01,Serero01,Mueggenburg02,Geng03,Moukarzel04}.
Much like in the case of piles, the cited and other experiments seem to be
sending mixed messages concerning the `correct' description of stress
transmission in granular slabs, as some render support to the elliptic
descriptions whereas others are compatible with hyperbolic models.  When the
response to the application of an external force is linear in the force (for
sufficiently small forces), the problem is tantamount to the study of the
Green's function of the system~\cite{Geng03}.

The present article is devoted to a detailed study of the response of
two-dimensional vertical slabs comprising polydisperse frictional disks subject
to gravity and a localized, compressive external load applied at the top of the
slab; it expands upon the results of~\cite{Goldenberg05} in both content and
detail.

The dependence of the response on the magnitude of the external force as well
as the interaction parameters and degree of disorder (or polydispersity) is
investigated.  The results suggest a resolution of the above mentioned
controversy as well as the seemingly mutually incompatible experimental
results.  First, noticing that strongly anisotropic elasticity can exhibit
hyperbolic-like features (see more below) it is convenient to consider the
response within the framework of elasticity. Note that it is not claimed that
the process leading to a given state is elastic (often it is not) but rather
that the excess stress field induced by an external force in a given
(``prestressed'') static state can be described by the equations of elasticity
and often even by linear isotropic elasticity (for an isotropic reference
configuration), even in the presence of friction (see also the recent elastic
model reviewed in~\cite{Jiang07}).  More specifically, large systems subject to
``small'' applied forces (which is likely to be the case in many engineering
applications), can be described by isotropic elasticity except in the near
vicinity of the point of application of the force (where one expects strong
induced anisotropy and irreversible rearrangements).  Sufficiently far from the
point of application the external force leads to practically no sliding or
rearrangement of the grains.  The latter events are rarer the larger the
friction and therefore, as shown below, the linearity and isotropy of the
response are {\em enhanced} by friction.
 
Nonlinear/irreversible effects such as changes in the contact network and
frictional sliding can naturally give rise to  large {\em anisotropy},
yielding hyperbolic-like response (as already suggested
in~\cite{Savage98b,Cates99}). In relatively small systems subject to
sufficiently large forces, as in some
experiments~\cite{Silva00,Geng01,Geng03,Mueggenburg02,Moukarzel04}, the
size of the nonlinear and anisotropic domain induced by the external forcing can be
comparable to the system's size, hence the stress transmission in such systems
can be well described by hyperbolic or near-hyperbolic (see below) equations.
Similar qualitative behavior is observed in both ordered (lattice) and
disordered (polydisperse) configurations, the main difference being
the range of magnitudes of external forces for which one observes elastic
response. 

Note that qualitatively similar results were obtained in a study of the {\em
  displacement} response rather than  stress~\cite{Kasahara04}, using a
similar numerical simulation. This  study focused on the effect
of the mean coordination number, controlled by the particle stiffness, rather
than the effects of the frictional properties and the disorder which are the
focus of the present  work. Interestingly, an approach using a force
network ensemble~\cite{Ostojic06} also seems to yield qualitatively similar
results concerning the effect of friction, although a full  physical interpretation
of this approach is still lacking. 

The paper is organized as follows. Section II presents an outline of the
elliptic (elastic) and hyperbolic descriptions of static granular matter and a
comparison between the two. Results of some  relevant experiments are reviewed 
in the same context. 
 Sec.~\ref{sec:sim} provides a description of the  simulation method. 
Sec.~\ref{sec:nofriction} reports the results obtained for
frictionless systems, and the subsequent sections describe more realistic
simulations incorporating friction (Sec.~\ref{sec:friction}) and disorder
(polydisperse systems, Sec.~\ref{sec:disorder}). The results of the simulations
and their analysis enable the interpretation of the findings in different
experiments:
 this comprises the content of Sec.~\ref{sec:interpretation}.
Sec. VIII offers a brief summary of the main results.\\
\section{Elliptic vs.\ Hyperbolic Descriptions}
\label{sec:elliptic-vs-hyperbolc}
\subsection{Theoretical description}
\label{sec:theoretical_description}
\begin{figure}
  \subfigure[][``Elliptic'']{\includegraphics[clip,width=0.48\hsize]{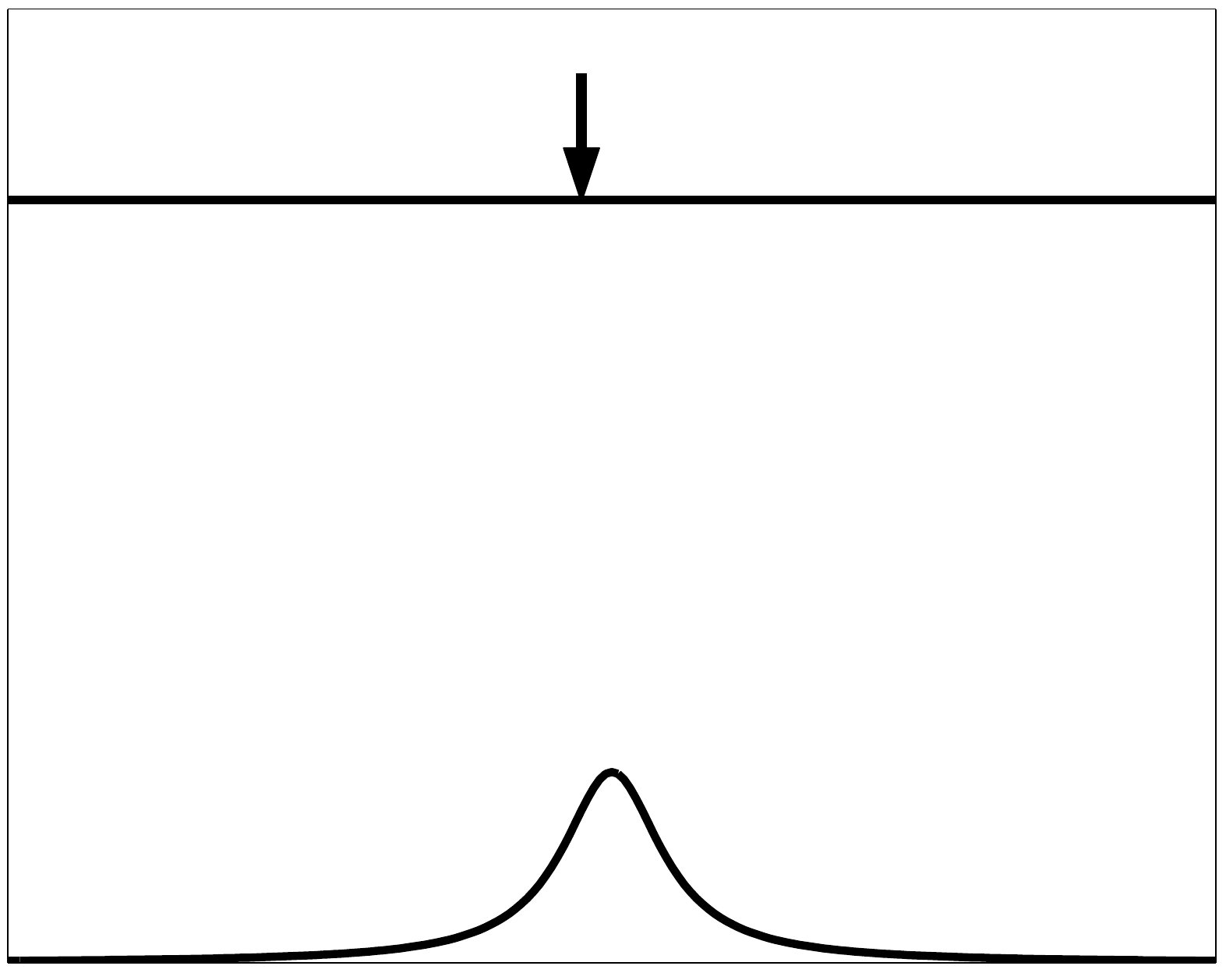}\label{fig:elliptic}}
  \subfigure[][``Hyperbolic'']{\includegraphics[clip,width=0.48\hsize]{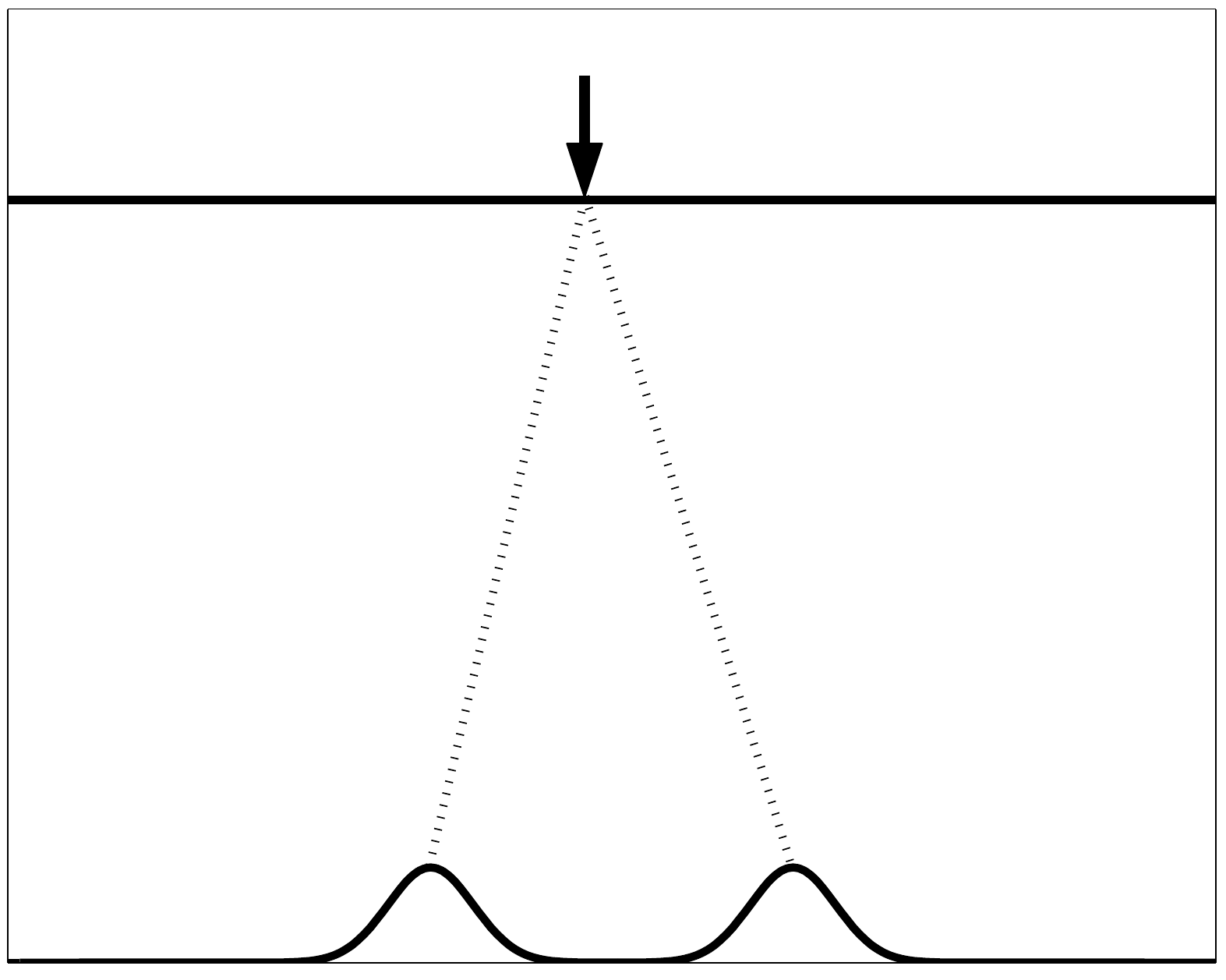}\label{fig:hyperbolic}}
\caption{Predictions of different models for the response of a granular slab to
  a localized vertical force applied to its top: pressure distribution on the floor supporting the slab.}\label{fig:elliptic+hyperbolic}
\end{figure}

As mentioned, there are two main classes of models that have been proposed for
the description of the response of granular assemblies. Broadly known as the
``elliptic'' and ``hyperbolic'' descriptions, their predictions differ both
qualitatively and quantitatively.  These respective predictions for the case of
the response (excess pressure on the floor due to the external force, i.e., the
pressure due to gravity in the otherwise unforced system is subtracted) to the
application of an external vertical force at the center top of a slab are
schematically illustrated in Fig.~\ref{fig:elliptic+hyperbolic}. The elliptic
case is represented here by the results of {\em isotropic}
elasticity~\cite{Landau86}. In this case, a single peaked response on the floor
is expected [Fig.~\ref{fig:elliptic}], its width being proportional to the
depth (or height) of the slab. The shape of the peak is determined by the
equations of elasticity, and depends on the boundary conditions at the
floor~\cite{Savage98b,Serero01,Goldenberg02} (e.g., its rigidity and
roughness). In contrast, hyperbolic
descriptions~\cite{Wittmer97,Tkachenko99,Head01,Edwards01,Ball02,Blumenfeld04}
dictate that the stress propagates along characteristic directions, and two
peaks are expected in 2D (a ring in 3D), their widths determined by diffusive
broadening due to disorder~\cite{Claudin98b} and (consequently) proportional to
the square root of the depth of the slab.  This type of description has been
shown~\cite{Tkachenko99,Head01,Edwards01,Ball02,Blumenfeld04} to apply to
(frictionless) isostatic systems~\cite{Moukarzel98c,Roux00,Moukarzel01}, i.e.,
systems in which Newton's first and third laws (and the boundary conditions)
provide a number of equations that equals the number of unknowns (force
components) thereby enabling the determination of the forces from these laws;
it turns out these conditions determine an average coordination number for the
particle contacts and that the latter is the minimum required to maintain the
stability of the packing. This notion cannot be directly extended to
hyperstatic systems (i.e., with a higher coordination number). The isostatic
case is a marginally stable configuration, which may correspond to a plastic
material which is everywhere at incipient yield (for which classical plasticity
results in hyperbolic equations~\cite{Savage98b}).  We also note that recent
models based on the description of the stress in terms of propagating force
chains which may split and merge~\cite{Bouchaud01,Socolar02,Otto03,Roichman04}
have been shown to correspond to elastic-like equations at large
scales~\cite{Bouchaud01,Socolar02,Otto03}; a physical interpretation of the
fields satisfying these equations is still lacking.

It is important to note that, as already mentioned
in~\cite{Savage98b,Cates99,Goldenberg02}, elastic systems can exhibit
hyperbolic-like behavior. In~\cite{Goldenberg02}, we used a simple 2D model in
order to demonstrate  this property, i.e., a triangular lattice of harmonic springs in which
the spring constant for the horizontal springs, $K_1$, is different from that
of the oblique ones, $K_2$. This model 
corresponds (in the continuum limit) to an
anisotropic elastic system (the response of an anisotropic elastic infinite
half-plane (2D) with uniaxial symmetry has been analyzed in detail
in~\cite{Otto03}). The qualitative nature of the response depends on the
degree of anisotropy: it is single peaked for values of $K_2/K_1$ near 1
($K_2/K_1=1$ corresponds to an isotropic system). The response is narrower than
the isotropic one for $K_2/K_1<1$ and  wider for $K_2/K_1>1$. For sufficiently
large values of $K_2/K_1$, the response becomes double peaked, i.e., in  qualitative
agreement with the predictions of the hyperbolic models. 
However, the equations of anisotropic
elasticity are elliptic, except in the extreme anisotropic limit
$K_2/K_1\to\infty$ in which the equations do become hyperbolic. This
is consistent with the fact that the limit $K_1\to 0$ corresponds
to the absence of horizontal springs, and is therefore {\em isostatic}.  This
limit is similar to the case of a square (or cubic) lattice of springs which
(to linear order in the displacement) has no shear rigidity along the lattice
directions~\cite{Alexander98}. Although this anisotropic model seems quite
artificial, we show below that anisotropy arises quite naturally in more
realistic descriptions of granular materials.
\subsection{A Review of Experimental Results}
\label{sec:response-experiments}
This Section reviews some of the experiments in which the response of a
granular slab to a localized force was measured. 

In the experiments described in~\cite{Geng01,Geng03}, a localized vertical
force was applied to a two-dimensional (2D) packing of photoelastic particles.
Recall that  stressed 
photoelastic solids  viewed through cross-polarizers  exhibit  fringes
whose density is proportional to the difference between
 the two principal values of
the internal particle  stress~\cite{Howell99}. In the described experiment 
the applied forces had  to be sufficiently large in
order to observe the fringes; in practice forces that were  about 150 times the
particle weight were used. The stress is concentrated at the
interparticle contacts. Post-processing 
techniques were used to extract the magnitudes of the  interparticle  forces
from  the experimental
images~\cite{Howell99,Geng01,Geng03}. Force chains were  prominent in 
all studied realizations. 

Three different types of packings were studied in~\cite{Geng01,Geng03}. The
most ordered was that of a triangular lattice of nominally monodisperse (equal)
disks; less order was found in a packing of bidisperse disks (two different
sizes), and a packing of pentagonally shaped particles was the most disordered
of the three.  The images from 50 configurations were
averaged~\cite{Geng01,Geng03} for each type of packing. The force profiles as a
function of the horizontal (orthogonal to gravity) coordinate were plotted for
different depths in each of the slabs~\cite{Geng01,Geng03}.  These profiles
exhibited strong dependence on the particle shapes: the ordered lattice of
monodisperse disks exhibited two prominent ``force chains'' along the lattice
directions ($60^\circ$ with respect to the horizontal), a result which appears
to be consistent with a hyperbolic model of propagating forces (note, however,
that a vertical force chain can be observed as well, and it is not anticipated
by these models). As the disorder is increased, the force chains ``fade out'',
and for the random configuration of pentagonal particles in 2D the measured
force profile possessed a single peak, the width of which varied linearly in
the depth; the latter result is compatible with the predictions of elasticity.
It was proposed that granular materials may experience a crossover from a
hyperbolic to an elliptic behavior as the degree of disorder
increases~\cite{Geng01}.

In another  experiment, in which a rather 
small system was studied (of about  10 diameters in height), the slab comprised
 round-edged  rectangular 2D particles~\cite{Silva00}. In this
case, the response appeared to exhibit a parabolic envelope. This type  of 
force
profile was predicted on the basis of 
models that  assume an
uncorrelated, diffusive propagation of the forces, such as the
q-model~\cite{Liu95,Coppersmith96} (which was originally proposed to describe
the statistics of force fluctuations). However, as mentioned, this system is
quite small in terms of the number of constituents; furthermore, even in this
case, it appears that the envelope may be narrowing down near the bottom of the
slab.  We shall return to the issue  of system size below.

In the simulations reported in~\cite{Moukarzel04} the {\em displacement}
(rather than the force) response of a 2D packing of disks of three different
diameters was measured (in response to a small displacement of a particle at
the bottom of the packing).  The displacement response is expected to be
qualitatively similar to the force response (in isostatic systems the two are
equivalent~\cite{Moukarzel98c,Roux00}). The measured response (averaged over
several hundred configurations) is single peaked, and the width of the peak
appears to grow as the square root of the height for small system heights, and
cross over to a linear dependence at larger heights (the total height of the
packing was about 10 diameters, which is quite small).  The response function
(scaled by the width) is fit quite well by a Gaussian. The same
paper~\cite{Moukarzel04} presents simulations of isostatic systems which show,
as expected, a double peaked response.

In the experiments described in~\cite{Reydellet01,Serero01}, an external
localized force was applied at the top of 3D disordered slabs of sand (and
other materials), and the response force profiles (i.e., the actual forces minus
the corresponding forces without the external load) on the floor were measured. 
In this case
the applied force was rather small (a few particle weights), and the response
was verified to be linear in the applied force. The measured response was
single peaked, the profile width being proportional to the slab depth, as
expected for an elastic system. However, the shape of the response was found to
depend on the preparation method: it was wider for loose packings (obtained by
pulling a sieve through the packing) than for dense packings (obtained by
filling the container layer by layer, pressing the packing after each layer is
deposited).  The response could not be fitted by isotropic elastic solutions
for  finite slabs~\cite{Serero01}.  

Unlike  the highly disordered
systems used in~\cite{Geng01,Geng03}, the experiments reported
in~\cite{Mueggenburg02} concerned  ordered 3D close-packed systems of spherical
particles, arranged on FCC (Face-Centered Cubic) and HCP (Hexagonal
Close-Packed) lattices~\cite{Kittel56}. The forces on the floor were measured
using pressure marks on carbon paper~\cite{Mueggenburg02}; the applied forces
used in these experiments were quite large (a few {\em thousand} times the
particle weight).  The results were averaged over several packings of nominally
equal spheres. For shallow systems, three distinct peaks were observed in the
case of the FCC lattice, and a sharp ring for the HCP lattice. This is
consistent with a description based on propagating forces~\cite{Mueggenburg02}
(which would correspond to a hyperbolic continuum description). However, for
larger systems the peaks were considerably less distinct, the response being
`smoother'. This further indicates that a hyperbolic-like response may apply
only to relatively small systems. For disordered (amorphous)
packings~\cite{Mueggenburg02}, a single peak was obtained (a force impulse was
used, since a large persistent force resulted in major rearrangements in the packing).

In summary, qualitatively different types of response to an external force 
were observed in experiments on granular assemblies,  some of which
seem to render support to  elliptic models, others to hyperbolic or 
parabolic models of stress transmission. The degree of disorder seems to play an important
role, as do the size of the system and the magnitude of the applied force.
 The observed differences also
suggest that other parameters, such as
the coefficient of friction, may be important.  In order to clarify the
role of these parameters and the associated physical mechanisms, we performed
extensive simulations of 2D granular slabs, as described below.
\section{Simulation Method}
\label{sec:sim}
In order to study the response of granular packings and its dependence on the
particle properties in more detail, we employ the discrete element
method~\cite{Cundall79,Cundall82} (DEM), a particle-based simulation method for
granular materials also known as the molecular dynamics (MD) simulation
method~\cite{Allen87,Frenkel96,Rapaport97,Sadus99}. This method has been
employed in studies of atomic and molecular assemblies as well as granular
systems, see, e.g., \cite{Wolf96,Herrmann98,Zhu07}) for recent reviews.  In the
present work, DEM simulations have been employed to study 2D systems composed
of disks. The equations of motion for each particle are integrated using the
Beeman algorithm~\cite{Frenkel96}, which provides more accurate velocities than
the commonly used Verlet algorithm~\cite{Frenkel96}. We verified that the use
of a Gear 5-value predictor-corrector algorithm~\cite{Sadus99} does not result
in any significant changes in the results.

\subsection{The Force Model}
\label{sec:forcemodel}
Models for the interactions among solid grains are usually based on contact
mechanics~\cite{Gladwell80,Johnson85}.  Following the work of Hertz (see,
e.g.,~\cite{Landau86}) and others, it is customary to assume that the forces
exerted by solid particles on each other are pairwise additive. Since for
relatively rigid particles the typical deformation of a particle is a very
small perturbation to its shape and size, it is common to model these
interactions as follows: the particles are fixed in shape and can overlap and
the corresponding normal forces (perpendicular to the plane of contact among
particles) are taken to be functions of the degree of this imaginary overlap.
The tangential interparticle forces are taken to depend on the particles'
rotations as well.

Most natural or industrial granular materials comprise non-spherical particles.
However, the latter are difficult to treat theoretically, and only few models
for simulating their behavior have been suggested and studied
(e.g.,~\cite{Poschel93,Walton93,Poschel95,Matuttis00,Vu-Quoc00,Schinner01}).
Therefore the present paper specializes to disk shaped particles in two
dimensions.  The force model we use is essentially the same as that employed
in~\cite{Cundall79} (for some other examples of force schemes commonly used in
simulations, see, e.g.,~\cite{Schafer96,Sadd93,Walton95}).

For spherical or disk-shaped particles, the overlap of two particles is defined by:
\begin{equation} \xi_{ij}\equiv
  R_{i}+R_{j}-|\vec{r}_{ij}|,\label{eq:overlap}\end{equation} where
$R_{i},R_{j}$ are the radii of the particles, $\vec{r}_i$ is the position of
the center of mass of particle, $i$, and $\vec{r}_{ij} \equiv \vec{r}_i-
\vec{r}_j$. In noncohesive granular materials, the particles are assumed to
interact only when they overlap, i.e., when $\xi_{ij}>0$.  For two frictionless
elastic spheres, a classical result by Hertz (see, e.g.,~\cite{Landau86}) is
that the force is proportional to $\xi_{ij}^{3/2}$, while for cylinders, it is
linear in $\xi_{ij}$ (up to a logarithmic correction, see,
e.g.,~\cite{Schwartz71,Schwager07}). In the simulations whose results are
presented below the normal component (parallel to $\vec{r}_{ij}$) of the force
is taken to be linear in the overlap (linear spring); this choice is not due to
the fact that we consider cylindrical particles, but rather in order to
simplify the application of theoretical considerations. It may be further
justified by the fact that we consider small deformations, in which case one
may linearize the force-displacement law around a reference
configuration~\cite{Majmudar07S}.

Even for frictionless particles there can be dissipation, e.g., due to
viscoelastic normal forces that depend on the normal relative velocity
$\dot{\xi}_{ij}$ (see, e.g.,~\cite{Brilliantov96,Schwager07}). While hysteretic,
rate independent dissipation models (see,
e.g.,~\cite{Walton95,Vu-Quoc99,Pournin01}) may be more realistic than
viscoelastic models, their implementation is more complicated. Since we focus
on the static response and small deformations the results should not be
sensitive to the precise choice of dissipation mechanism.  We therefore use a
damping term which is linear in the relative velocity of interacting particles
(linear dashpot). All in all the normal interparticle force exerted by particle
$j$ on particle $i$, $\vec{f}_{ij}^{\mathrm{N}}$, can be expressed as:
\begin{equation}
\vec{f}_{ij}^{N}=\left(k_{N}\xi_{ij}+\nu_{\mathrm{N}}v_{ij}^{N}\right)H(\xi_{ij})\hat{\vec{r}}_{ij},\label{eq:normal_force}\end{equation}
 where  $\hat{\vec{r}}_{ij}$ is the
unit vector in the direction of $\vec{r}_{ij}$, $v_{ij}^{N}\equiv(\vec{v}_{i}-\vec{v}_{j})\cdot\hat{\vec{r}}_{ij}\equiv\vec{v}_{ij}\cdot\hat{\vec{r}}_{ij}$
is the relative normal velocity of the particles,  $\nu_{\mathrm{N}}$ is a fixed 
damping constant, and $H(x)$ is the
Heaviside function, \begin{equation}
H(x)\equiv\left\{ \begin{array}{cl}
1 & x>0\\
1/2 & x=0\\
0 & x<0\end{array}\right..\label{eq:heaviside}\end{equation}

Next, consider the friction-induced tangential forces.  A simple model for
friction~\cite{Cundall79} includes tangential springs for modeling static
friction, often with velocity-dependent damping to facilitate relaxation to a
static configuration. The springs are generated at zero length, which is also
their rest length, when two particles come into contact and ``snap'' when the
resulting frictional force, $f^{T}$, satisfies $f^{T}>\mu_{S}f^{N}$, i.e., when
the Coulomb limit is exceeded.  Once a tangential spring is severed, the
corresponding particles are allowed to slip with respect to each other and
experience dynamic friction, $f^{T}=\mu_{D}f^{N}$.  Note that the description
of static friction in terms of a tangential spring is physically reasonable,
and consistent with both contact mechanics on the macroscopic
scale~\cite{Gladwell80,Johnson85} and microscopic approaches to the description
of friction (see, e.g.,~\cite{Krim02} for a list of references on friction).
The tangential springs are of course quite stiff, giving rise to very small,
but measurable~\cite{Rabinowicz51} displacement prior to slip. Tangential
springs have often been used to model interparticle interactions in mean-field
derivations of granular elasticity~\cite{Bathurst88,Chang96,Jenkins01,Gay02}.
More realistic descriptions of frictional contacts account for the history of
contact deformations (see, e.g.,~\cite{Schafer96,Sadd93,Walton95,Vu-Quoc99}).

The above tangential spring-dashpot model in conjunction with Coulomb's law of
friction serve as the basic tangential force law in the present work, as
in~\cite{Cundall79}. For simplicity, the coefficients of static and dynamic
friction are taken equal each other:  $\mu_{D}=\mu_{S}=\mu$ (the value of
$\mu_{D}$ does not directly affect the static configurations which are the main
concern in the present work, although it may influence the evolution towards
these states). In order to employ a tangential spring-dashpot force model, a
relative tangential displacement has to be defined at each contact. To this
end, note that the relative tangential velocity of two disks (in 2D) is given by:
\begin{equation}
  \vec{v}_{ij}^{T}=\vec{v}_{ij}-v_{ij}^{N}\hat{\vec{r}}_{ij}-\hat{\vec{t}}_{ij}(R_{i}\omega_{i}+R_{j}\omega_{j}),\label{eq:tan_velocity}\end{equation}
where $\omega_{i}\equiv\dot{\theta_{i}}$ is the angular velocity
and $\hat{\vec{t}}_{ij}\equiv\hat{\vec{z}}\times\hat{\vec{r}}_{ij}$.
In the model used in this work the force exerted by a tangential spring depends on the relative tangential
displacement, $\vec{s}_{ij}$, and   is given
by:\[
\vec{f}_{ij}^{T}=\left(-k_{T}\vec{s}_{ij}-\nu_{T}\vec{v}_{ij}^{T}\right)H(\xi_{ij}),\]
where $k_T$ is the tangential spring constant and $\vec{s}_{ij}$ obeys the following equation:\[
\dot{\vec{s}}_{ij}=\left[\vec{v}_{ij}^{T}+\left(\dot{\vec{s}}_{ij}\cdot\hat{\vec{t}}_{ij}\right)\dot{\hat{\vec{t}}}_{ij}\right]H\left(\left|\mu\vec{f}_{ij}^{N}\right|-k_{T}\left|\vec{s}_{ij}\right|\right),\]
which ensures that the spring is always tangent to the plane of contact
(a line in 2D), and  does not exceed the length that corresponds
to the Coulomb limit. In addition, $\vec{s}_{ij}$ is set to zero
if there is no overlap (i.e., $\xi_{ij}<0$). In practice, the tangential
displacement is obtained by integrating the relative tangential
velocity from the time the particles establish contact. When $\mu
\vec{f}_{ij}^{T}>\vec{f}_{ij}^{N}$, the length of the tangential spring is
kept fixed in order to prevent the frictional force from exceeding the Coulomb
limit. This  is important in order to avoid unrealistic
behavior when the contacts cease ``sliding'' (i.e., $f^{T}=\mu f^{N}$), and
revert to ``sticking'' ($f^{T}<\mu f^{N}$)~\cite{Brendel98}: if
the integration of the tangential velocity is continued for a sliding contact,
and the result used to further stretch the tangential spring, the result will be
an unrealistically large force when the contact reverts to sticking. Note that
the torques exerted on the particles determine their angular accelerations; the
corresponding equations are part of the system of equations that are solved. 

The interactions of the particles with the walls are similar to the
interparticle interactions. Since the walls considered here are rigid and fixed straight lines
(in 2D), the overlaps are calculated accordingly, with the same force models.
The force constants for particle-wall interactions
($k_{N}^{\mathrm{wall}},k_{T}^{\mathrm{wall}},\nu_{N}^{\mathrm{wall}},\nu_{T}^{\mathrm{wall}},\mu^{\mathrm{wall}}$)
may be specified to be different from those used for particle-particle
interactions (which are taken to be the same for all particle pairs).

In addition to interparticle forces and particle-wall forces, the force laws
account for gravity. External forces and torques may also be applied to
specific particles (see below).

\subsection{Simulation Parameters}
\label{sec:sim-params}
The systems studied here consist of collections of polydisperse disks, whose
radii are uniformly distributed in the interval $[R-\delta\cdot R,R]$.  In
order to facilitate the use of parameters from experiments performed on short
cylinders,
the cylinder thickness, $W$, and its volume density, $\rho$, are specified, 
and the
particle masses are given by $m_{i}=\pi R_{i}^{2}W\rho$.  We consider the case
of homogeneous disks, for which the axial moment of inertia is given by
$I_{i}=\frac{1}{2}m_{i}R_{i}^{2}$.  The parameters used in this work correspond
to experiments performed by the Duke group using 2D photoelastic
disks~\cite{GengPC}: $R=3.75\cdot10^{-3}\,\mathrm{m}$;
$W=6.6\cdot10^{-3}\,\mathrm{m}$; $\rho=1.15\cdot10^{3}\,\mathrm{kg/m^{3}}$.

The simulation results are presented in terms of non-dimensional quantities
defined as follows. The length unit is the mean particle radius $\bar{R}$, the
time unit is $\tau=\sqrt{\bar{R}/g}$ (where $g$ is the magnitude of the
acceleration of gravity) and the mass unit is the mean particle mass $\bar{m}$.
In simulations which account for gravity, we use $g=9.8\mathrm{m/s^{2}}$.

The normal spring constant used for particle-particle interactions is
$k_{N}=3000\bar{m}g/\bar{R}$. This value is based on force-displacement
measurements performed on the photoelastic disks mentioned
above~\cite{Majmudar07S}. Although the particles are cylindrical, measurements
showed a force proportional to the $\frac{3}{2}$ power of the displacement
(possibly indicating that the contact area is elliptic rather than
rectangular). For the typical deformations obtained in the experiments, a
linear fit to the particle force law provides quite a good
description~\cite{Majmudar07S}, which further justifies our choice of modeling
the normal force as that exerted by a linear spring with the above mentioned
effective spring constant. For particle-wall interactions, we use
$k_{N}^{\mathrm{wall}}=2k_{N}$.

The damping coefficient is typically chosen to correspond to half the critical
damping value for an individual contact, $\nu_c=2\sqrt{k}$ (recall that $k$ and
$\nu$ are non-dimensional), i.e., we use $\nu_{N,T}=\sqrt{k_{N,T}}$.  This
value was found to produce the fastest relaxation of the system towards static
equilibrium and is irrelevant in the state of mechanical equilibrium itself.

The time step used in the simulations was \mbox{$\Delta
  t=5\cdot10^{-4}\tau=5\cdot10^{-4}\sqrt{\bar{R}/g}$}.  We verified that
decreasing the time step did not affect the results of the simulations.

Other simulation parameters such as the ratio $k_{T}/k_{N}$ (and
$k_{T}^{\mathrm{wall}}/k_{N}^{\mathrm{wall}}$, which is taken to
equal  it), the coefficients of friction $\mu,\mu^{\mathrm{wall}}$
and the polydispersity $\delta$ were varied in different runs, and
their influence is discussed below.

\subsection{Initial Conditions and Simulation Procedure}
\label{sec:simulation-procedure}
The initial configuration is produced by placing the particles on a triangular
lattice with lattice constant $2R$. The side walls and the floor are placed at
a distance $R$ from the centers of the particles closest to the walls (so that
for a monodisperse packing, $R_i=R$, these particles are tangent to the side
walls and the floor). The initial (translational and angular) velocities for
all particles are set to zero. The system is then allowed to relax under
gravity until it reaches static equilibrium, as described below.

In order to study the response to a localized force, the resulting
configuration (with all velocities, translational and angular, reset to zero)
is used as an initial configuration for a second run of the simulation, in
which an additional external force (and/or an external torque in some
simulations) is applied to a particle at the center of the top row of
particles. The force is increased linearly in time from zero to its desired
value.  The system is then relaxed again to static equilibrium.

The main criterion used for testing whether the system has reached static
equilibrium (at which time the simulation is stopped) is the kinetic energy per
particle. We found that in order to obtain an accuracy of less than 1\% for the
forces on the floor in a given realization, which are used here to define
the response of the system to a localized force, the system has to be relaxed
to a kinetic energy of $E_{k}^{\mathrm{stop}}=10^{-13}\bar{m}g\bar{R}$ per
particle, which is significantly smaller than that used in previous studies (e.g.,
in~\cite{Landry04}, the systems are relaxed to
$E_{k}^{\mathrm{stop}}=2\cdot10^{-8}\bar{m}g\bar{R}$ per particle). For
polydisperse systems the response was coarse grained and averaged over a number
of different configurations (Sec.~\ref{sec:disorder}), since the forces exhibit
strong fluctuations. In this case, sufficient accuracy was obtained with
$E_{k}^{\mathrm{stop}}=10^{-9}\bar{m}g\bar{R}$. We verified that several other
criteria for static equilibrium (see~\cite{Atman03}) were satisfied: the
contact network was fixed for at least several hundred time steps (and each
particle had at least two contacts), there were no sliding contacts, and the
mean particle acceleration was less than $10^{-5}\bar{m}g$.
\begin{figure}[h!]
  \subfigure[Forces: gravity only.]{%
    \label{fig:sim-forces-gravonly}%
    \includegraphics[width=\hsize]{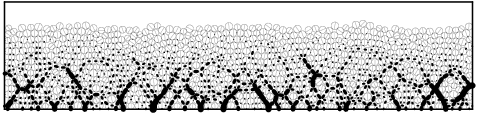}}
  \subfigure[Forces with a vertical applied force $F_{\rm ext}=5\bar{m}g$ at
    $\downarrow$ (gravity subtracted).]{%
      \label{fig:sim-forces-appforce}\includegraphics[width=\hsize]{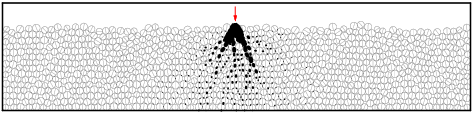}}%
\caption{Forces obtained in a DEM simulation of a polydisperse ($\delta=0.25$)
  frictional system with $k_{T}/k_{N}=0.8$, $\mu=\mu^{\mathrm{wall}}=0.2$. The
  line widths and lengths are proportional to the force magnitudes.
  \subref{fig:sim-forces-gravonly} After relaxation under gravity,
  \subref{fig:sim-forces-appforce} After relaxation with an additional vertical
  force $F_{\rm ext}=5\bar{m}g$, where $\bar{m}$ is the mean particle mass,
  with the forces obtained in~\subref{fig:sim-forces-gravonly} subtracted. The
  lines drawn inside the particles  indicate the rotation angles: in the initial configuration for the
  simulation (before the relaxation under gravity) the lines are vertical.
An arrow denotes the position of the externally applied force.}
\label{fig:sim-forces}
\end{figure}
For frictionless systems ($\mu=\mu^{\mathrm{wall}}=0)$, we found that reaching
such small energies in a reasonable simulation time was very difficult due to
the existence of slow global modes with a low dissipation rate, so that the
following algorithm was used for accelerating the relaxation: the MD simulation
was stopped at a higher energy
($E_{k}^{\mathrm{stop}}=10^{-6}\bar{m}g\bar{R}$).  The configuration obtained
at this stage was used as a reference state around which (since the
interactions are harmonic in the static case) the linearized equations of
motion were used iteratively to obtain a system which is closer to static
equilibrium. In each iteration the equations were solved by matrix inversion,
and the connectivity of the particles was updated to ensure that there was no
tension. The iterations were stopped when the maximal relative particle
displacement was less than $10^{-14}\bar{R}$ (which is essentially the
numerical accuracy). The typical kinetic energy obtained when using the
resulting configuration as an initial configuration for the full DEM simulation
was less than $E_{k}=10^{-22}\bar{m}g\bar{R}$ per particle. This configuration
was then used for calculating the interparticle forces.

An example of the interparticle forces obtained in a typical simulation run is
presented in Fig.~\ref{fig:sim-forces}. In order to calculate the response to an
applied force, the forces due to gravity alone (without the applied force) are
subtracted (vectorially) at each contact. 
\section{Response of Frictionless Slabs}
\label{sec:nofriction}
Consider first frictionless particles, interacting by unilateral
(``one-sided'') springs, which apply force only when compressed, modeling
repulsive-only particle interactions~\cite{Goldenberg02}, see Eq.~(\ref{eq:normal_force}). As shown
in~\cite{Goldenberg02}, the application of a localized force at the top of the
packing can lead to rearrangements in the contact network: horizontal springs
in the region below the point of application of the force are opened (as also
observed in~\cite{Luding97} for a pile geometry). The force chains in this
system are qualitatively similar to those obtained with regular (``two-sided'')
springs. However,  the force magnitudes vs.\ the horizontal coordinate at different
depths are expectedly  in better agreement with experiment in the case of one-sided springs~\cite{Geng01,Geng03} than that of  regular springs~\cite{Goldenberg02}, since the former 
model  is more realistic  for granular systems.  The stress
distribution obtained for unilateral springs~\cite{Goldenberg02} is 
anisotropic, and shows two peaks at the floor. This anisotropy is obviously
related to the existence of a region of open horizontal  contacts where the anisotropy
is large and the hyperbolic limit applies~\cite{Goldenberg02}. This
anisotropy is not present in the system without the application of the external
forces; it is induced (through the changes in the contact network) by the
applied force~\cite{Goldenberg02}. The induced changes in the contact network
obviously qualify as a nonlinear process.

These changes in the contact network may be modeled as a nonlinear extension of
the linear elastic continuum behavior obtained in a network of harmonic
springs. While the (possibly position-dependent) elastic moduli in linear
elasticity are time-independent material properties, a possible extension is to
introduce a stress history dependence of the elastic moduli (i.e., the
anisotropy induced by the opening of contacts in certain regions may be
considered a result of an attempted tensile stress in those regions).  A similar type of
stress-induced anisotropy has been suggested in the context of plastic models
of soil mechanics~\cite{Oda93} as well as in nonlinear elastic
models~\cite{Jiang07}.  If the particle positions do not change significantly,
so that only the contact network is modified in response to the applied stress,
the behavior can possibly be modeled as ``incrementally elastic''.  Under certain
conditions (corresponding to plastic yield), the system is no longer able to
support the applied stress without a major rearrangement of the particles.
Incipient plastic yield may  be related to a local extreme anisotropy
typical of a marginally stable isostatic configuration.

\subsection{Dependence on the Applied Force: Crossover from Hyperbolic-Like to
  Elliptic Response}
\label{sec:nofric-crossover}
\begin{figure}[h!]
\includegraphics[clip,width=\hsize]{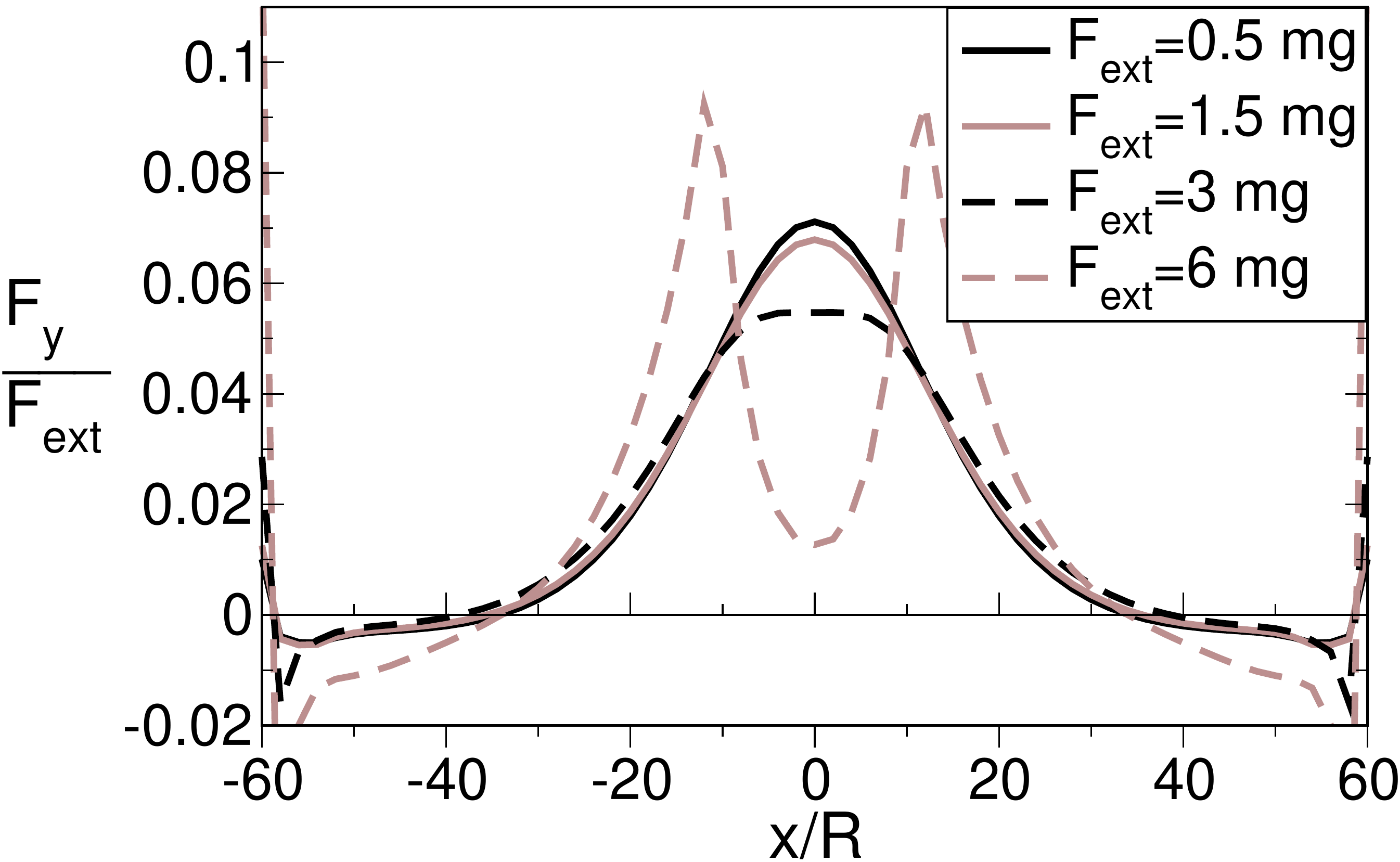}
\caption{The response of frictionless ordered
    systems for different applied forces, $F_{\rm ext}$.
}\label{fig:response-diff-fext-nofric}
\end{figure}

\begin{figure}[h!]
  \begin{tabular}{ccc}
    \subfigure[$F_{\rm ext}=1.5mg$.]{\includegraphics[clip,width=0.3\hsize]{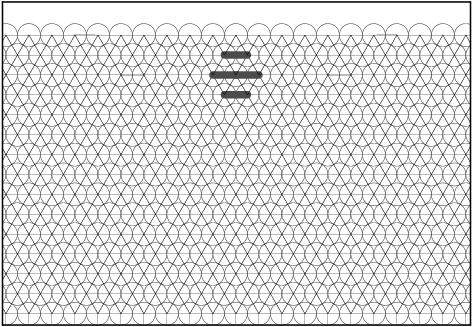}\label{fig:netdiff_f1.5}}&
      \subfigure[$F_{\rm ext}=3mg$.]{\includegraphics[clip,width=0.3\hsize]{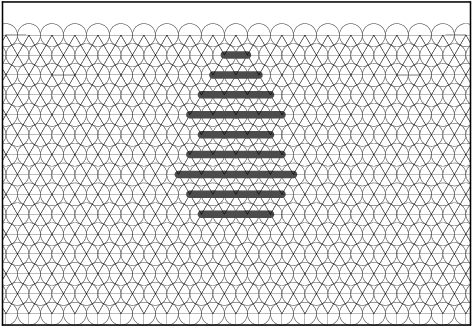}\label{fig:netdiff_f3}}&
      \subfigure[$F_{\rm ext}=6mg$.]{\includegraphics[clip,width=0.3\hsize]{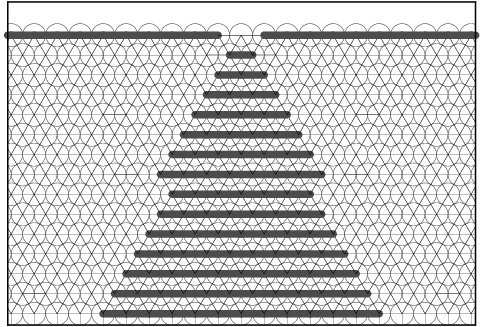}\label{fig:netdiff_f6}}
    \end{tabular}
    \caption{Changes in the contact network in an ordered frictionless system for
      different applied forces. The central third of the system is shown. Thick
      lines connecting particle centers indicate contacts opened due to the
      application of the force; thin black lines represent contacts which are
      unaffected by the external force.  }\label{fig:netdiff-frictionless}
\end{figure}

In order to examine the changes in the contact network in more detail, we
performed DEM simulations, as described in Sec.~\ref{sec:sim}, of a system
similar to those discussed in~\cite{Goldenberg02}, with different applied
forces. We focused on systems of 15 layers of 60 particles each; the effect of
system size is discussed below. The force response on the floor (subtracting
the effect of gravity) is shown in Fig.~\ref{fig:response-diff-fext-nofric}. A
crossover from a single peaked to a double peaked response occurs as the applied
force is increased.
The changes in the contact network corresponding to the systems of
Fig.~\ref{fig:response-diff-fext-nofric} are shown in
Fig.~\ref{fig:netdiff-frictionless}.  For a sufficiently small force (not
shown), the contact network is unchanged, and the response is fully elastic
(see Sec.~\ref{sec:friction} for a discussion of the linearity of the
response).  As the force is increased, horizontal contacts are opened in a
teardrop shaped region below the point of application of the force, whose size
increases with the force and decreases with friction. As mentioned above, in
this region the extreme anisotropic limit of elasticity applies. When the
``teardrop'' is sufficiently far from the floor
[Fig.~\ref{fig:netdiff_f1.5}-\subref{fig:netdiff_f3}], the response at the
floor is single peaked, as the changes induced by the force can be considered
to be localized. Otherwise, the anisotropy induced by the external force
reaches the floor (Fig.~\ref{fig:netdiff_f6}, inducing a double peaked response
(the crossover actually occurs at slightly smaller forces, at which the
``teardrop'' almost reaches the floor).
\begin{figure*}
\begin{tabular}{cc}
\subfigure[Depth $L$.]{\includegraphics[clip,width=0.45\hsize]{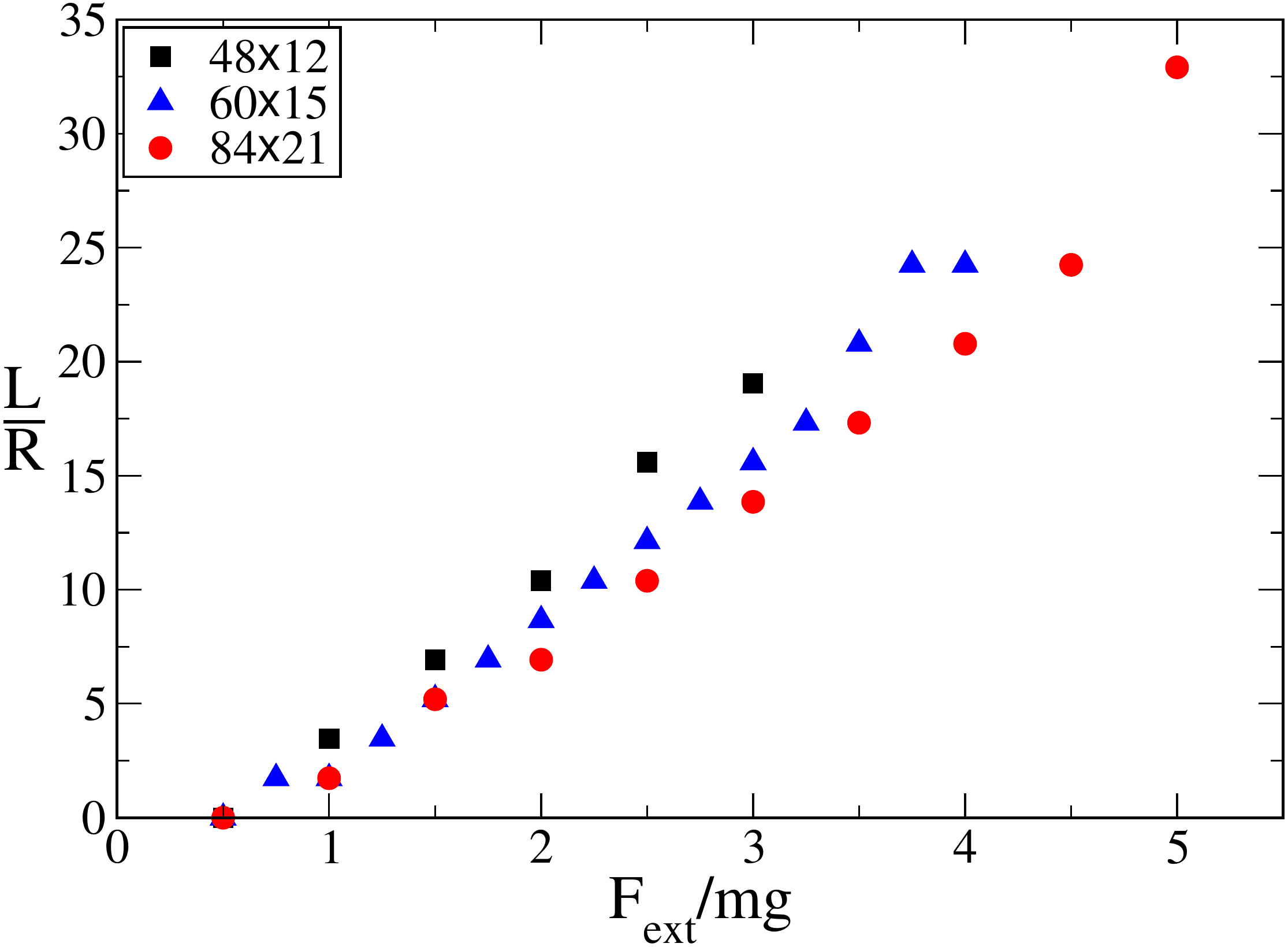}\label{fig:teardrop_length}}&
\subfigure[Maximum width $W$.]{\includegraphics[clip,width=0.45\hsize]{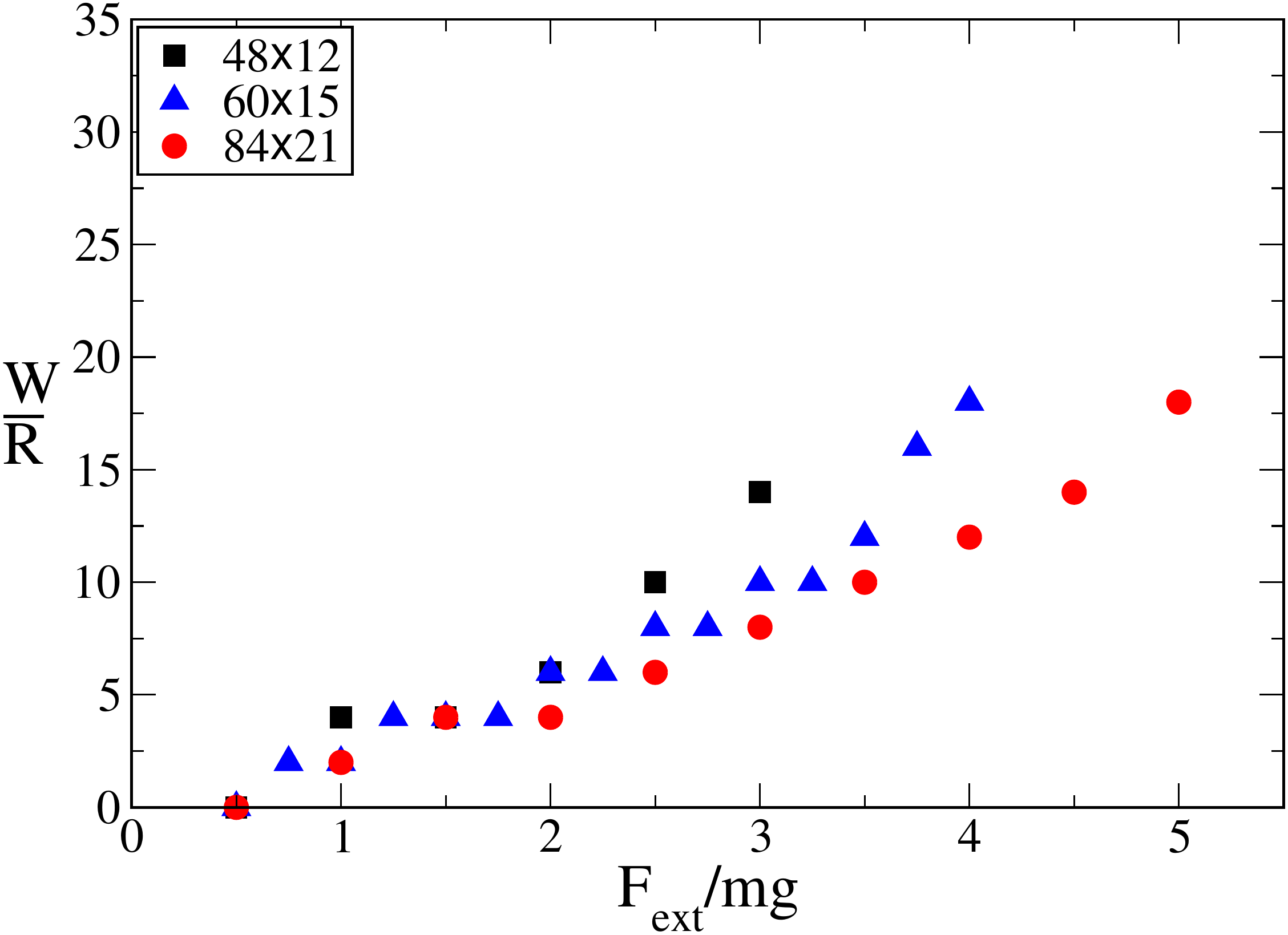}\vspace{-0.04in}\label{fig:teardrop_width}}
\end{tabular}
\caption{The depth and maximal width of the
  ``teardrop'' in units of the particle radius, $R$, 
 in ordered (monodisperse)  frictionless systems
  (Fig.~\ref{fig:netdiff-frictionless}), as a function of the applied force,
normalized by the  weight of a particle,
for different system sizes.
  \label{fig:teardrop_length_width}}
\end{figure*}

\subsection{Dependence on System Size}
\label{sec:nofric-syssize}

The dependence of the crossover on the size of the system is important for the
interpretation of experimental results. The depth and maximal  width of the
``teardrop'' as a function of the applied force, for several system sizes, are
plotted in Fig.~\ref{fig:teardrop_length_width}.  As the ``process'' of opening
of contacts with increasing applied force is nonlinear (since the effective
elastic properties of the system are  modified in the process), it is difficult to model
it analytically. However, Fig.~\ref{fig:teardrop_length_width} clearly
demonstrates that the size of the ``teardrop'' only slightly  decreases as the  depth of the
slab increases. 

This indicates that for large systems (as typically found in nature and in
engineering application), this change in the contact network gives 
a finite size correction to the response~\cite{Goldenberg05}. For identical external loads,
very deep systems should exhibit a single peaked response whereas relatively
shallow ones, as studied in some experiments
(Sec.~\ref{sec:response-experiments}), should exhibit two peaks. This is the
main reason that different experiments, using differently sized samples, yield
qualitatively different results. This observation also applies when frictional
interactions are accounted for, as described in Sec.~\ref{sec:friction}.

\subsection{Dependence on Particle Stiffness}
\label{sec:nofric-stiffness}
It is clearly important to consider the effect of the stiffness
of the particles. We performed simulations with different values of $k_{N}$,
and found that the size of the ``teardrop'' did increase with the stiffness
of the particles, but appeared  to saturate for $k_{N}\gtrsim 2000mg/R$. The 
results presented here pertain to $k_{N}=3000mg/R$. The response is also quite insensitive to the
choice of stiffness for $k_{N}\gtrsim 1000mg/R$: at $x=0$, it only changes by
about $2\%$ in the range \mbox{$1000mg/R < k_{N} < 10^5 mg/R$}. We conclude
that the crossover from a single peaked to a double peaked response is essentially
independent of the particle rigidity (provided that the particles are
sufficiently stiff; otherwise the particle ``overlaps'' may be appreciable and
the localized force would lead to significant rearrangements of the particles).

The independence on the particle stiffness may be understood as follows: in the
reference configuration, contacts are compressed due to gravity as well as the
rigid walls and floor, i.e., the stress, and in particular its component
parallel to the horizontal contacts, $\sigma_{xx}$, is compressive (the same
condition may of course be obtained by applying an external pressure to the
system). The external force acts in the opposite direction, i.e., it attempts
to give rise to tensile forces.  Therefore, the opening of contacts is due to a
``competition'' between the compression due to gravity and the tensile (excess)
forces due to the applied force, and is therefore determined directly by the
stress, rather than the strain (which does depend on the particle stiffness, of
course), provided that the geometry is not significantly affected by the
applied force (as it would for particles which are sufficiently soft to allow
significant overlap). This justifies our choice of the particle weight as the
unit of force (rather than a scale based on the particle stiffness and size).
The same behavior may persist even in the limit $k_{N}\to \infty$ under the
same boundary conditions provided that the particles in the reference state are
still in contact along the horizontal direction. However, as the stiffness
increases, this requirement may be hard to comply with, as the particles must
fit exactly between the walls in order to remain in contact: in the limit
$k_{N}\to \infty$ most horizontal contacts may be absent already in the
reference state, rendering it isostatic (therefore, extremely anisotropic), and
this would lead to a hyperbolic-like response even for an infinitesimal applied
force, as suggested in,
e.g.,~\cite{Tkachenko99,Head01,Edwards01,Ball02,Blumenfeld04}. Therefore the
above `limit of infinite stiffness' should be understood as `large but finite'
stiffness else isostaticity comes into play. Furthermore, one must bear in mind
that frictionless particles are a rather artificial idealization in the context
of granular materials. Indeed, as we show in the next section, the presence of
friction affects the behavior of the system very significantly and, in
particular, renders it much less sensitive to the contact network.
\section{Effects of Friction}
\label{sec:friction}
\subsection{Dependence on the Applied Force: Friction Increases the Linear Range}
\label{sec:fric-linearity}
\begin{figure}
\includegraphics[clip,width=\hsize]{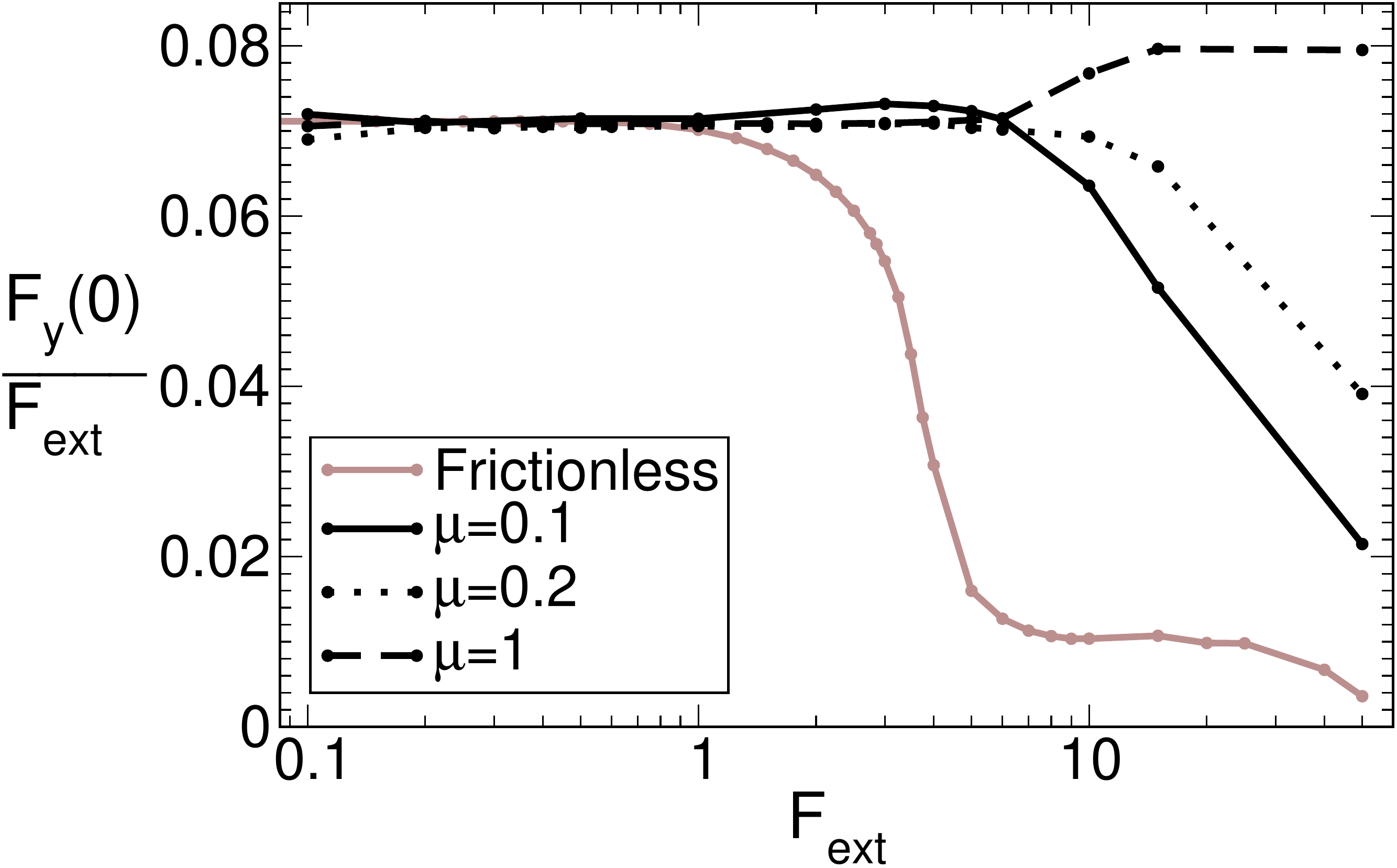}
\caption{The response at $x=0$ of ordered systems, normalized
  by the applied force, $F_{\rm ext}$, vs.\ the applied force (given in units of the particle weight,
  $mg$), for different coefficients of friction $\mu$,
  frictionless walls ($\mu^{\rm wall}=0$) and $k_{T}/k_{N}=0.8$.
\label{fig:linearity-friction}} 
\end{figure}

As mentioned in Sec.~\ref{sec:nofriction}, for sufficiently small external
forces no
changes are induced in the contact network, so that the response is expected to
be linear in the applied force~\cite{Goldenberg04}. This is shown in more detail in
Fig.~\ref{fig:linearity-friction}, which presents the response at the floor 
(vertical force on the floor) at
$x=0$ (below the point of application of the force) as a function of the
applied force, in both frictionless and frictional systems (with
$k_{T}/k_{N}=0.8$; the effect of this parameter is discussed below).  The
results shown here were obtained with frictionless walls ($\mu^{\rm wall}=0$;
see Sec.~\ref{sec:sim}), since for weak applied forces, frictional walls can
support some of the force, inducing nonlinearity in the response at the floor.
A linear ({\em elastic}) range is observed for a sufficiently small applied
forces {\em even for frictionless particles}, due to the fact that the particles
are slightly deformed by gravity, and a small force does not cause the contacts
to open.

As shown in Fig.~\ref{fig:linearity-friction}, friction has a significant
effect on the linearity of the response: The extent of the linear range is
significantly larger (by almost an order of magnitude; notice the logarithmic
scale on the horizontal axis in Fig.~\ref{fig:linearity-friction}) in
frictional systems, so that elasticity {\em is enhanced by friction}.

\begin{figure*}
  \begin{tabular}{cc}
      \subfigure[${\mu}=0.2$.]{\includegraphics[clip,width=0.45\hsize]{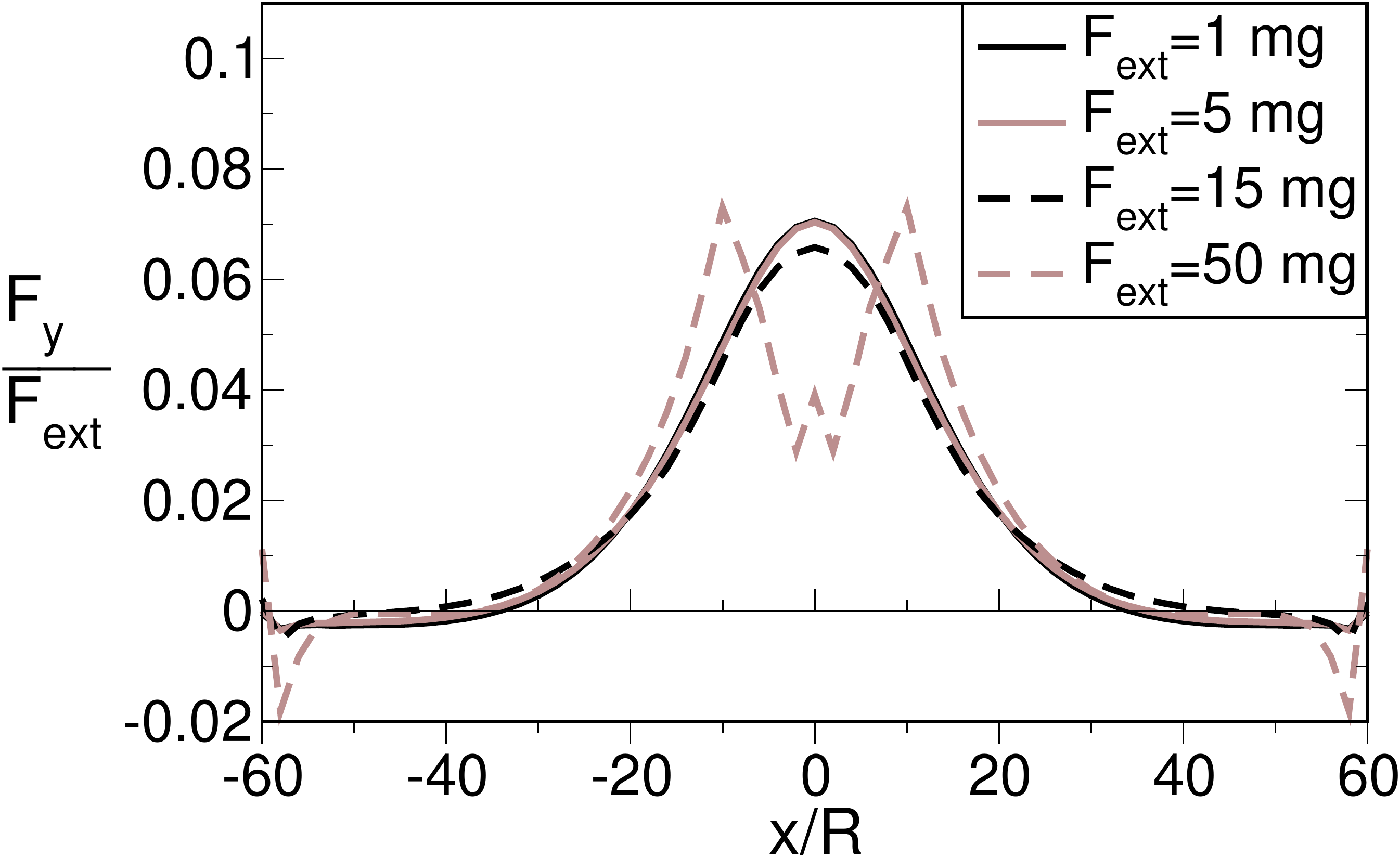}\label{fig:response-diff-fext-mu0.2}}&
      \subfigure[${\mu}=1$.]{\includegraphics[clip,width=0.45\hsize]{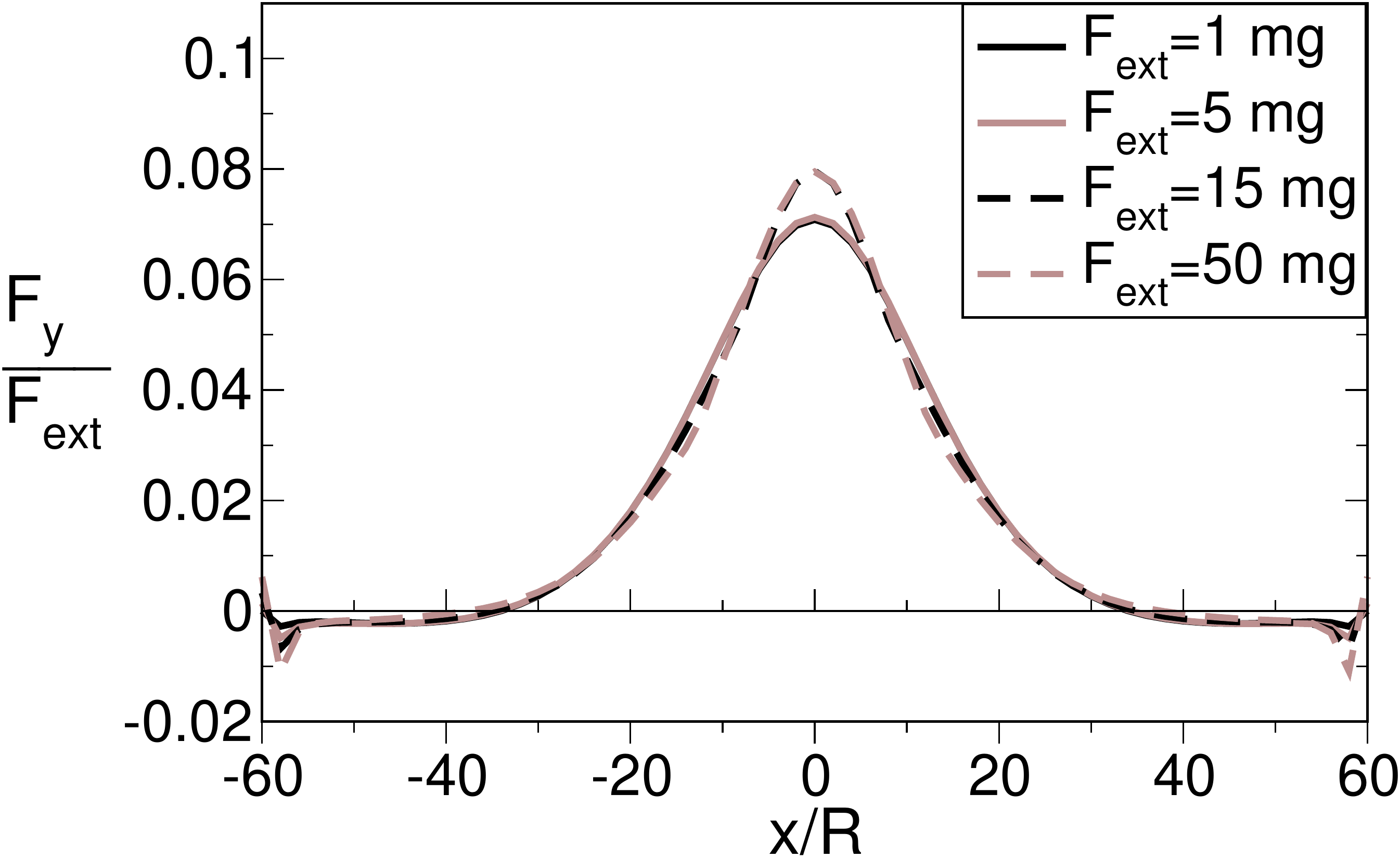}\label{fig:response-diff-fext-mu1}}
    \end{tabular}
  \caption{The response of ordered systems, with different coefficients
    of friction $\mu$, to  applied forces ($F_{\rm ext}$) of different
    magnitudes, frictionless walls ($\mu^{\rm wall}=0$) and
    $k_{T}/k_{N}=0.8$ (compare to Fig.~\ref{fig:response-diff-fext-nofric}).
    \label{fig:response-diff-fext-fric}}
\end{figure*}

The effect of friction on the response is shown in more detail in
Fig.~\ref{fig:response-diff-fext-fric}, which presents the response profile on
the floor for different applied forces, for $\mu=0.2$ and $\mu=1$ (compare to
the frictionless case presented in Fig.~\ref{fig:response-diff-fext-nofric}).
Notice that the force for  which the crossover from a single peaked to a double
peaked response occurs  rapidly
 increases with friction (this is also observed in
Fig.~\ref{fig:linearity-friction}), so that friction renders the response
closer to that expected from {\em isotropic} elasticity (this is further discussed  below).
For $\mu=1$ no such crossover is observed even
for the largest force shown, $F_{\rm ext}=50mg$ (a different type of crossover
occurs for $\mu=1$ for  large forces, as described in Sec.~\ref{sec:fric-largemu}).
Note that much larger forces may induce major rearrangements (i.e., plastic
flow), while our focus here is on the solid-like regime, in which the particle
displacements are small.

\subsection{The Effect of Friction on the Contact Network}
\label{sec:fric-contacts}
To gain a better understanding of the effect of friction on the response, we
examine the changes in the contact network. We first recall that such changes
occur at significantly larger applied forces in the frictional case (by almost
an order of magnitude): the first contact is opened when $F_{\rm ext}=0.75mg$
in the frictionless case, $F_{\rm ext}=4mg$ for $\mu=0.2$, and $F_{\rm
  ext}=6mg$ for $\mu=1$. This is the origin of the extended linear range
observed in frictional systems.
  
\begin{figure}
  \begin{tabular}{ccc}
    \subfigure[$F_{\rm ext}=6mg$.]{\includegraphics[clip,width=0.3\hsize]{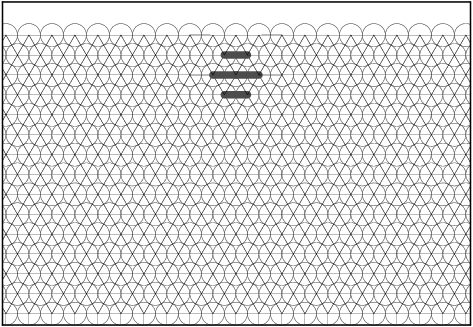}\label{fig:netdiff_f6_mu0.2}}&
    \subfigure[$F_{\rm ext}=15mg$.]{\includegraphics[clip,width=0.3\hsize]{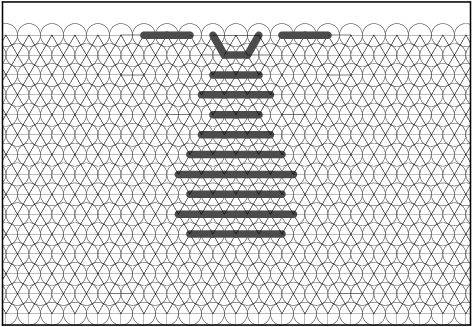}\label{fig:netdiff_f15_mu0.2}}&
    \subfigure[$F_{\rm ext}=50mg$.]{\includegraphics[clip,width=0.3\hsize]{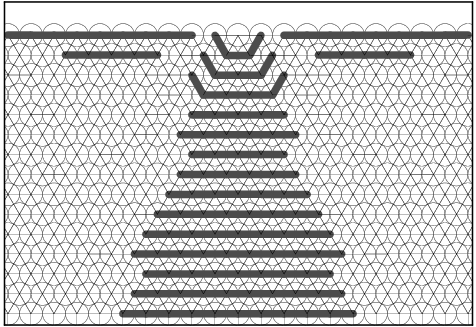}\label{fig:netdiff_f50_mu0.2}}
    \end{tabular}
    \caption{Changes in the contact network in a frictional
    system with $\mu=0.2$, with different applied forces $F_{\rm ext}$,
    frictionless walls ($\mu^{\rm wall}=0$) and $k_{T}/k_{N}=0.8$. The various lines are explained in the caption of Fig.~\ref{fig:netdiff-frictionless}.
    \label{fig:netdiff-mu0.2}}
\end{figure}

The effect of the applied force on the changes in the contact network in a
system with $\mu=0.2$ is shown in Fig.~\ref{fig:netdiff-mu0.2} (compare to
Fig.~\ref{fig:netdiff-frictionless}). For the same force, the region of open
contacts is considerably smaller in the frictional case than in the
frictionless case [compare Fig.~\ref{fig:netdiff_f6} and
Fig.~\ref{fig:netdiff_f6_mu0.2}]. In addition, this region reaches the floor
only for a much larger force [compare Fig.~\ref{fig:netdiff_f6} and
Fig.~\ref{fig:netdiff_f50_mu0.2}]. This may explain the fact that the crossover
to a double peaked response occurs at larger forces in the frictional case,
since this type of response is related
to the anisotropy in this domain. 

\begin{figure}[h!]
  \begin{tabular}{ccc}
    \subfigure[$F_{\rm ext}=6$.]{\includegraphics[clip,width=0.3\hsize]{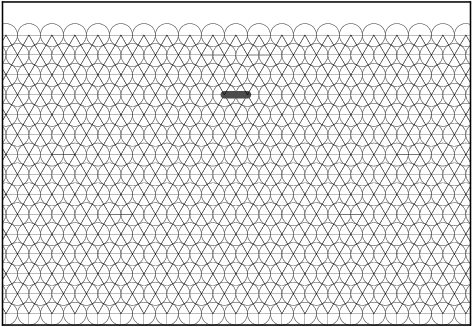}\label{fig:netdiff_f6_mu1_b}}&
    \subfigure[$F_{\rm ext}=15$.]{\includegraphics[clip,width=0.3\hsize]{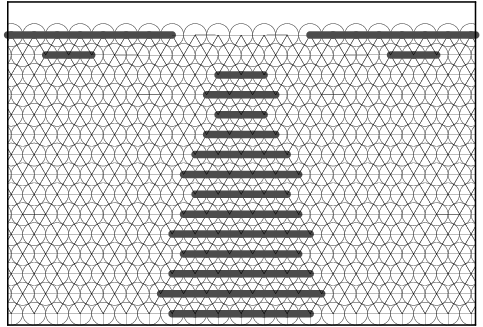}\label{fig:netdiff_f15_mu1}}&
    \subfigure[$F_{\rm ext}=50$.]{\includegraphics[clip,width=0.3\hsize]{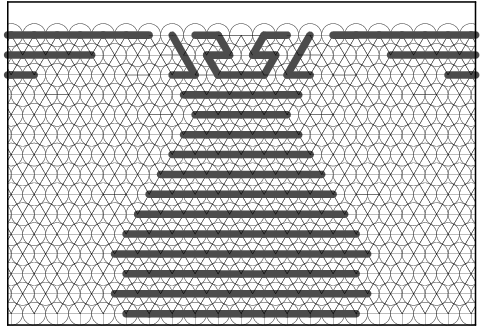}\label{fig:netdiff_f50_mu1}}
    \end{tabular}
    \caption{Changes in the contact network in a frictional
    system with $\mu=1$, with different applied forces $F_{\rm
      ext}$, frictionless walls ($\mu^{\rm wall}=0$) and
    $k_{T}/k_{N}=0.8$.
    \label{fig:netdiff-mu1}}
\end{figure}

The above observations of the changes in the contact network seem to explain
the effect of friction on the range of linearity and   magnitude of the crossover force.
However, the source of this effect, namely, the reason for the smaller changes
in the contact network in the presence of friction, still needs to be
elucidated (see below). In addition, for a larger coefficient of friction, the
obtained response does not seem to be compatible with the changes in the
contact network: as mentioned above [see
Fig.~\ref{fig:response-diff-fext-mu1}], for $\mu=1$ the response remains single
peaked (and actually its shape becomes sharper) even for rather large forces
(see however Sec.~\ref{sec:fric-largemu}), while for the same force, the
response for $\mu=0.2$ is double peaked. The changes in the contact network for
$\mu=1$ are shown in Fig.~\ref{fig:netdiff-mu1}. While the region of open
contacts does reach the floor (even for a smaller force than in the case of
$\mu=0.2$), the response remains single peaked.

\begin{figure}
\includegraphics[clip,width=\hsize]{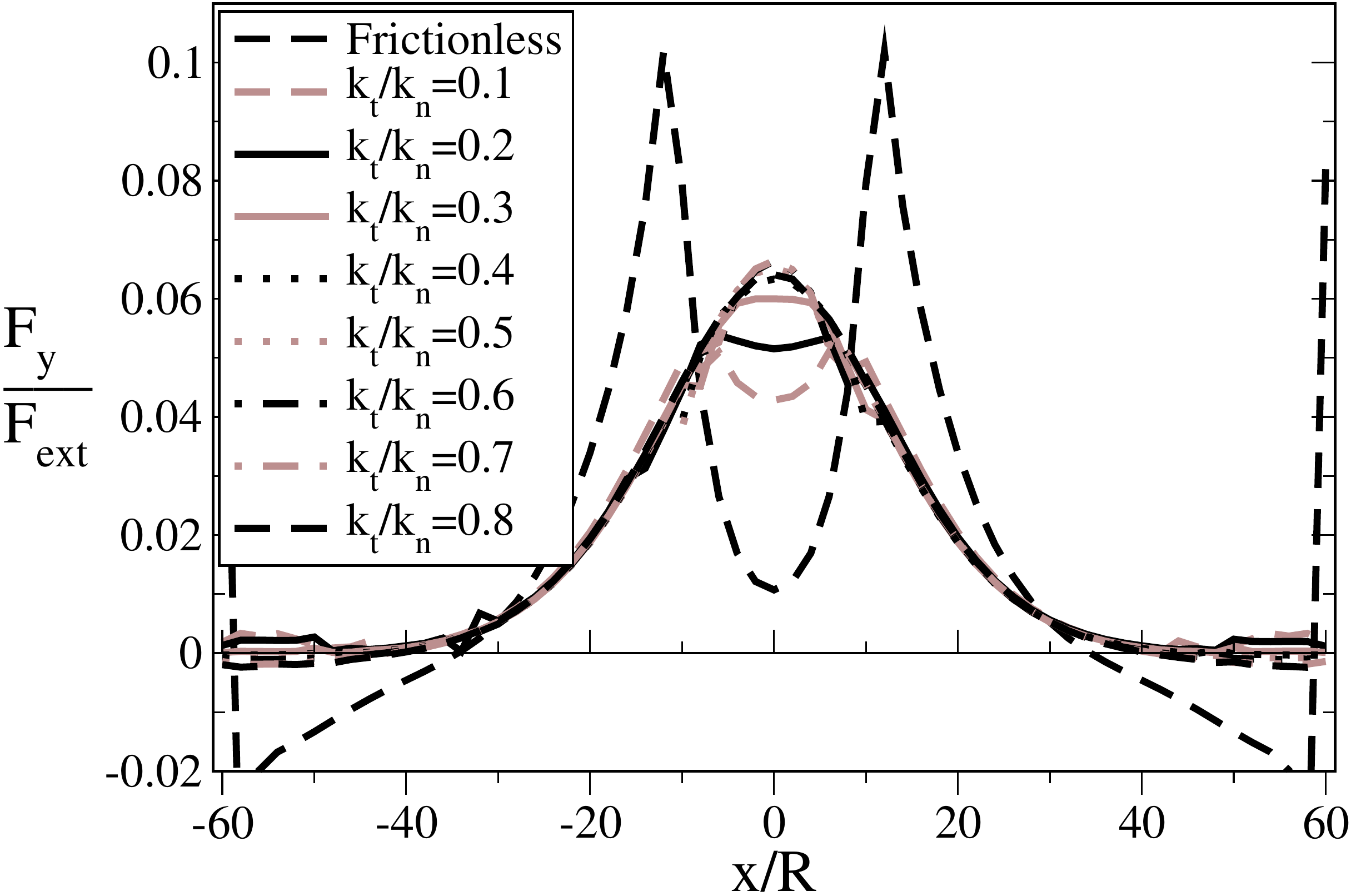}\caption{The response of ordered systems
  for a large coefficient of friction ($\mu=10$) and different values of the ratio
  of tangential to normal stiffness, $k_{T}/k_{N}$, with $F_{\rm ext}=15 mg$.
\label{fig:response-largemu-diffkt}} 
\end{figure}

The results obtained for $\mu=1$ suggest that in the presence of friction, the
anisotropy of the region of open contacts is greatly reduced (compared to the
frictionless case). When $\mu$ is sufficiently large (and $F_{\rm ext}$
sufficiently small), its value is essentially immaterial, since sliding is
never reached. In this case, the ratio of tangential to normal stiffness,
$k_{T}/k_{N}$ (see Sec.~\ref{sec:forcemodel}), determines the response:
Fig.~\ref{fig:response-largemu-diffkt} presents the response obtained with
$F_{\rm ext}=15 mg$ for $\mu=10$ (which is practically equivalent to
$\mu=\infty$, as used in~\cite{Kasahara04}, since no sliding occurs), with
different values of $k_{T}/k_{N}$.  A crossover from a single peaked to a
double peaked response occurs with {\em decreasing} $k_{T}/k_{N}$ (the
crossover occurs at $k_{T}/k_{N}\simeq 0.3$).  For larger values of
$k_{T}/k_{N}$, the response is nearly independent of $k_{T}/k_{N}$. This phenomenon
is further studied immediately below.

\subsection{A Model for Large Friction: the Role of Tangential Stiffness}
\label{sec:fric-model}
\begin{figure}
\includegraphics[clip,width=0.75\hsize]{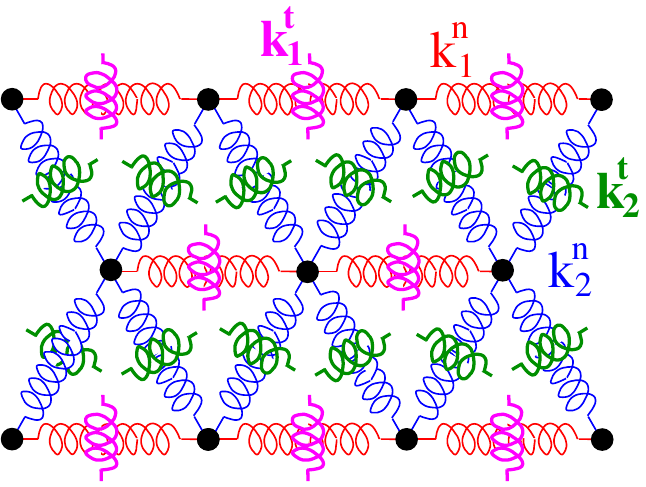}
\caption{A spring model with normal and tangential springs.
  \label{fig:normtan-2d-network}}
\end{figure}

In order to understand the dependence of the response on $k_{T}/k_{N}$, we
consider a 2D spring network model similar to the one described
in~\cite{Goldenberg02} (i.e., a triangular lattice with different spring
constants for horizontal springs, $k_1^n$, and oblique springs, $k_2^n$).  In
addition, this model (see Fig.~\ref{fig:normtan-2d-network}) contains
tangential springs (as used in our DEM simulations; Sec.~\ref{sec:forcemodel}),
with spring constants which are different for horizontal, $k_1^t$, and oblique
contacts, $k_2^t$~\cite{Goldenberg04}. Note that if one considers Hertzian
interactions, the existence of anisotropic prestress (e.g., due to gravity) may
indeed result in different spring constants in the horizontal and oblique
directions. This model does not incorporate sliding, so that it corresponds to
$\mu=\infty$.  Similar models were studied
in~\cite{Duffy57,Bathurst88,Chang92}, and more recently
in~\cite{Gay02,daSilveira02,Otto03}, where equal spring constants were used in
all lattice directions.

The elastic moduli corresponding to the long-wavelength limit of this model can
be calculated as follows: to leading order in the relative particle
displacements $\vec{u}_{ij}$ (which is the relevant order for linear
elasticity), the elastic energy of the system is given by:
\begin{equation}
  \label{eq:mic_elastic_energy_normtan}
  E^{\rm el}=\frac{1}{2}\sum_{<ij>} k^n_{ij}
  \left[\hat{r}_{ij}\cdot\vec{u}_{ij}\right]^2+
  k^t_{ij} \left[\vec{u}_{ij}
    -\left(\hat{r}_{ij}\cdot\vec{u}_{ij}\right)\hat{r}_{ij}\right]^2,
\end{equation}
where the sum is over nearest neighbors, and the values of $k^n_{ij}$ and
$k^t_{ij}$ are taken according to Fig.~\ref{fig:normtan-2d-network} and the
above description. In order to obtain the elastic moduli corresponding to the
continuum limit of the considered system, an affine deformation (which is
appropriate for a lattice configuration) is defined by a symmetric, uniform
strain field: \mbox{$u_{ij\alpha}(\vec{r},t)=\epsilon_{\alpha\beta}
  r_{ij\beta}$}.  Note that when tangential forces are present, the stress is
not necessarily symmetric (as assumed in classical elasticity); however, the
micropolar terms due to this asymmetry are of higher order in the strain. We
verified (in the DEM simulations) that they are very small, even near the point
of application of the external force, where the magnitude of the antisymmetric
part of the stress is only a few percent of the pressure.

Using the notation of~\cite{Otto03}:
\begin{equation}
  \label{eq:elastic_moduli_otto}
E^{\rm el}=\frac{1}{2}
  \left(\begin{array}{c} 
\displaystyle \epsilon_{xx}\\
\displaystyle \epsilon_{zz}\\
\displaystyle \epsilon_{xz}
\end{array}\right)^{\!\!\!\rm T}
\left(\begin{array}{ccc}
a&c&0\\
c&b&0\\
0&0&d\end{array}\right)
  \left(\begin{array}{c} 
\displaystyle \epsilon_{xx}\\
\displaystyle \epsilon_{zz}\\
\displaystyle \epsilon_{xz}
\end{array}\right),
\end{equation}
where the superscript, $\rm T$, denotes the transpose, one obtains the following
elastic moduli:
\begin{eqnarray}
  \label{eq:elastic_moduli_solution}
  a&=&\frac{R^2}{2A}\left(8k_1^n+k_2^n+3k_2^t\right)\\
  b&=&\frac{R^2}{2A}\left(9k_2^n+3k_2^t\right)\\
  c&=&\frac{R^2}{2A}\left(3k_2^n-3k_2^t\right)\\
  d&=&\frac{R^2}{2A}\left(6k_2^n+4k_1^t+2k_2^t\right),
\end{eqnarray}
where $A=2\sqrt{3}R^2$ is the area of the unit cell. These results are
consistent with those obtained in~\cite{Otto03} for normal springs only, but
different from the model with bending interactions introduced in~\cite{Otto03}.

In a domain  of open horizontal
contacts one has $k_1^n=k_1^t=0$. It was already shown in
Sec.~\ref{sec:nofriction} that in the absence of tangential forces
($k_2^t=0$), this system corresponds to the extreme anisotropic limit. However,
since the oblique tangential springs apply forces which have horizontal
components, they can (at least partially) compensate for the absence of normal
horizontal springs, and therefore significantly decrease the anisotropy.

Otto et al.~\cite{Otto03} present continuum elastic solutions for an
anisotropic infinite half plane subject to a localized force. They found a
criterion for the elastic moduli at which a crossover occurs from a single
peaked response to a double peaked one~\cite{Otto03}: two peaks are expected
for $r\equiv \frac{1}{bd}\left[ab-c(d+c)\right]<0$.  Note that this criterion
refers to an infinite half-plane rather than a slab of finite width, which is
more appropriate for describing our simulations. In addition, this model is
homogeneous, while the region of open contacts (even when it reaches the floor)
is roughly a triangle below the point of application of the force.  Nevertheless,
the estimate obtained using this criterion fits our result for the finite slab
quite well: for the model used here (with no horizontal springs),
$r=\frac{\beta^2+10\beta-3}{\beta^2+6\beta+9}$, where $\beta\equiv
k_2^t/k_2^n$. Hence (in the physically relevant range $\beta>0$), two peaks are
expected for $\beta\lesssim 0.2915$, which is consistent with the value
$k_{T}/k_{N}\simeq 0.3$ obtained in the simulation
(Fig.~\ref{fig:response-largemu-diffkt}).

For an infinitesimal tangential load applied to a system composed of elastic
spheres, the Cattaneo-Mindlin model~\cite{Johnson85} yields
$k_{T}/k_{N}=\frac{2(1-\nu)}{2-\nu}$, where $\nu$ is the Poisson ratio of the
spheres. For the range of positive Poisson ratios ($0\leq\nu\leq 0.5$; notice
that this is a 3D Poisson's ratio), this implies that $\frac{2}{3}\leq
k_{T}/k_{N} \leq 1$, so that the minimum value of $k_{T}/k_{N}$ is well above
the crossover from two peaks to one. In most of the simulations presented in
this work, we use $k_{T}/k_{N}=0.8$, which corresponds to a realistic Poisson
ratio of about 0.3.

\subsection{More on the effects of friction}

\begin{figure}
\includegraphics[clip,width=\hsize]{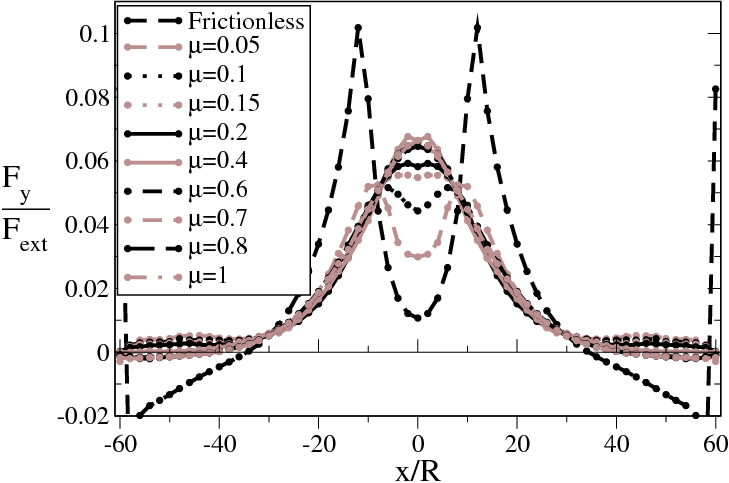}
\caption{The response of ordered systems for
  $k_{T}/k_{N}=0.8$ and  different values of the coefficient of friction, with
  $F_{\rm ext}=15 mg$.
\label{fig:response-krat0.8-diffmu}} 
\end{figure}

The above results show that static friction acts to retain an ``effective
connectivity'' between the grains when the horizontal contacts are
disconnected, and renders the system more isotropic than one would have naively
anticipated. However, the frictional forces are limited by the Coulomb
condition ($f^{T}\leq\mu f^{N}$; see Sec.~\ref{sec:forcemodel}), so that if
$\mu$ is too small (or $F_{\rm ext}$ large), sliding occurs. This increases the
anisotropy, since not all the tangential springs can exert forces as large as
predicted by the above model (in which $\mu=\infty$) and leads to a crossover
to a double peaked response.  This crossover is shown as a function of $\mu$
(for $F_{\rm ext}=15mg$) in Fig.~\ref{fig:response-krat0.8-diffmu}.

In considering the effect of the coefficient of {\em static} friction, note
that unlike the coefficient of {\em dynamic} friction, which is typically
smaller than $1$, the effective coefficient of static friction (which
determines the onset of sliding) may be significantly larger than $1$ even for
rough spherical particles, and certainly for irregularly shaped particles
(which can interlock and prevent their relative rotation).

\begin{figure}
\begin{tabular}{ccc}
\subfigure[$\mu=0$]{\includegraphics[clip,width=0.3\hsize]{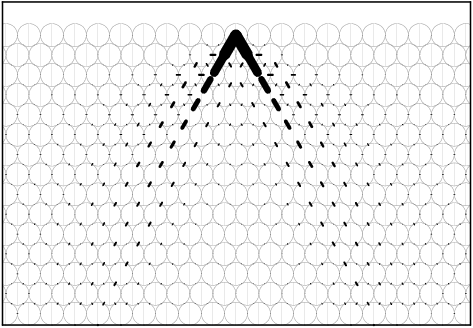}}&
\subfigure[$\mu=0.2$]{\includegraphics[clip,width=0.3\hsize]{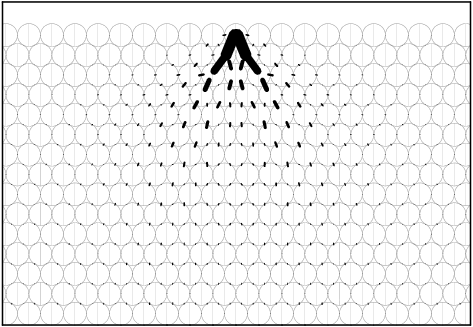}}&
\subfigure[$\mu=1$]{\includegraphics[clip,width=0.3\hsize]{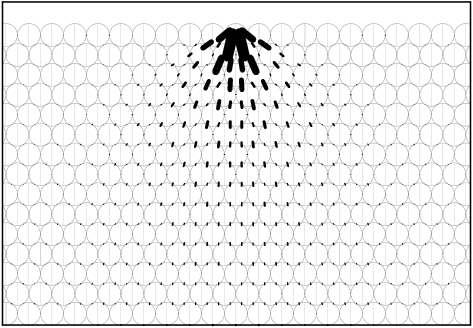}}
\end{tabular}
\caption{The interparticle forces in ordered systems with different
  coefficients of friction ($\mu^{\rm wall}=\mu,k_{T}/k_{N}=0.8$), with $F_{\rm
    ext}=15mg$.  The effect of gravity has been subtracted. The central third
  of the system is shown.  Line widths and lengths are proportional to the
  force magnitude.
\label{fig:forcechains-diffmu}}
\end{figure}

\begin{figure}
\begin{tabular}{ccc}
\subfigure[$\mu=0$]{\includegraphics[clip,width=0.3\hsize]{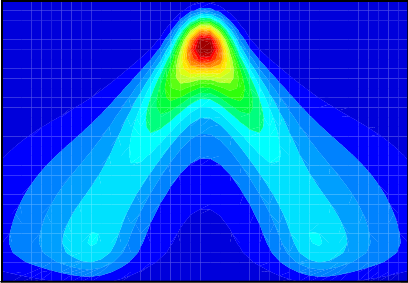}}&
\subfigure[$\mu=0.2$]{\includegraphics[clip,width=0.3\hsize]{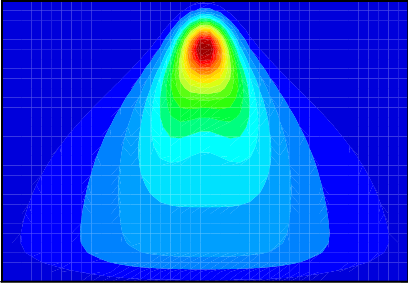}}&
\subfigure[$\mu=1$]{\includegraphics[clip,width=0.3\hsize]{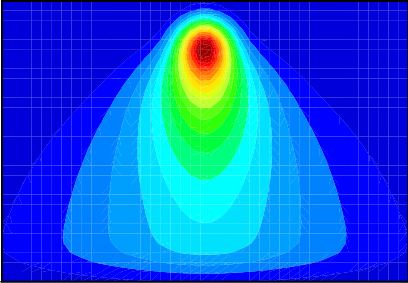}}
\end{tabular}
\caption{The vertical stress $\sigma_{zz}$ in the systems shown in
  Fig.~\ref{fig:forcechains-diffmu}, calculated using a Gaussian coarse
  graining function with a coarse graining width $w=2R$. The effect of gravity
  has been subtracted.
\label{fig:stress-diffmu}}
\end{figure}

The effects of friction are also evident in the forces
(Fig.~\ref{fig:forcechains-diffmu}) as well as the vertical stress component
$\sigma_{zz}$ (Fig.~\ref{fig:stress-diffmu}), calculated using the expression
presented in~\cite{Goldhirsch02,Goldenberg02}. In particular, the reduced
anisotropy is apparent in the stress field (the stress field for $\mu=1$ is
quite similar to that obtained in the case of an isotropic harmonic
lattice~\cite{Goldenberg02}).

\begin{figure*}
  \begin{tabular}{cc}
      \subfigure[${\mu}=0$.]{\includegraphics[clip,width=0.49\hsize]{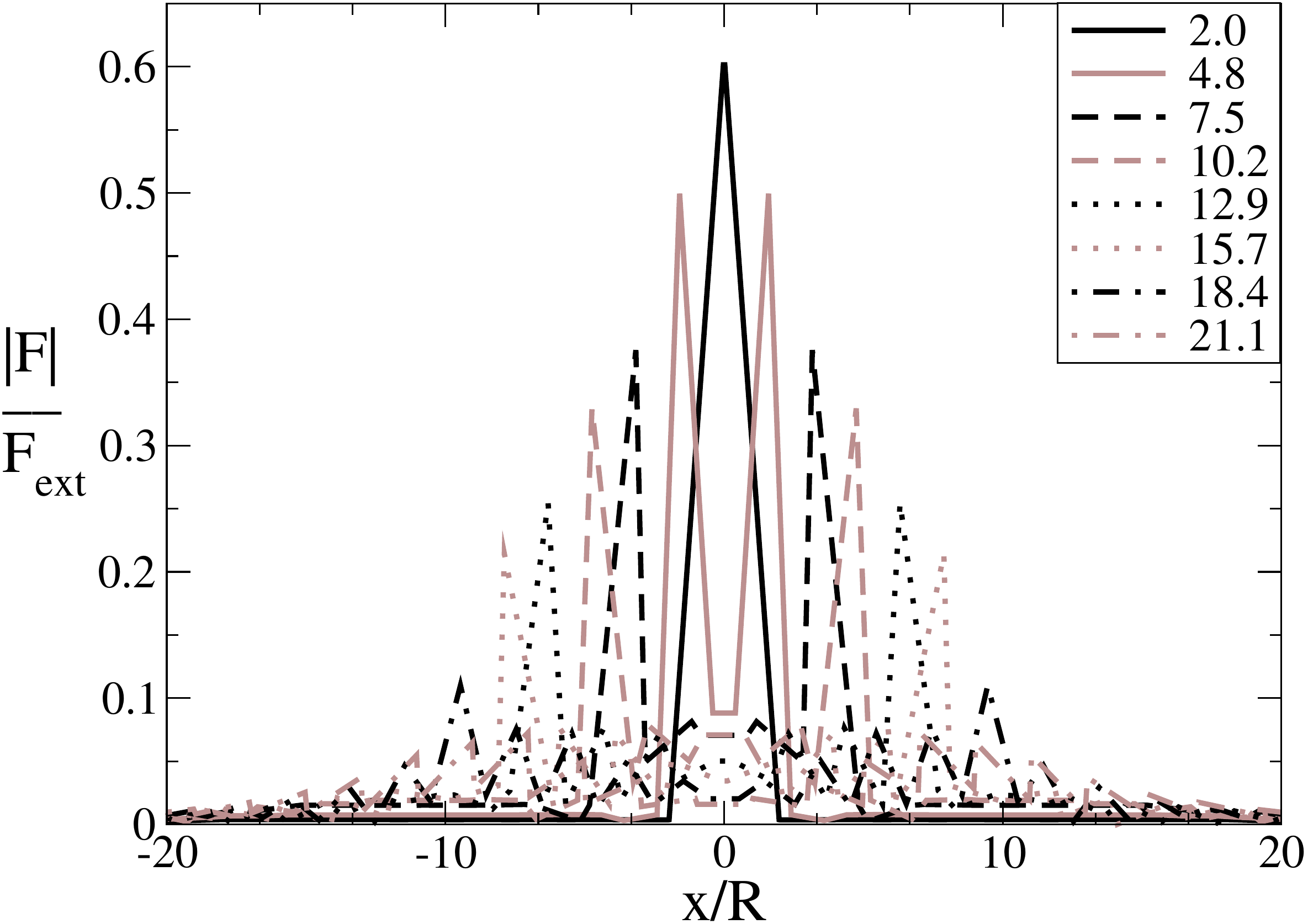}\label{fig:forces-depth-dem-nofric}}&
      \subfigure[${\mu}=1$.]{\includegraphics[clip,width=0.49\hsize]{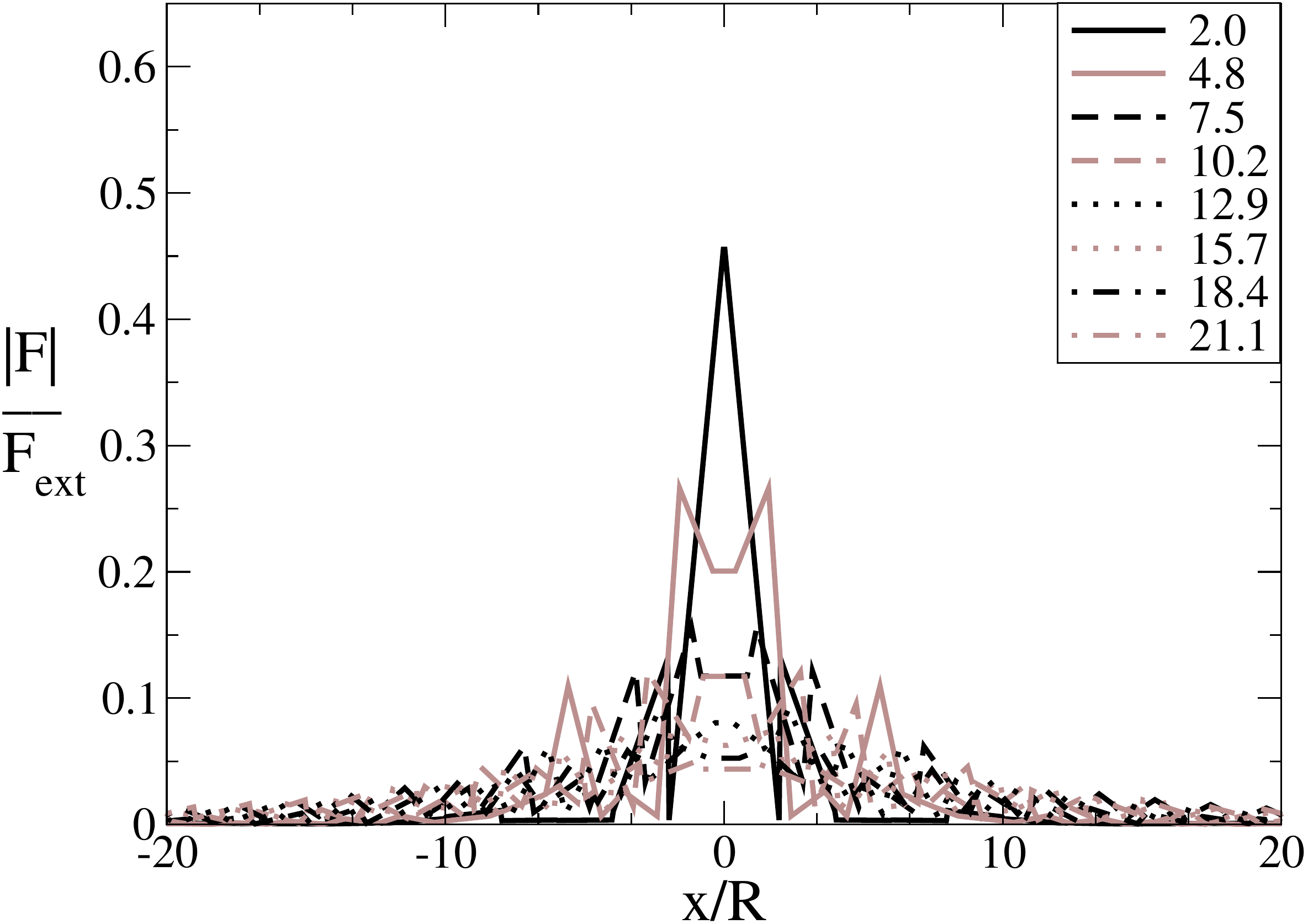}\label{fig:forces-depth-dem-mu1}}
    \end{tabular}
    \caption{The norms of the interparticle
      forces, $|\mathbf{f}|$, vs.\ the horizontal position, $x$, in ordered
      systems, with applied force $F_{\rm ext}=150mg$.  The legend indicates
      the depth measured from point of application of the force, in particle
      radii (compare to the experimental measurements shown
      in~\cite{Geng01,Geng03}).  \subref{fig:forces-depth-dem-nofric}
      Frictionless particles, \subref{fig:forces-depth-dem-mu1} Frictional
      particles with $\mu^{\rm wall}=\mu=1$, $k_{T}/k_{N}=0.8$.
\label{fig:forces-depth-dem}} 
\end{figure*}

The most prominent effect of friction on the force chains
(Fig.~\ref{fig:forcechains-diffmu}) is that the forces (and hence the chains)
no longer need to be aligned with the (here, triangular) lattice direction, as
they are in the absence of tangential forces. In particular, a vertical force
chain is evident for $\mu=1$. Such a chain is actually observed in the
experiment~\cite{Geng01,Geng03}. The inclusion of friction does provide a
better fit to the experimental results obtained in~\cite{Geng01,Geng03} for the
monodisperse ordered packing: the force magnitudes vs.\ the horizontal
coordinate at different depths in DEM simulations with frictionless and
frictional ($\mu=1$; for the particles used in the experiments,
$\mu\simeq 0.94$~\cite{GengPC}) are shown in Fig.~\ref{fig:forces-depth-dem}.
We have also been able to reproduce~\cite{Goldenberg04b} force chains
along non-lattice directions observed in experiments with an applied force at
oblique angles~\cite{Geng03}, using similar simulations of frictional particles
(with some polydispersity; see Sec.~\ref{sec:disorder} below).
\subsection{Symmetry Breaking for $\mu=1$}
\label{sec:fric-largemu}
\begin{figure*}
  \begin{tabular}{cc}
      \subfigure[${\mu}=0.2$.]{\includegraphics[clip,width=0.45\hsize]{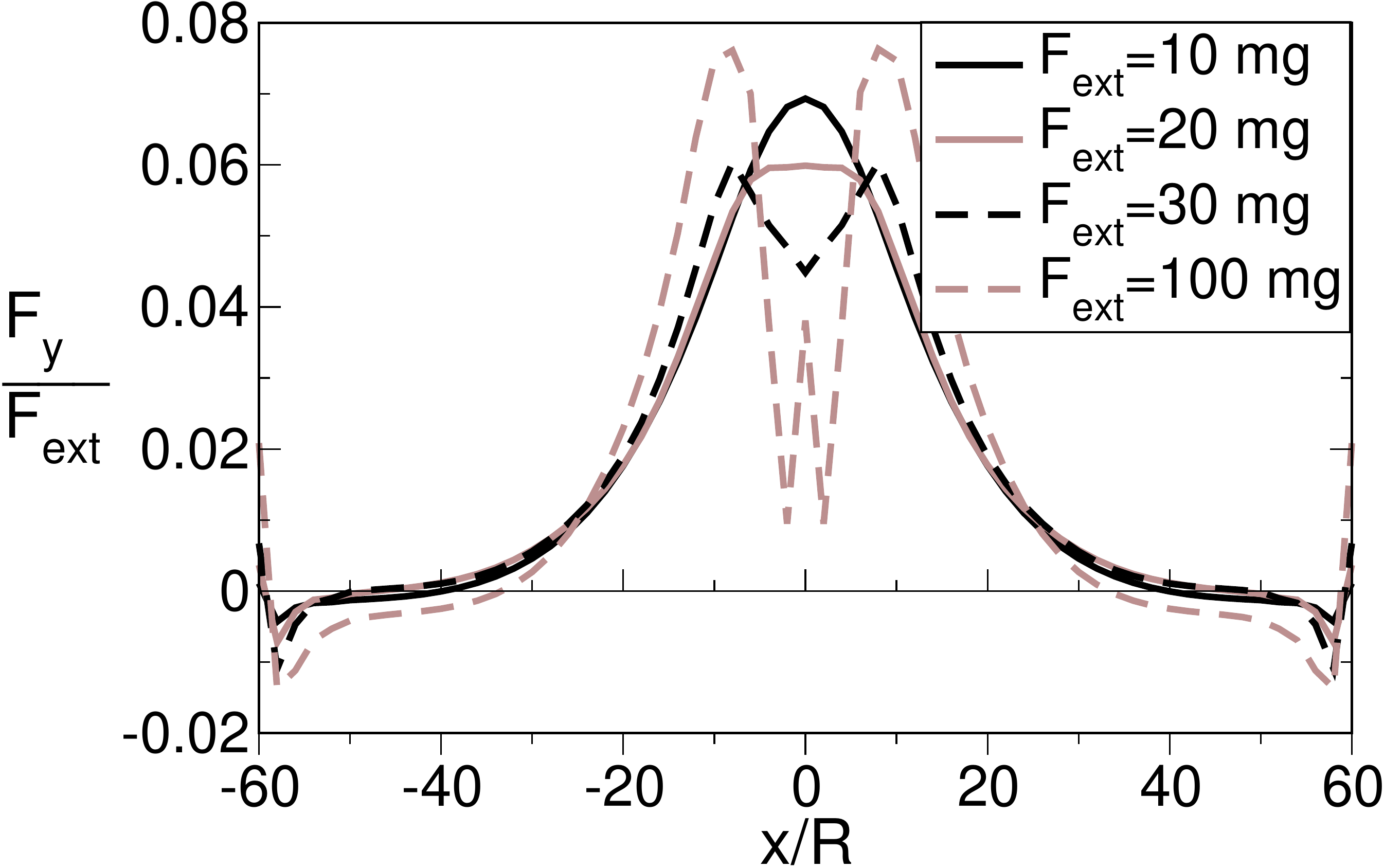}\label{fig:response-diff-fext-largef-mu0.2}}&
      \subfigure[${\mu}=1$.]{\includegraphics[clip,width=0.45\hsize]{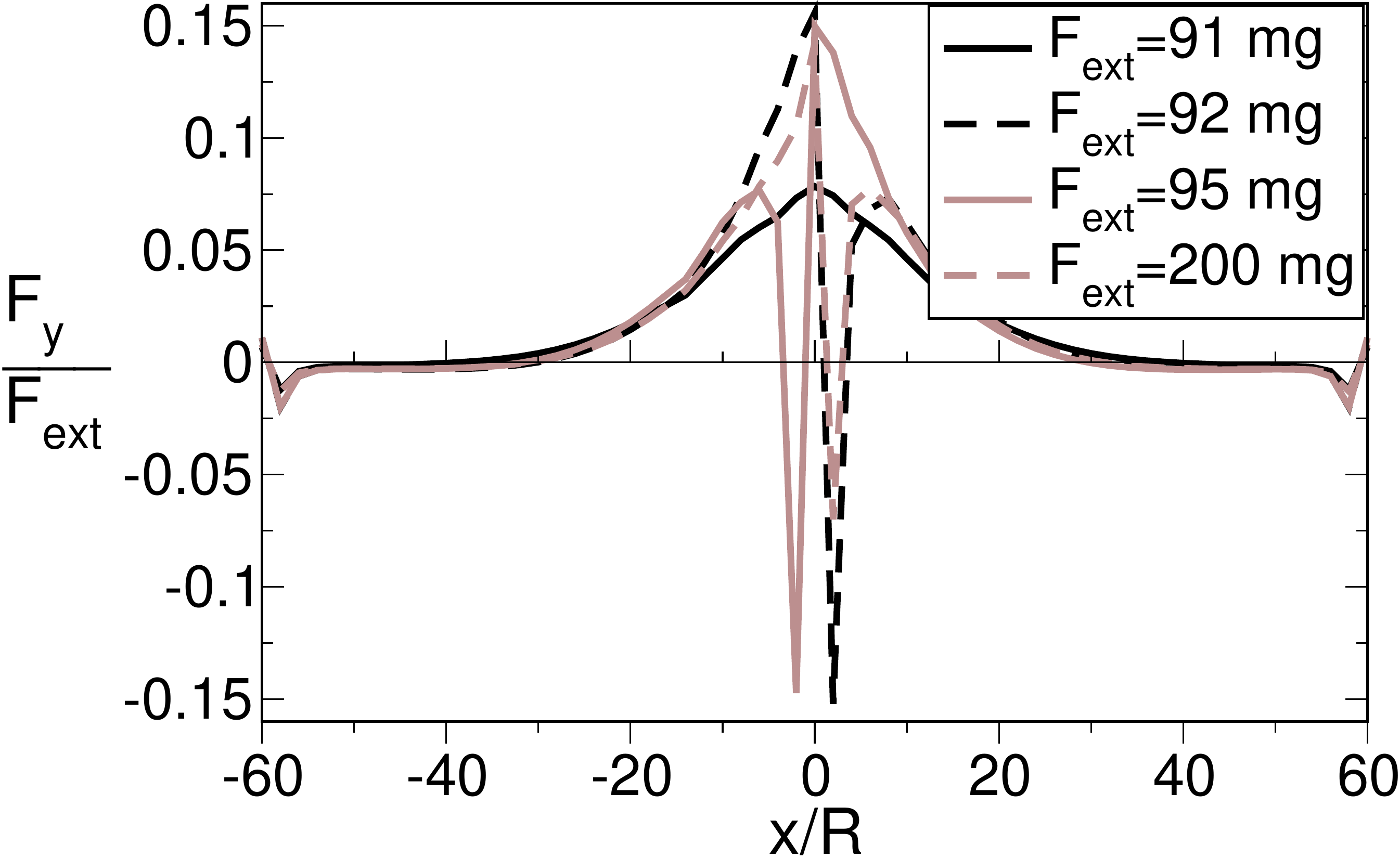}\label{fig:response-diff-fext-largef-mu1}}
    \end{tabular}
  \caption{The response of ordered systems, with different coefficients
    of friction $\mu$, to an applied force ($F_{\rm ext}$) of different
    magnitudes, frictionless walls ($\mu^{\rm wall}=0$) and
    $k_{T}/k_{N}=0.8$.
    \label{fig:response-diff-fext-largef-fric}}
\end{figure*}

As described above, for a relatively small coefficient of friction we observe a
gradual crossover from a single peaked to a double peaked response as a function of the 
magnitude of the externally applied force, much like in 
frictionless systems, the main difference being that it
occurs at larger values of the applied force [compare
Figs.~\ref{fig:response-diff-fext-nofric} and
\ref{fig:response-diff-fext-mu0.2}], although the shape of the response is not
identical. We reiterate that  for the frictional case, this crossover can be
explained as follows: in the limit $\mu \to \infty$, the induced anisotropy due
the changes in the contact network (the opening of contacts; see
Figs.~\ref{fig:netdiff-frictionless} and \ref{fig:netdiff-mu0.2}) is
compensated by the tangential forces, which renders the response single peaked
for $k_{T}/k_{N}\gtrsim 0.3$. For finite $\mu$, this compensation is limited by
the Coulomb condition, a limitation which becomes more significant as the
applied force is increased: for $\mu=0.2$, the region of open contacts (the
``teardrop'') reaches the floor for $F_{\rm ext}\simeq 20mg$, which in this
case seems to be the crossover force [see
Fig.~\ref{fig:response-diff-fext-largef-mu0.2}]. Sliding does occur in the
system (mainly near the point of application) even for smaller forces, and
sliding contacts spread gradually downward with increasing $F_{\rm ext}$
(possibly due to the fact that the displacements induced by the applied force
decay with the distance from it). It therefore appears that the crossover for
$\mu=0.2$ is associated with the gradual reduction of the compensating effect
of the tangential forces due to sliding.  For larger forces ($F_{\rm
  ext}\gtrsim 40mg$), a small third peak appear at $x=0$ (below the point of
application), which appears to be related to a reorganization of the sliding
contacts within the ``teardrop''.

For $\mu=1$, however, we observe quite a different behavior. As described in
Sec.~\ref{sec:fric-linearity}, the response remains single peaked (and even
becomes sharper) for rather large forces, much beyond those for which the
crossover is obtained for $\mu=0.2$ [see
Fig.~\ref{fig:response-diff-fext-mu1}].  However, when the force exceeds
$F_{\rm ext}= 91mg$, we obtain a sharp transition to a very different response
shape [see Fig.~\ref{fig:response-diff-fext-largef-mu1}], which is very
asymmetric (unlike the symmetric double peaked response obtained for lower
friction at large forces).  The sign of symmetry breaking appears to be related
to the transient vibrations which are excited by the application of the force;
for some very close values of the force [e.g., $F_{\rm ext}= 95mg$, shown in
Fig.~\ref{fig:response-diff-fext-largef-mu1}] we obtain almost exactly the same
rescaled response reflected with respect to the vertical axis through $x=0$.
The shape remains qualitatively similar for much larger $F_{\rm ext}$. The
transition to an asymmetric response is accompanied by a significant change in
the contact network (which becomes asymmetric), and by a marked {\em reduction}
in the number of sliding contacts.

This instability at large $\mu$ appears to be dominated by sliding, rather than
by changes in the contact network. Its source  (and in particular the
breaking of symmetry) is yet unclear, although it may be related to the
well-known phenomenon of shear banding. It also bears some similarities to
elastic buckling (however, it is clearly outside the linear elastic regime).
\subsection{A Pile Geometry}
\label{sec:fric-pile} 
While the work presented here is focused on the response of a granular slab, we
also examined the effect of friction in a pile composed of $11$ layers of
monodisperse disks arranged on a triangular lattice, prepared as described in
Sec.~\ref{sec:simulation-procedure} (similar to the frictionless piles studied
in~\cite{Luding97}). These piles are quite unrealistic, since the sides of the
pile are at $30^{\circ}$ to the horizontal, which is larger than the realistic
angle of repose for disks.  Therefore, the edges of the pile have to be
supported by side walls (the construction of stable piles on an
uncorrugated floor requires the introduction of {\em rolling friction}, or
rolling resistance~\cite{Zhou99}).

\begin{figure}
  \begin{tabular}{cc}
      \subfigure[Forces, ${\mu}=0$.]{\includegraphics[clip,width=0.45\hsize]{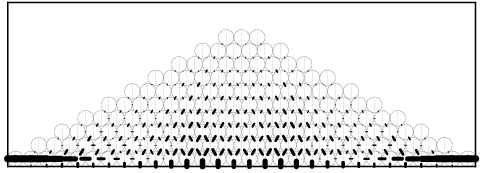}\label{fig:forces-pile-nofric}}&
      \subfigure[Forces, ${\mu}=0.94$.]{\includegraphics[clip,width=0.45\hsize]{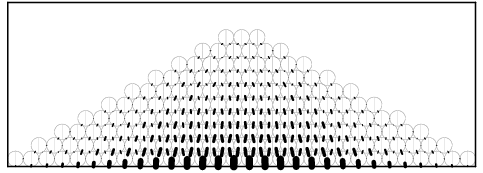}\label{fig:forces-pile-fric}}\\
      \subfigure[Vertical stress, ${\mu}=0$.]{\includegraphics[clip,width=0.45\hsize]{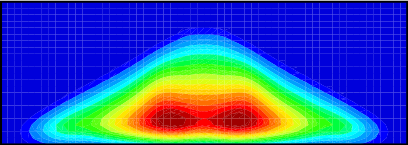}\label{fig:stress-pile-nofric}}&
      \subfigure[Vertical stress, ${\mu}=0.94$.]{\includegraphics[clip,width=0.45\hsize]{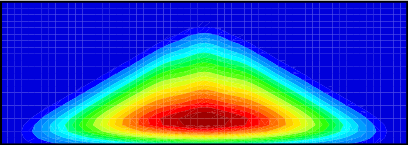}\label{fig:stress-pile-fric}}\\
      \subfigure[Contact network, ${\mu}=0$.]{\includegraphics[clip,width=0.45\hsize]{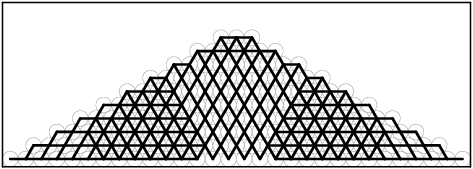}\label{fig:network-pile-nofric}}&
      \subfigure[Contact network, ${\mu}=0.94$.]{\includegraphics[clip,width=0.45\hsize]{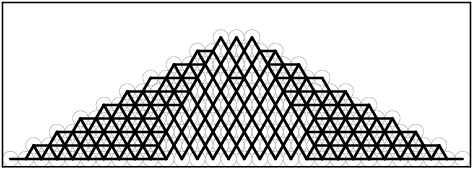}\label{fig:network-pile-fric}}
    \end{tabular}
    \caption{\subref{fig:forces-pile-nofric},\subref{fig:forces-pile-fric} the
      interparticle forces in a pile of monodisperse disks under gravity, with
      and without friction (in the frictional case, $\mu=0.94$, $\mu^{\rm
        wall}=0.35$ and $k_{T}/k_{N}=0.5$). Line widths and lengths are
      proportional to the force magnitude;
      \subref{fig:stress-pile-nofric},\subref{fig:stress-pile-fric} the
      vertical stress $\sigma_{zz}$ in the same systems, calculated using a Gaussian coarse
      graining function with a coarse graining width $w=2R$;
      \subref{fig:network-pile-nofric},\subref{fig:network-pile-fric} the
      contact network: particles in contact are connected by lines.
      \label{fig:pile}}
\end{figure}

The forces and the corresponding vertical stress component $\sigma_{zz}$ in the
pile are shown in
Figs.~\ref{fig:forces-pile-nofric}-\subref{fig:stress-pile-fric}. As in the
case of a slab (Figs.~\ref{fig:forcechains-diffmu}
and~\ref{fig:stress-diffmu}), the effect of friction is rather large. In
particular, as shown in
Figs.~\ref{fig:stress-pile-nofric},\subref{fig:stress-pile-fric}, the pile of
frictionless disks exhibits a dip in the stress under the apex, but the pile of
frictional disks does not. The dip in the frictionless case is due to the open
horizontal contacts in the central region of the pile; see
Fig.~\ref{fig:network-pile-nofric}. In the frictional case, the size of the
region of open contacts in the central region of the pile is quite similar, but
its shape is different
[Figs.~\ref{fig:network-pile-nofric},\subref{fig:network-pile-fric}]. The
absence of the dip results from the reduced anisotropy due to the frictional
forces, as discussed above for a slab geometry. We therefore conclude that the
presence or absence of a dip in a pile geometry is determined by the degree of
anisotropy in the {\em mechanical properties} of the pile (which is not always
simply related to the anisotropy of the contact network, as shown here in the
frictional case). These properties are expected to depend on the way the pile
is prepared, as observed in~\cite{Vanel99b}.
\subsection{Effects of an applied torque}
\label{sec:apptorque}
\begin{figure}
\begin{tabular}{ccc}
\subfigure[$M=0$]{\includegraphics[clip,width=0.3\hsize]{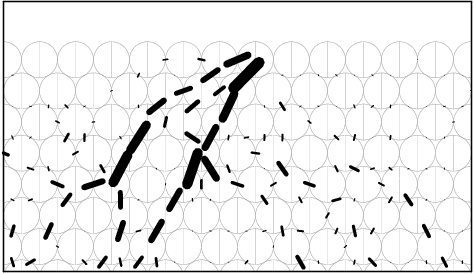}}&
\subfigure[$M=0.2F_{\rm ext}\bar{R}$]{\includegraphics[clip,width=0.3\hsize]{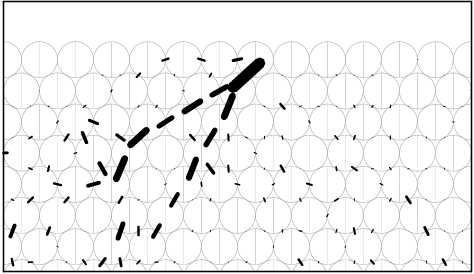}}&
\subfigure[$M=0.4F_{\rm ext}\bar{R}$]{\includegraphics[clip,width=0.3\hsize]{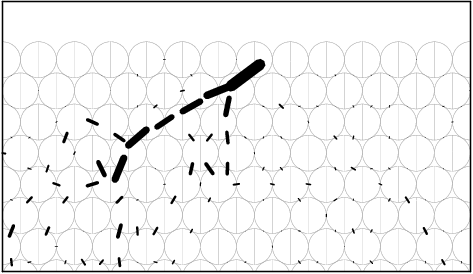}}
\end{tabular}
\caption{Force chains in 2D packings of slightly
  polydisperse frictional particles (\mbox{$\delta= 10^{-3}$}), with a force
  $F_{\rm ext}=10\bar{m}g$ applied at $30^\circ$ to the horizontal, and
  different applied torques in the clockwise direction (indicated below each
  figure). The effect of gravity has been subtracted. The same realization of
  the packing was used in all cases.  The region shown is the central third of
  the upper half of the system.
\label{fig:forcechains-difftorque}}
\end{figure}

It is interesting to note that the force chains are also sensitive to
additional applied torques. Fig.~\ref{fig:forcechains-difftorque} shows the
force chains obtained in systems of $25\times 13$ slightly polydisperse
particles (\mbox{$\delta= 10^{-3}$}) with $k_{T}/k_{N}=0.5$, $\mu=\mu^{\rm
  wall}=0.5$, for an applied force $F_{\rm ext}=10\bar{m}g$ at $30^\circ$ to
the horizontal, with different applied torques (as may occur if the line of action of
 an external
force applied to a particle does not pass through its center of mass).
As shown, additional torques can have quite a large effect on the obtained
force chains.  Although the effect on the stress field (in particular, its
asymmetric part) is typically confined to a region close to the point of
application, this may indicate that for some effects it is important to extend
the continuum description to incorporate torques and rotations, i.e., use a
micropolar or Cosserat continuum model~\cite{Kroner67,Jaunzemis67,Eringen68}).

\section{Effects of Disorder}
\label{sec:disorder}
While the study of ordered systems of monodisperse grains allows for simpler
theoretical modeling, real granular systems are never composed of perfectly
identical particles. Even disks of nominally equal sizes (as used, e.g.,
in~\cite{Geng01,Geng03}) exhibit a certain polydispersity. Most real granular
matter exhibits considerable polydispersity (as well as a variability in
particle shapes). It is therefore important to verify that the results obtained
for monodisperse system, as described above, are not limited to this
particular, idealized case. We therefore performed simulations with
polydisperse systems (with radii distributed uniformly in the interval
$[R-\delta\cdot R,R]$; see Sec.~\ref{sec:sim-params}, with $\delta=0.01, 0.1,
0.25$). In general, we find that polydisperse systems exhibit qualitatively
similar behavior to the ordered systems described in Sec.~\ref{sec:friction}.
The effects of polydispersity are described in detail in this section.

Since disorder induces fluctuations (which are not present in ordered
systems), the forces  were coarse grained using a Gaussian coarse-graining function
 in the horizontal
($x$) direction: $\frac{1}{w\sqrt{\pi}}e^{-(x/w)^2}$ with $w=3\bar{R}$ or
$w=6\bar{R}$, which amounts to calculating the  vertical stress
$\sigma_{zz}$ at the floor~\cite{Goldenberg06b}. The stress was then averaged over several
realization of the disorder (typically five).

\begin{figure}
  \begin{tabular}{cc}
\subfigure[${\mu}=0$.]{\includegraphics[clip,width=0.45\hsize]{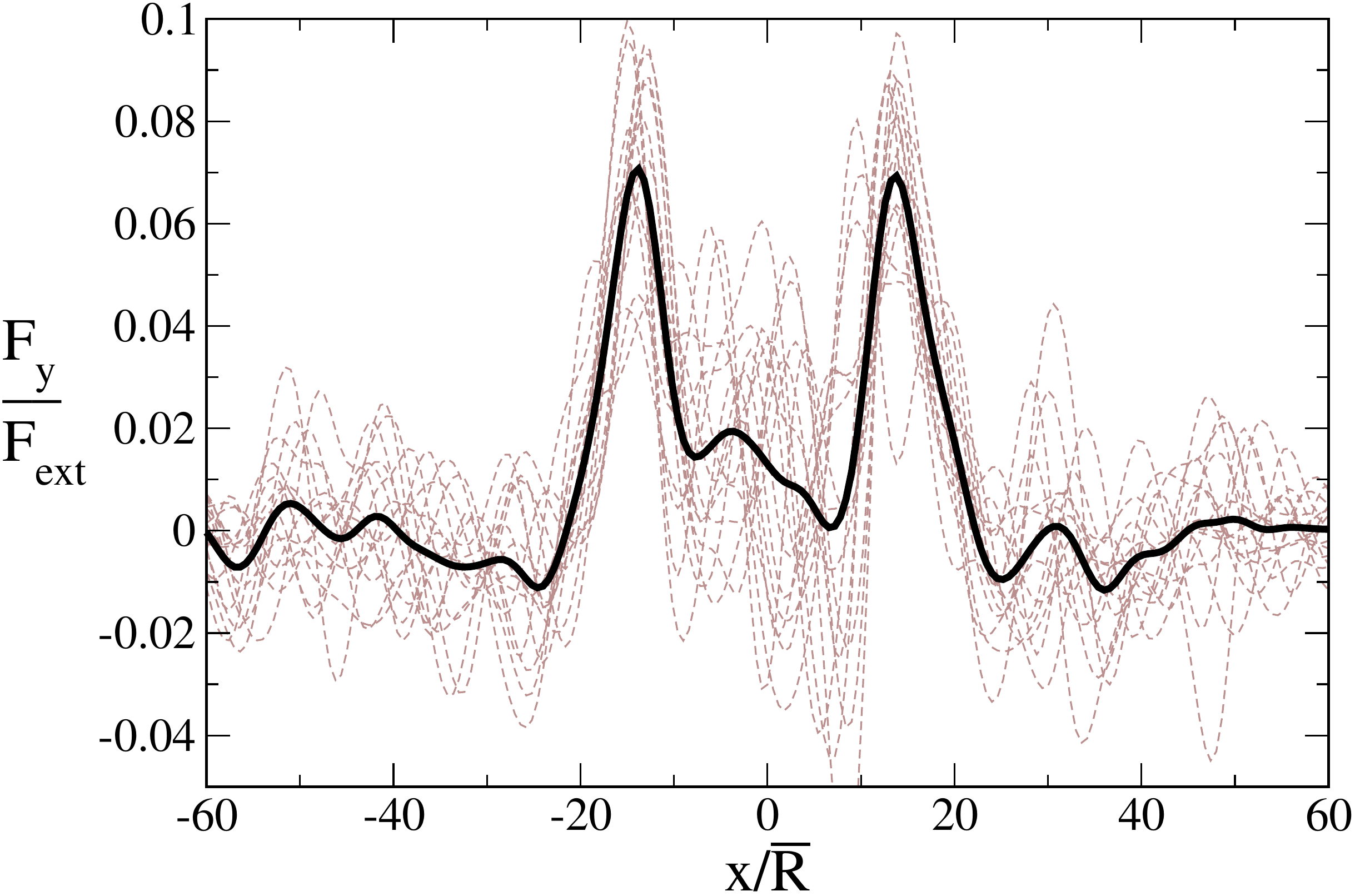}\label{fig:response-poly1p-nofric}}&
\subfigure[${\mu}=0.2$.]{\includegraphics[clip,width=0.45\hsize]{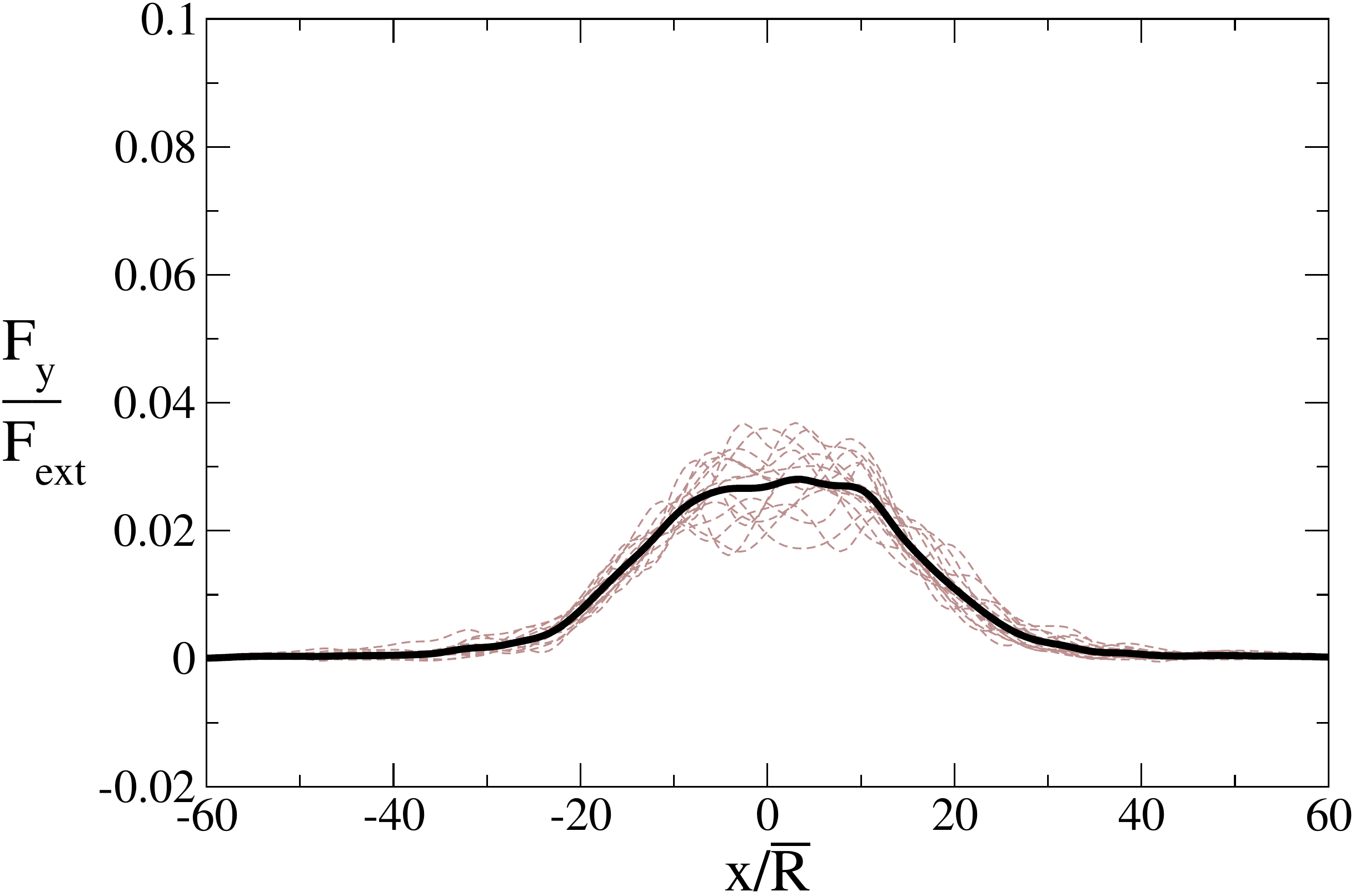}\label{fig:response-poly1p-mu0.2}}
    \end{tabular}
    \caption{The response of frictionless and frictional (\mbox{$\mu=\mu^{\rm
          wall}=0.2, k_{T}/k_{N}=0.8$}) disordered systems (with polydispersity
      \mbox{$\delta= 0.01$}) with $F^{\rm ext}=15\bar{m}g$ ($\bar{m}$ is the
      mean particle mass).  Thin gray lines correspond to the responses of 15
      individual realizations in each case, smoothed with a Gaussian of width
      $w=3\bar{R}$; the thick black line corresponds to an average over these
      15 configurations.
  \label{fig:response-poly1p-fricnofric}}
\end{figure}

The effect of friction on the response in disordered systems is qualitatively
similar to that found in the ordered lattices described in
Sec.~\ref{sec:friction}: Fig.~\ref{fig:response-poly1p-fricnofric} presents a
comparison of the response (averaged over 15 different realizations, coarse
grained with $w=3\bar{R}$) obtained in frictionless and frictional ($\mu=0.2$)
disordered packings with $\delta=0.01$, with an applied force $F^{\rm
  ext}=15\bar{m}g$ ($\bar{m}$ is the mean particle mass). As in the ordered
case, the response is very different: there are two very distinct peaks in the
frictionless case, but one peak in the frictional case. The fluctuations are
quite large even for this small degree of polydispersity (and larger for the
frictionless case), but for the magnitude of the applied force used in this
case the difference between the frictionless and frictional case is quite
evident even for individual realizations.
\subsection{Linearity and the Crossover Force}
\label{sec:disorder-linearity}
\begin{figure}
\includegraphics[clip,width=\hsize]{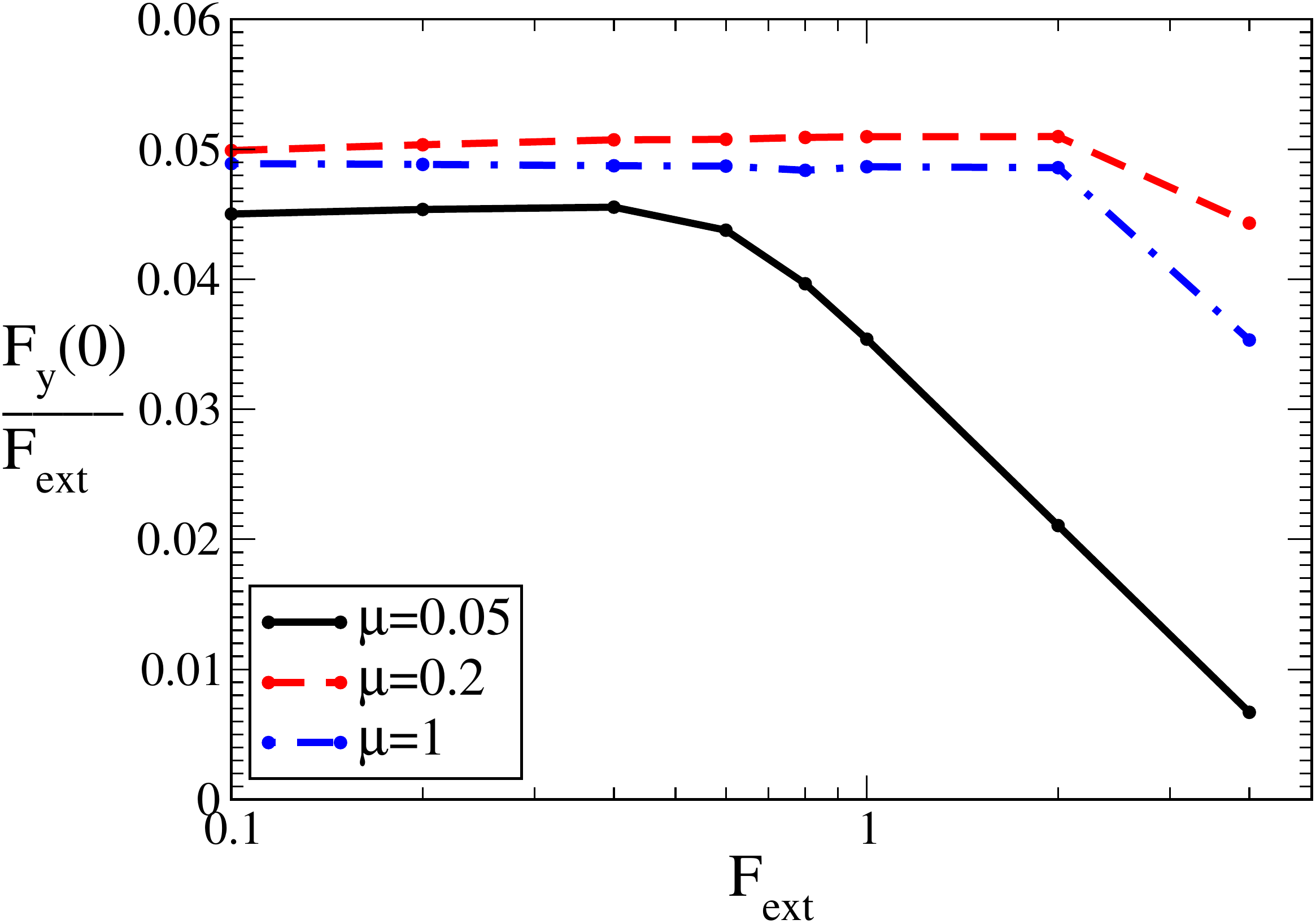}
\caption{The response at $x=0$ (coarse grained with
  $w=3\bar{R}$) of a single polydisperse systems with
  $\delta=0.01$, normalized by the applied force, $F_{\rm ext}$,  vs.\ the applied force
(given in units
  of the mean particle weight, $\bar{m}g$), for
  different coefficients of friction $\mu$, with frictionless walls ($\mu^{\rm
    wall}=0$) and $k_{T}/k_{N}=0.8$ (compare to
  Fig.~\ref{fig:linearity-friction}). 
  \label{fig:linearity-friction-poly1p}}
\end{figure}

A notable difference between monodisperse and polydisperse systems is that in
the latter, the range of linear response is smaller (see
Fig.~\ref{fig:linearity-friction-poly1p} compared to
Fig.~\ref{fig:linearity-friction}).

\begin{figure*}
\begin{tabular}{ccc}
\includegraphics[clip,width=0.3\hsize]{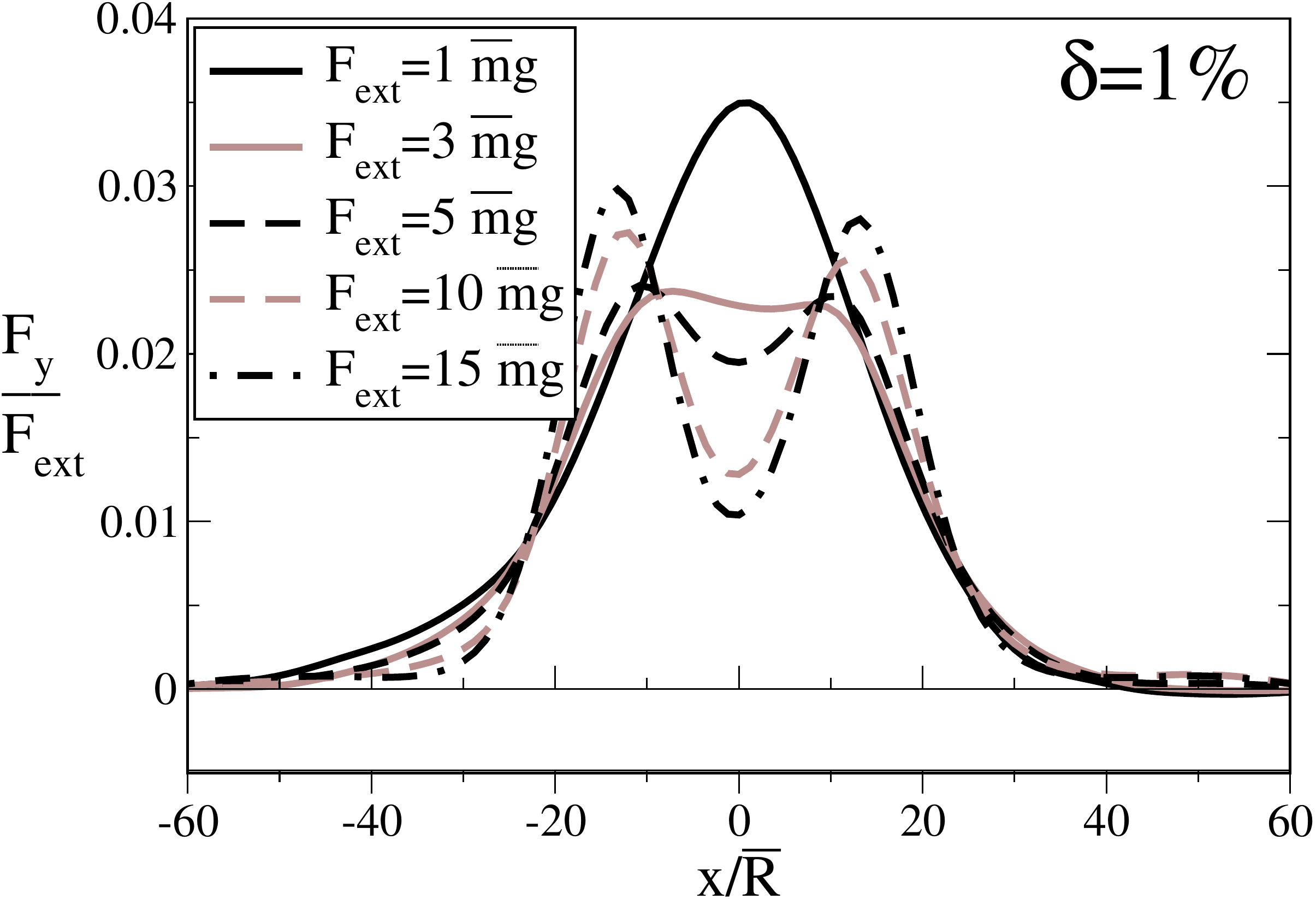}&
\includegraphics[clip,width=0.3\hsize]{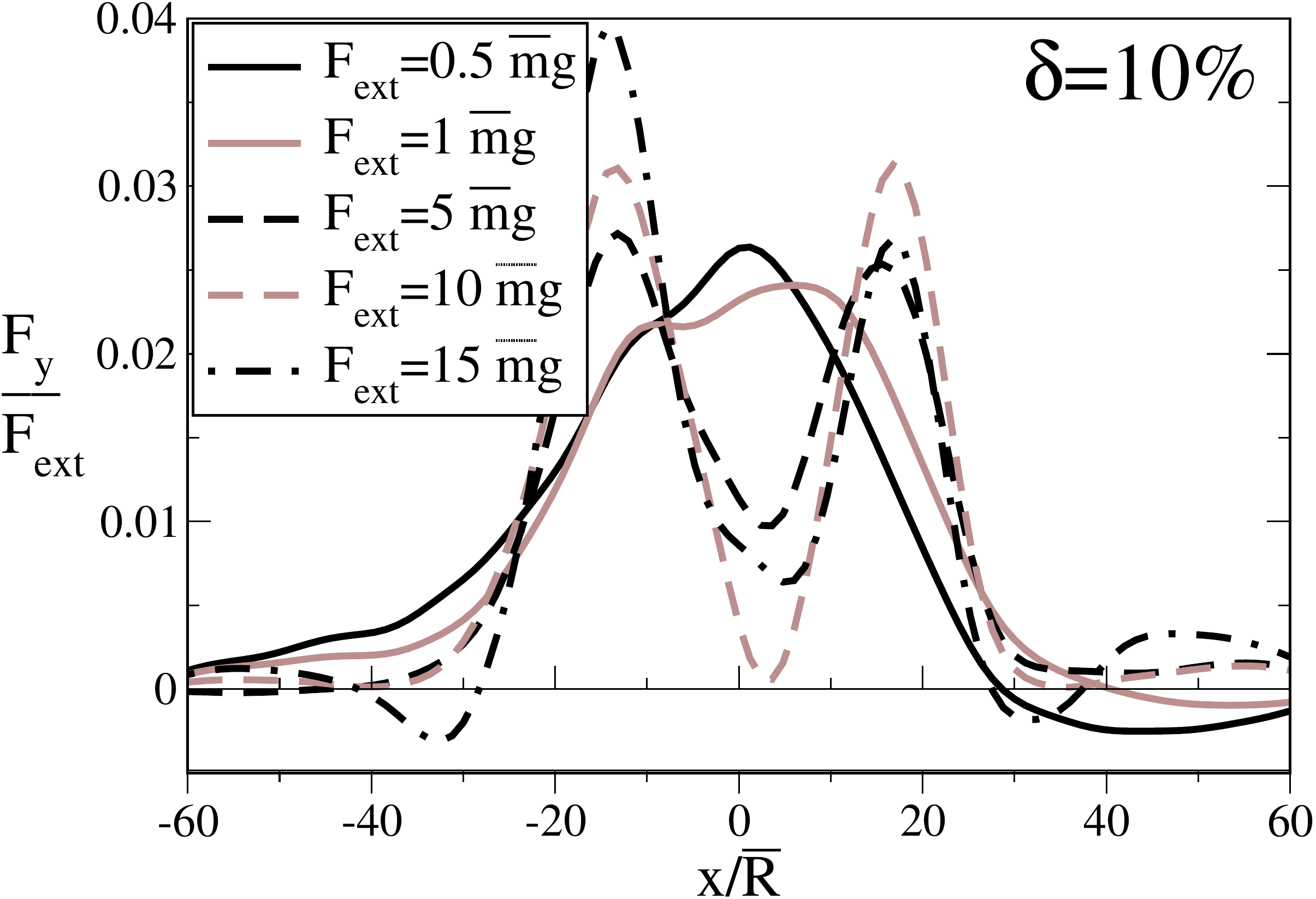}&
\includegraphics[clip,width=0.3\hsize]{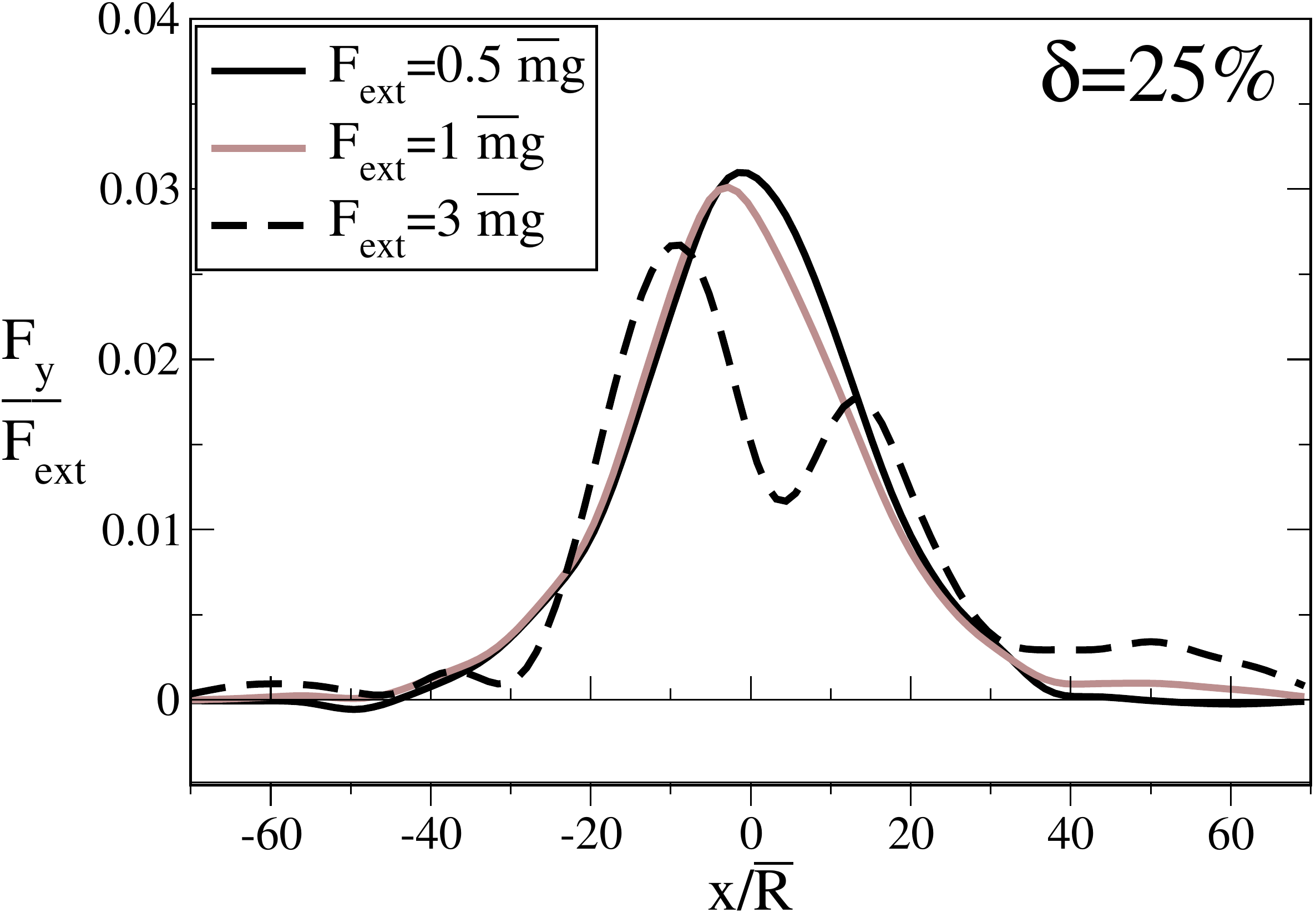}\\
\multicolumn{3}{c}{\large ${\mu}=0.05$}\\
&&\\
\includegraphics[clip,width=0.3\hsize]{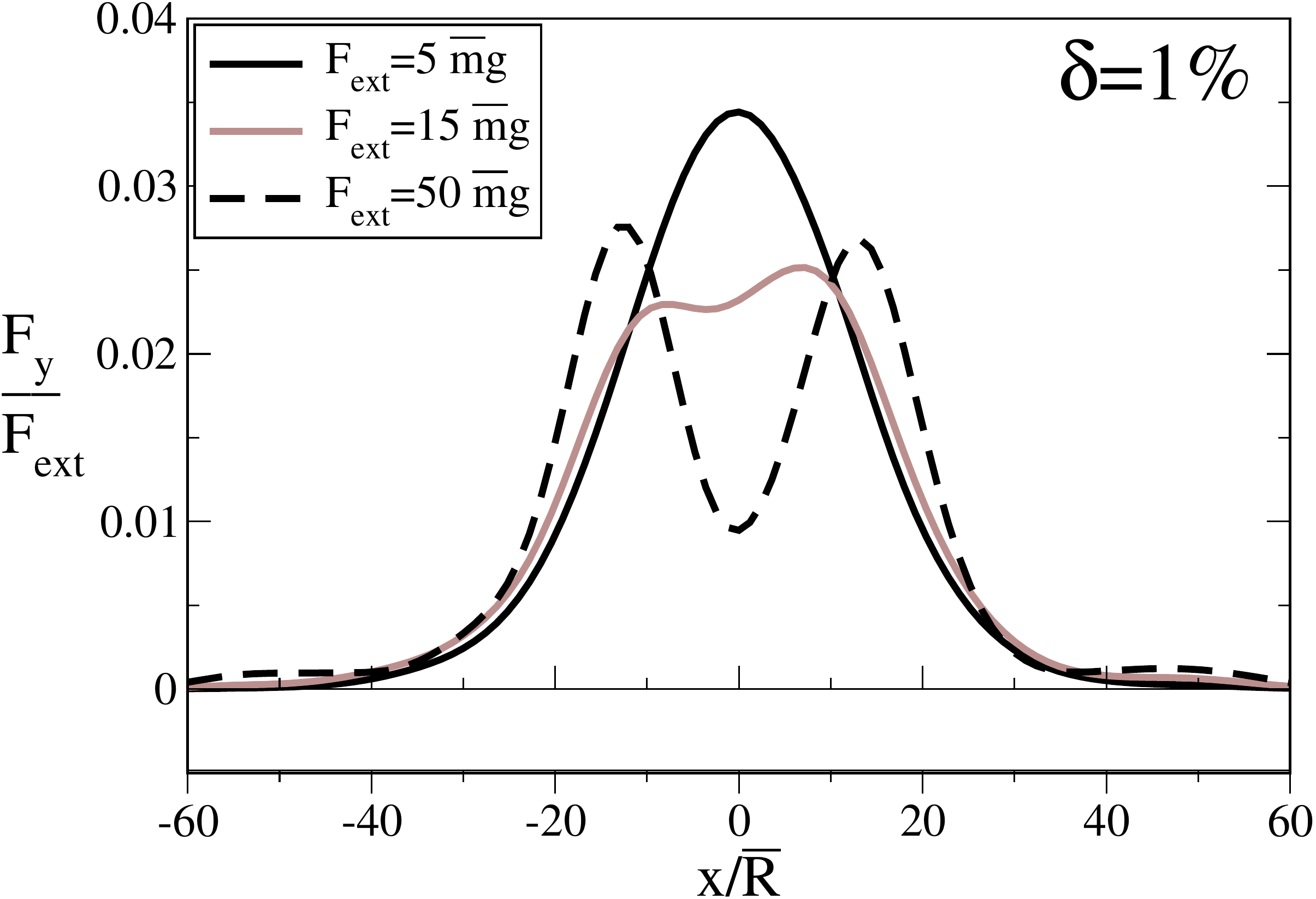}&
\includegraphics[clip,width=0.3\hsize]{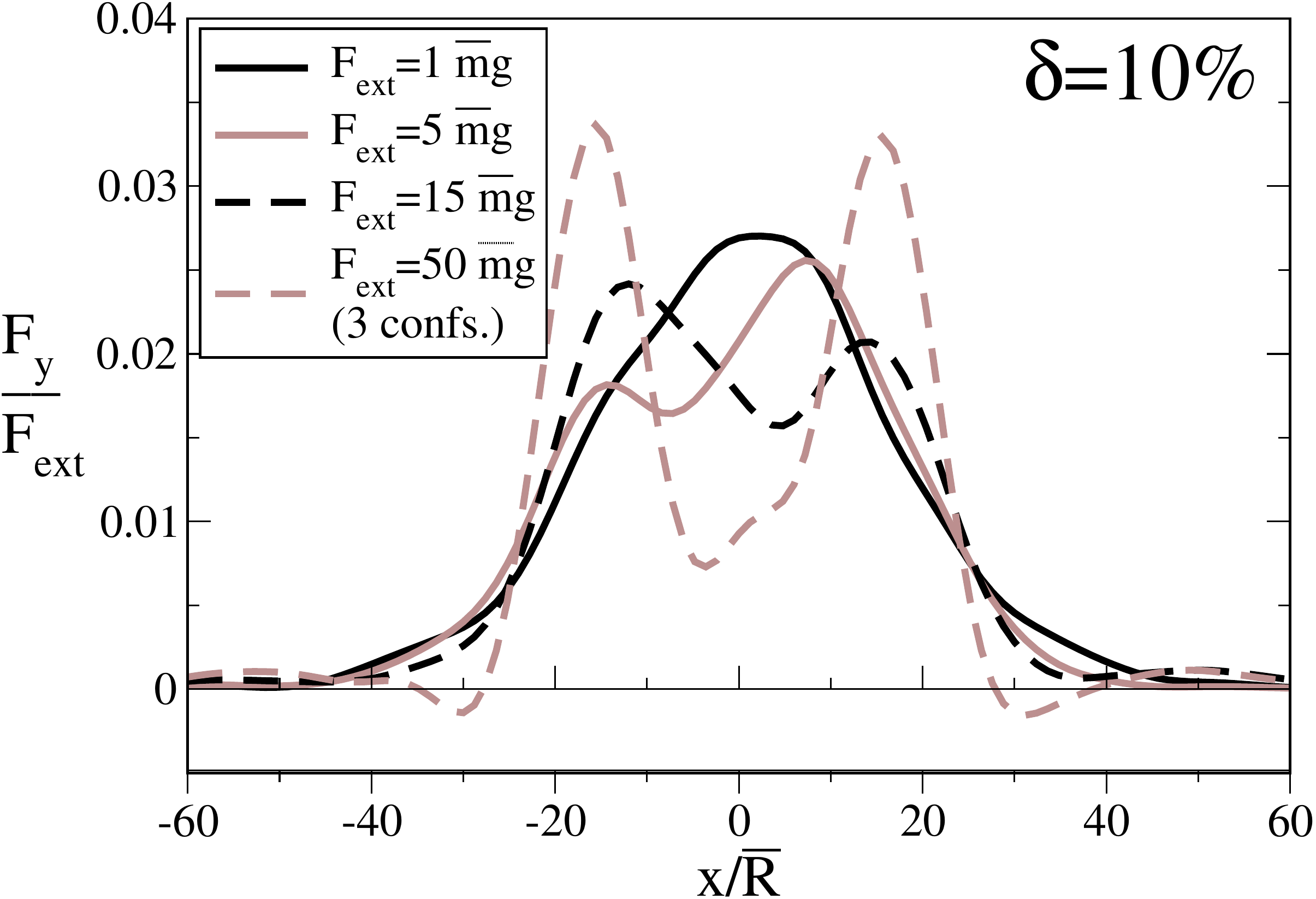}&
\includegraphics[clip,width=0.3\hsize]{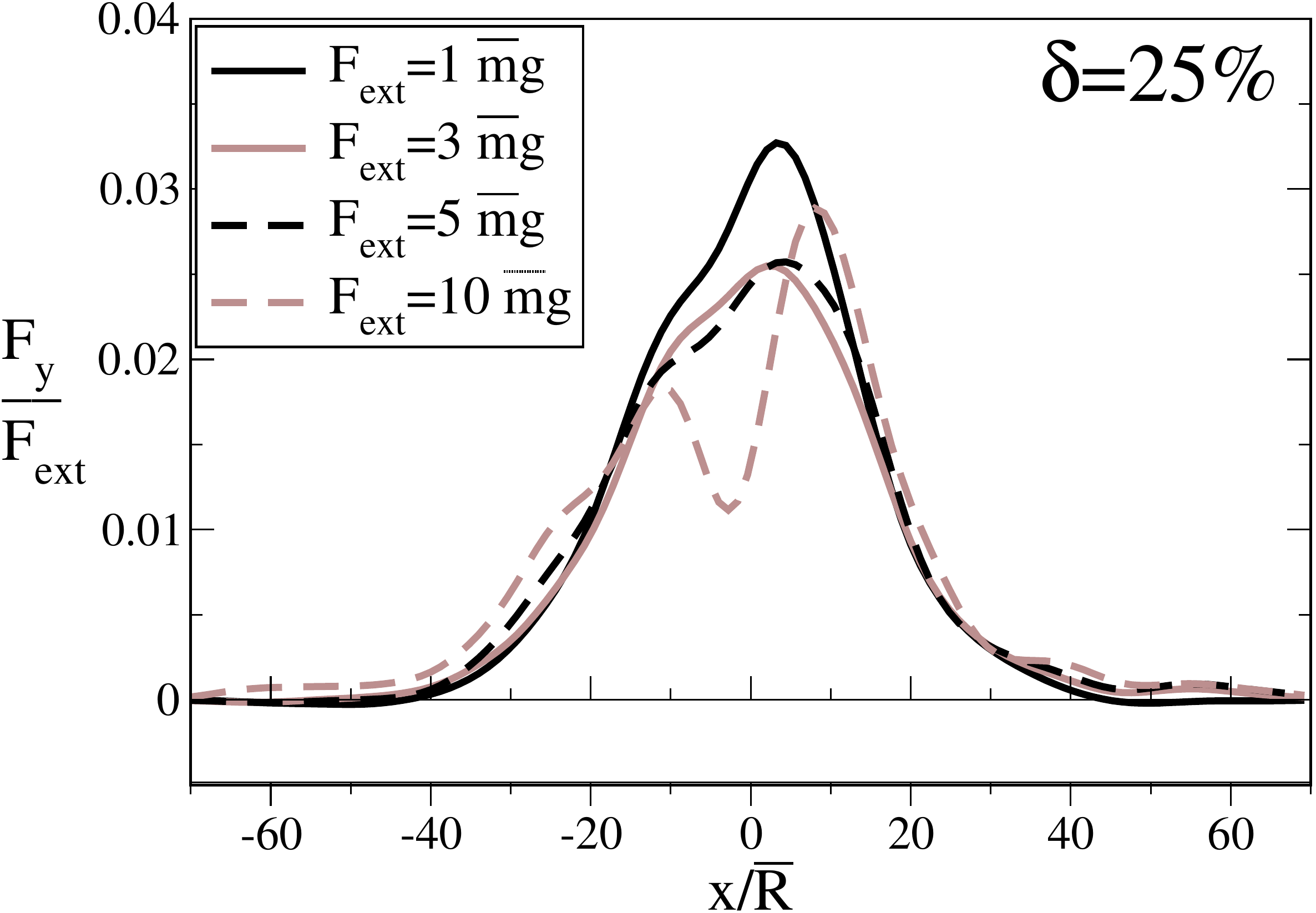}\\
\multicolumn{3}{c}{\large ${\mu}=0.2$}\\
&&\\
\includegraphics[clip,width=0.3\hsize]{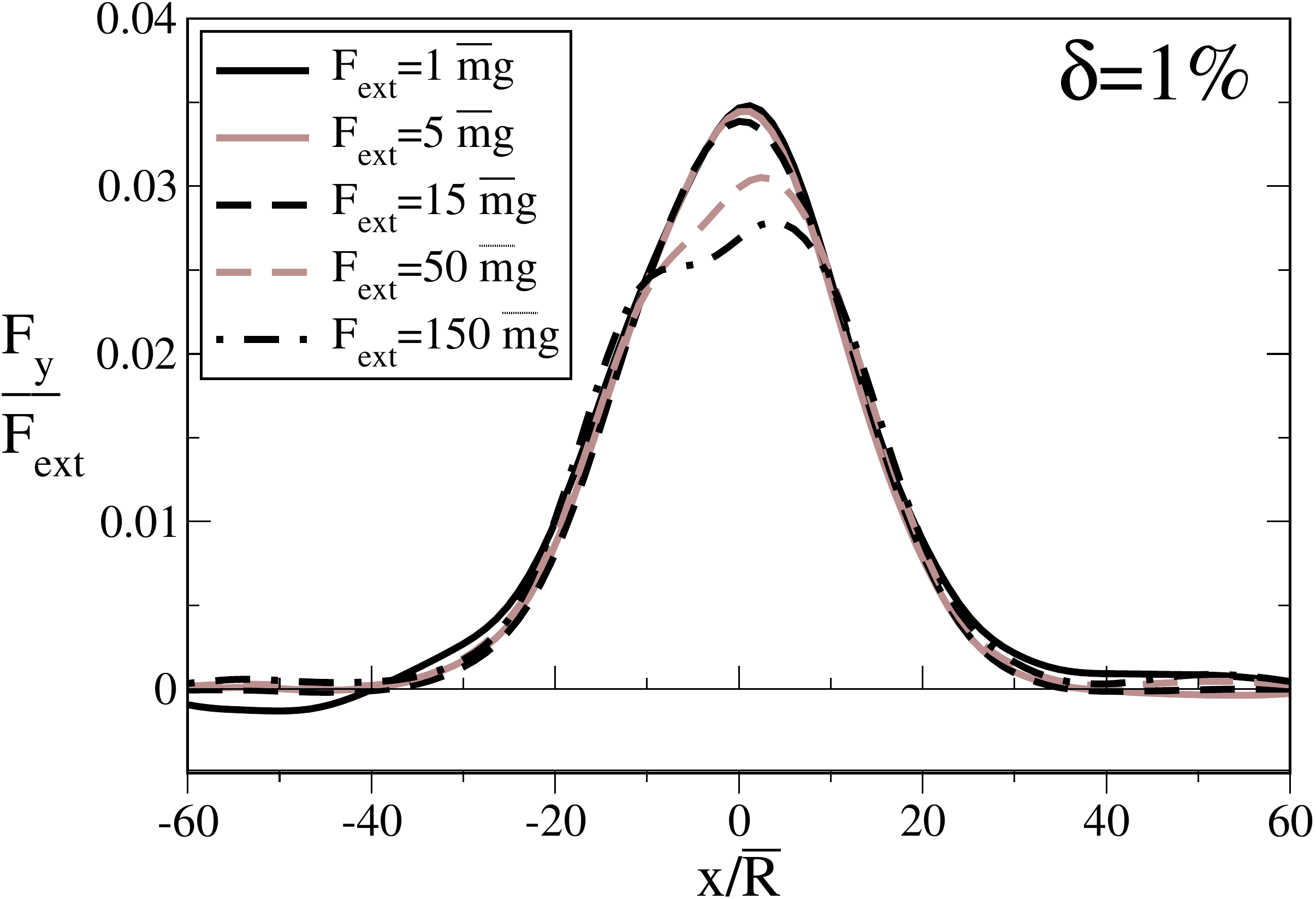}&
\includegraphics[clip,width=0.3\hsize]{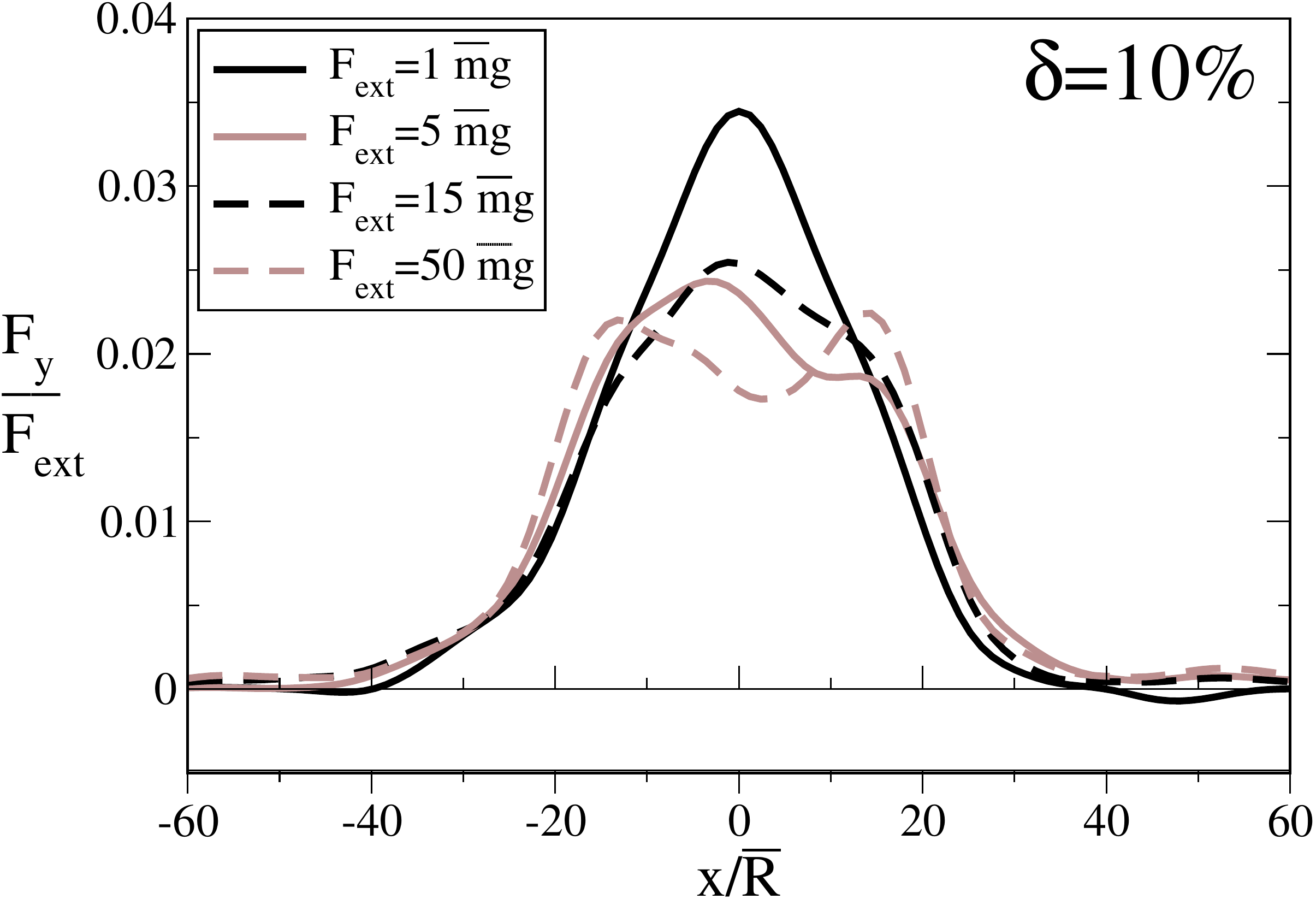}&
\includegraphics[clip,width=0.3\hsize]{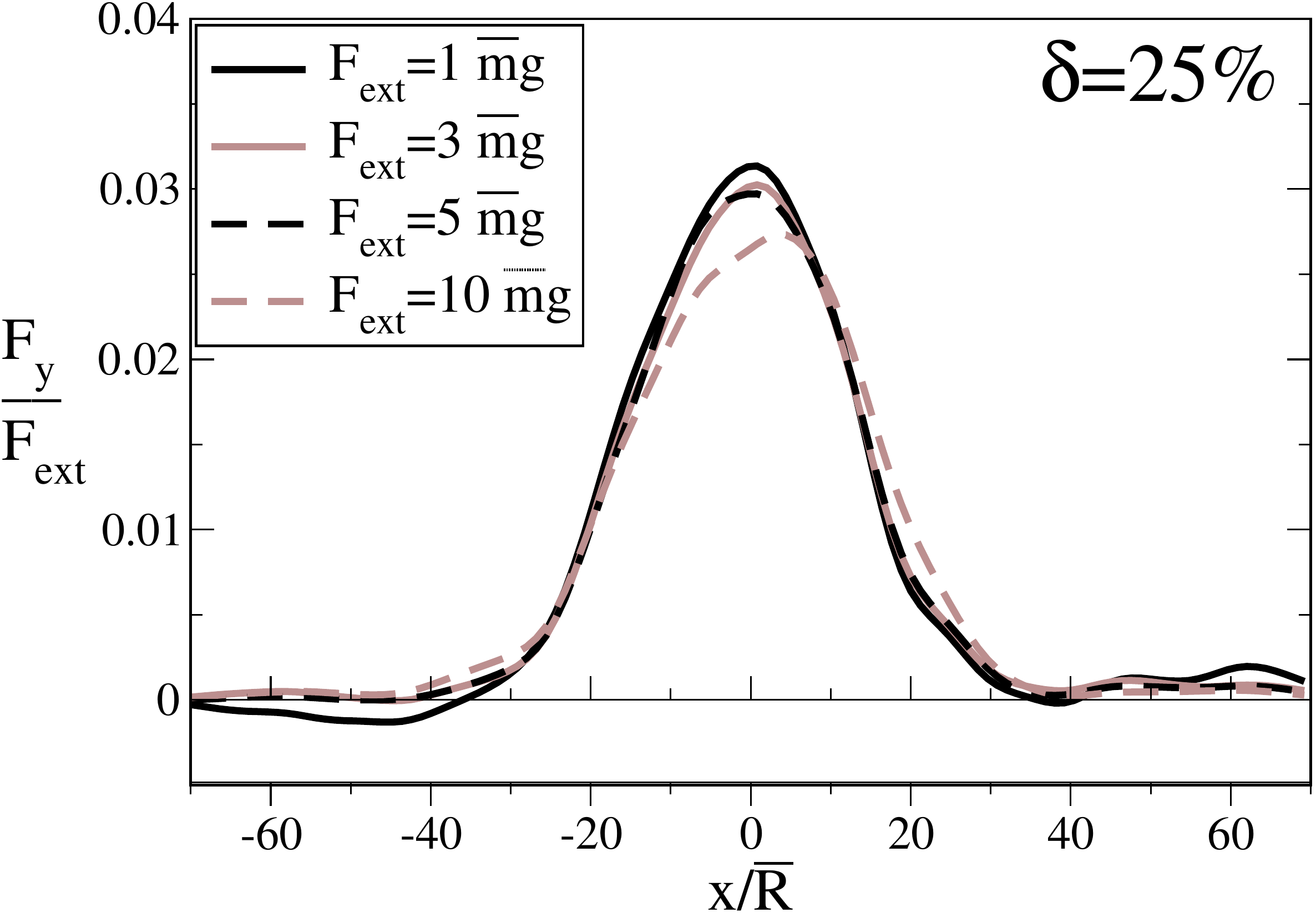}\\
\multicolumn{3}{c}{\large ${\mu}=1$}\\
\end{tabular}
\caption{The response of disordered systems with different degrees of
  polydispersity, $\delta$, and coefficients of friction, ($\mu=\mu^{\rm
    wall}$), with $k_{T}/k_{N}=0.8$, to different applied forces $F_{\rm ext}$.
  The response is averaged over 5 configurations (3 for the largest force used
  with $\delta=10\%$ and $\mu=0.2$, for which the system was unstable) and
  smoothed with a Gaussian of width $w=6\bar{R}$.
  \label{fig:response-diffpoly-diff_fric}}
\end{figure*}

In order to examine the effects of polydispersity and friction in more detail,
we performed similar simulations with different degrees of polydispersity and
coefficients of friction $\mu$, for different applied forces. The results are
depicted in Fig.~\ref{fig:response-diffpoly-diff_fric}. The larger the
coefficient of friction, the larger the applied force at which the crossover
from a single peaked to a double peaked response occurs, much like 
in ordered systems (Sec.~\ref{sec:fric-linearity}. As mentioned
above, the crossover force decreases with increasing polydispersity, or
disorder. For  polydisperse systems we observed a rather gradual crossover
even for $\mu=1$, unlike the sharp transition to a highly asymmetric response
in  lattice configurations [Sec.~\ref{sec:fric-largemu}]. A possible
explanation  may be that for a single disordered realization, the
response is not symmetric even for a small external force. In addition, it may be
impossible to apply very large forces to polydisperse systems
without causing  major rearrangements.

It is quite remarkable that an average over a very small ensemble of only five
realizations (with coarse graining) is sufficient to demonstrate the crossover
from a single peak to a double peak, and even results (in some cases) in nearly
symmetrical graphs. It appears that, at least when the force is not too close
to its crossover value, the fluctuations within the ensemble are rather small
for this choice of coarse graining length (see~\cite{Goldenberg06b} for a
detailed study of another set of simulations in the linear force regime), so
that the typical response of a realization resembles that of the ensemble
average.

\begin{figure}
  \includegraphics[clip,width=\hsize,clip]{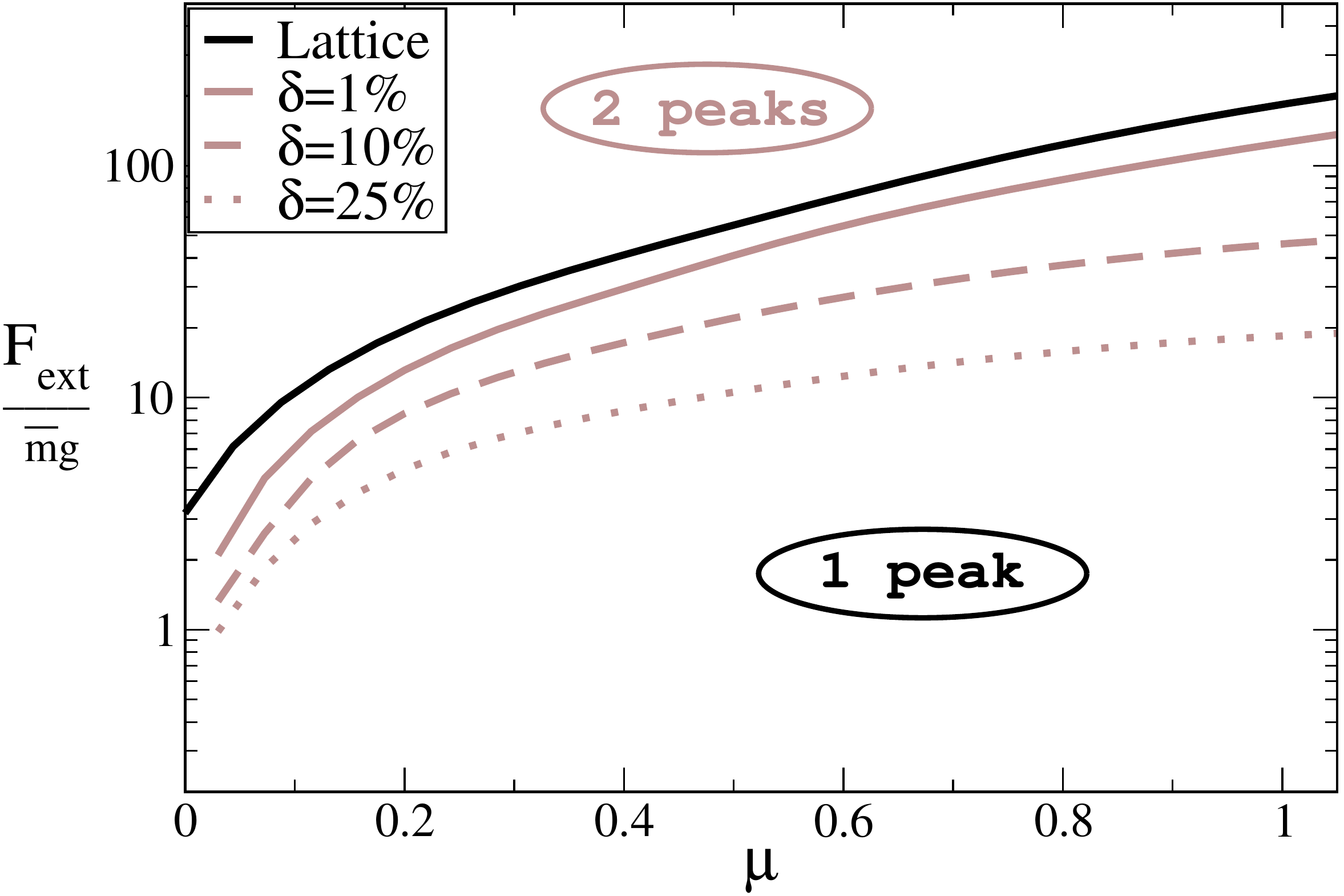}
  \caption{Schematic phase diagram (in the $F_{\rm ext}$-$\mu$ plane) for
  the crossover from a single peaked to a double peaked response, with
  different degrees of polydispersity.}
\label{fig:phase_diagram}
\end{figure}

The results for different degrees of polydispersity (including the case of a
lattice) can be summarized in a schematic phase diagram,
Fig.~\ref{fig:phase_diagram}~\cite{Goldenberg05}.  Note that in our simulations a specific
preparation method has been used: relaxation under gravity from a near-lattice
configuration, so that (at least for small $\delta$), the configuration retains
partial order. In the range of polydispersity studied here, more disordered
systems exhibit a smaller crossover force, i.e., they are more susceptible to
induced anisotropy. It is possible that this result pertains to small
polydispersity and  a larger polydispersity may actually stabilize the system. In the
systems studied here, the distribution of contact angles may not be quite
isotropic, exhibiting preferred direction close to those of the triangular
lattice, and therefore more susceptible to failure along certain direction.
Therefore, in more disordered systems, the degree of induced anisotropy may be
smaller, yielding a non-monotonic dependence of the crossover force on the
degree of polydispersity. In any case, it is likely that the crossover does not
depend on the polydispersity alone but also on the preparation method (as
observed in experiments on granular piles~\cite{Vanel99b} and
slabs~\cite{Serero01}). A more complete phase diagram would presumably further depend
on additional  parameters which characterize the geometry of the packing (in a
statistical way). A common approach is to use fabric tensors~\cite{Kanatani84},
often simplified by considering the distribution of contact angles. However,
this characterization is often employed in conjunction with a mean-field approach, which fails for
disordered granular materials~\cite{Chang94,Makse99} due to the large
non-affine component of the microscopic displacements (see
also~\cite{Goldhirsch02}).

\subsection{Superposition}
\label{sec:superposition}
Another issue related to the question of linearity, discussed in
Sec.\ref{sec:disorder-linearity}, is that of superposition: is the sum of
responses to several forces applied individually the same as the response to
all the forces applied at the same time?

\begin{figure*}
  \begin{tabular}{ccc}
    \subfigure[$d=8\bar{R}$.]{\includegraphics[clip,width=0.3\hsize]{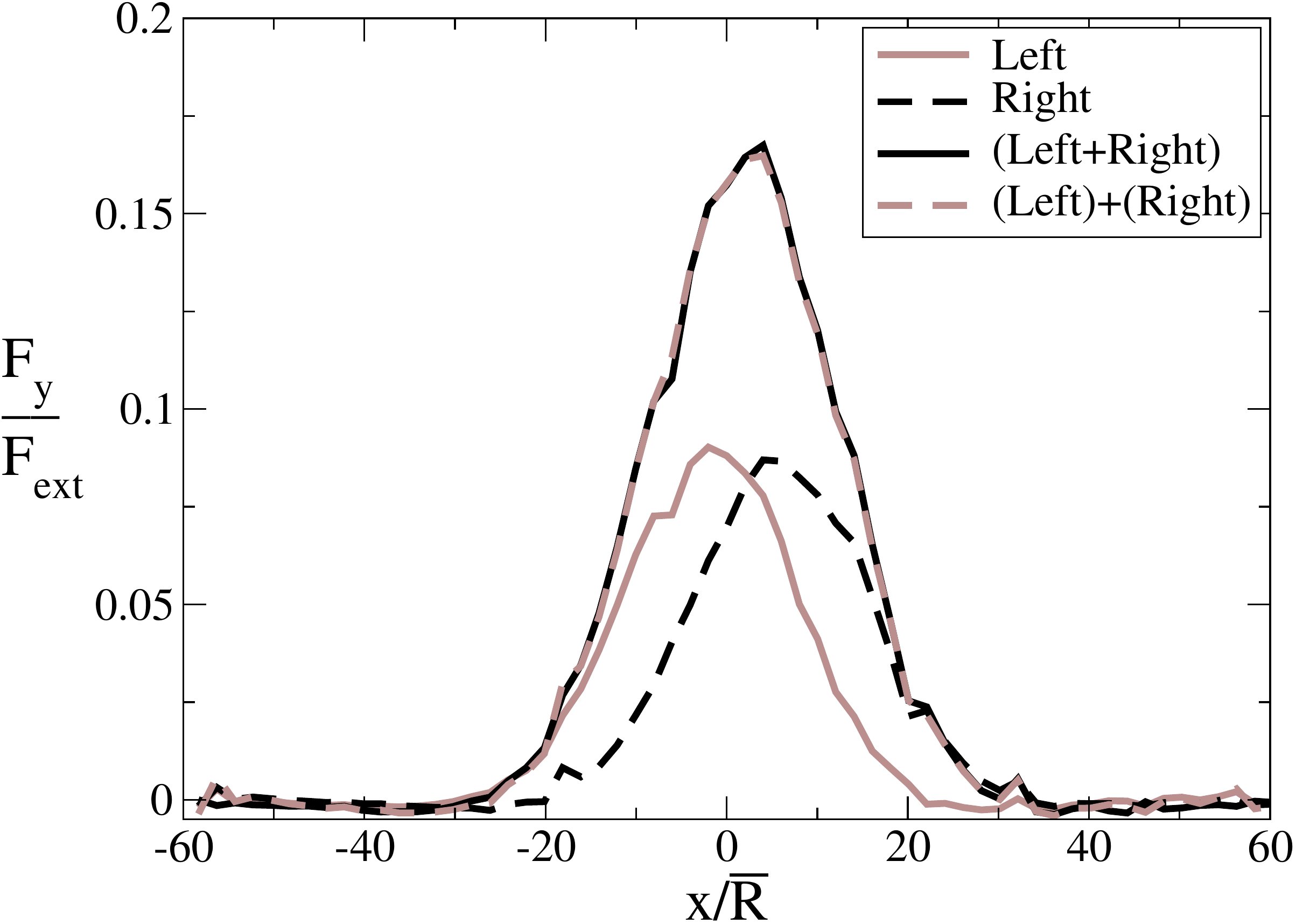}\label{fig:superp-poly1p-mu0.05_f0.2_dist8R}}&
    \subfigure[$d=16\bar{R}$.]{\includegraphics[clip,width=0.3\hsize]{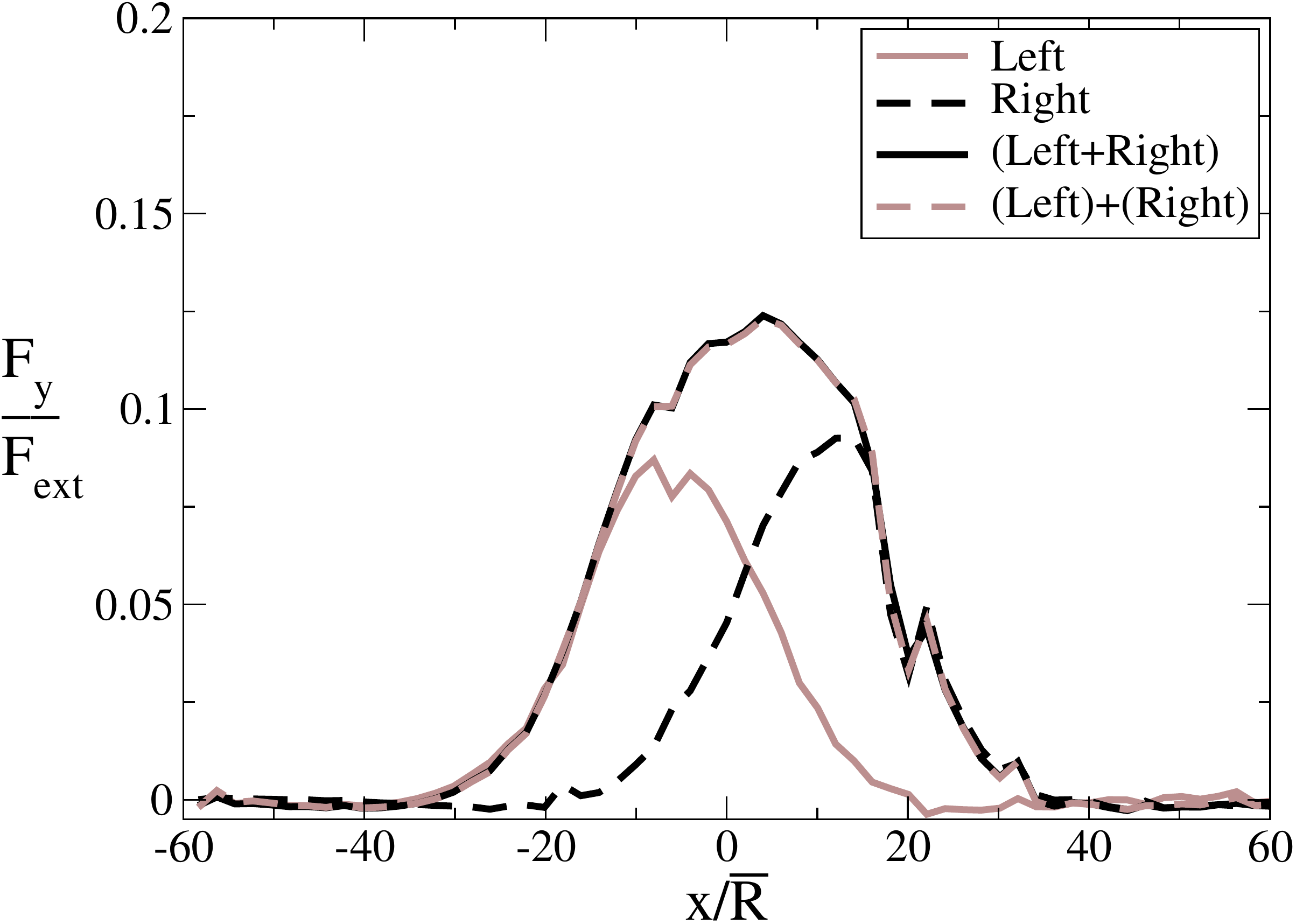}\label{fig:superp-poly1p-mu0.05_f0.2_dist16R}}&
    \subfigure[$d=24\bar{R}$.]{\includegraphics[clip,width=0.3\hsize]{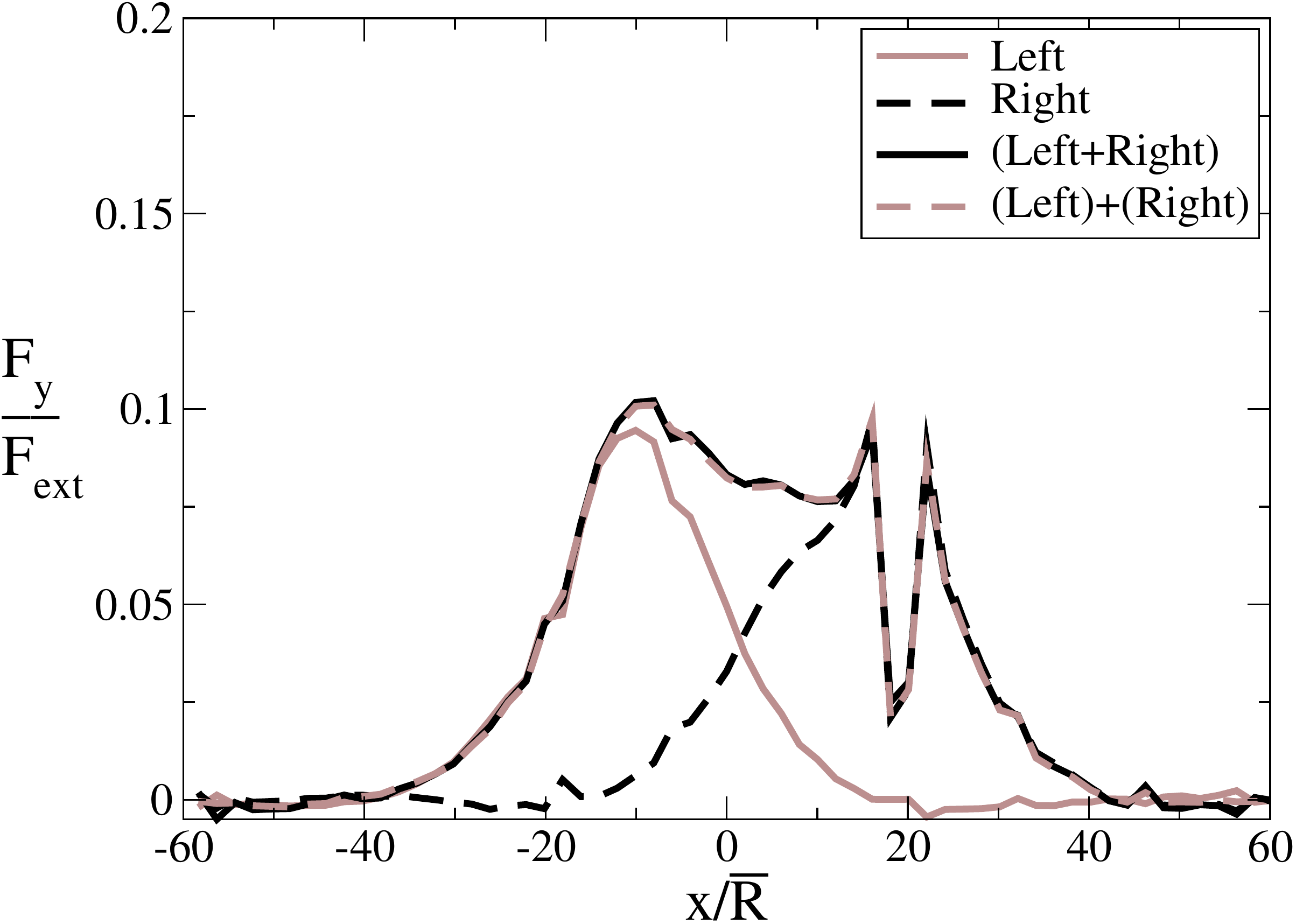}\label{fig:superp-poly1p-mu0.05_f0.2_dist24R}}
    \end{tabular}
    \caption{Superposition in a polydisperse system with $\delta=0.01$, 
      $\mu=0.05$,  and $F_{\rm ext}=0.2\bar{m}g$, frictionless walls ($\mu^{\rm
        wall}=0$) and $k_{T}/k_{N}=0.8$. $d$ indicates the distance between the
      points of application of the two forces (applied symmetrically with
      respect to the center top particle), given in mean particle diameters
      $\bar{R}$. In the legend, {\bf Left} denotes the response to a force
      applied to the left of the center particle, {\bf Right} to a force
      applied to its right, {\bf (Left+Right)} to the two forces applied
      together and {\bf (Left)+(Right)} the sum of {\bf Left} and {\bf Right}.
      The results shown are for the individual forces (no coarse graining) in a
      single realization.
      \label{fig:superp-poly1p-mu0.05-f0.2}}
\end{figure*}

Fig.~\ref{fig:superp-poly1p-mu0.05-f0.2} demonstrates a typical superposition
test. Here, two downward pointing vertical forces, each of magnitude $F_{\rm
  ext}=0.2\bar{m}g$, are applied at two points at the top of the slab,
otherwise characterized by $\delta=1\%$ and $\mu=0.05$, for which the linear
range is rather small ($F_{\rm ext}\lesssim 0.4\bar{m}g$).  We compared the
response of the system when each of the forces was applied separately and
simultaneously at different horizontal separations.  As shown in
Fig.~\ref{fig:superp-poly1p-mu0.05-f0.2}, superposition holds even on the {\em
  microscopic scale}, i.e., for individual forces without coarse graining.

\begin{figure*}
  \begin{tabular}{ccc}
    \subfigure[$d=8\bar{R}$.]{\includegraphics[clip,width=0.3\hsize]{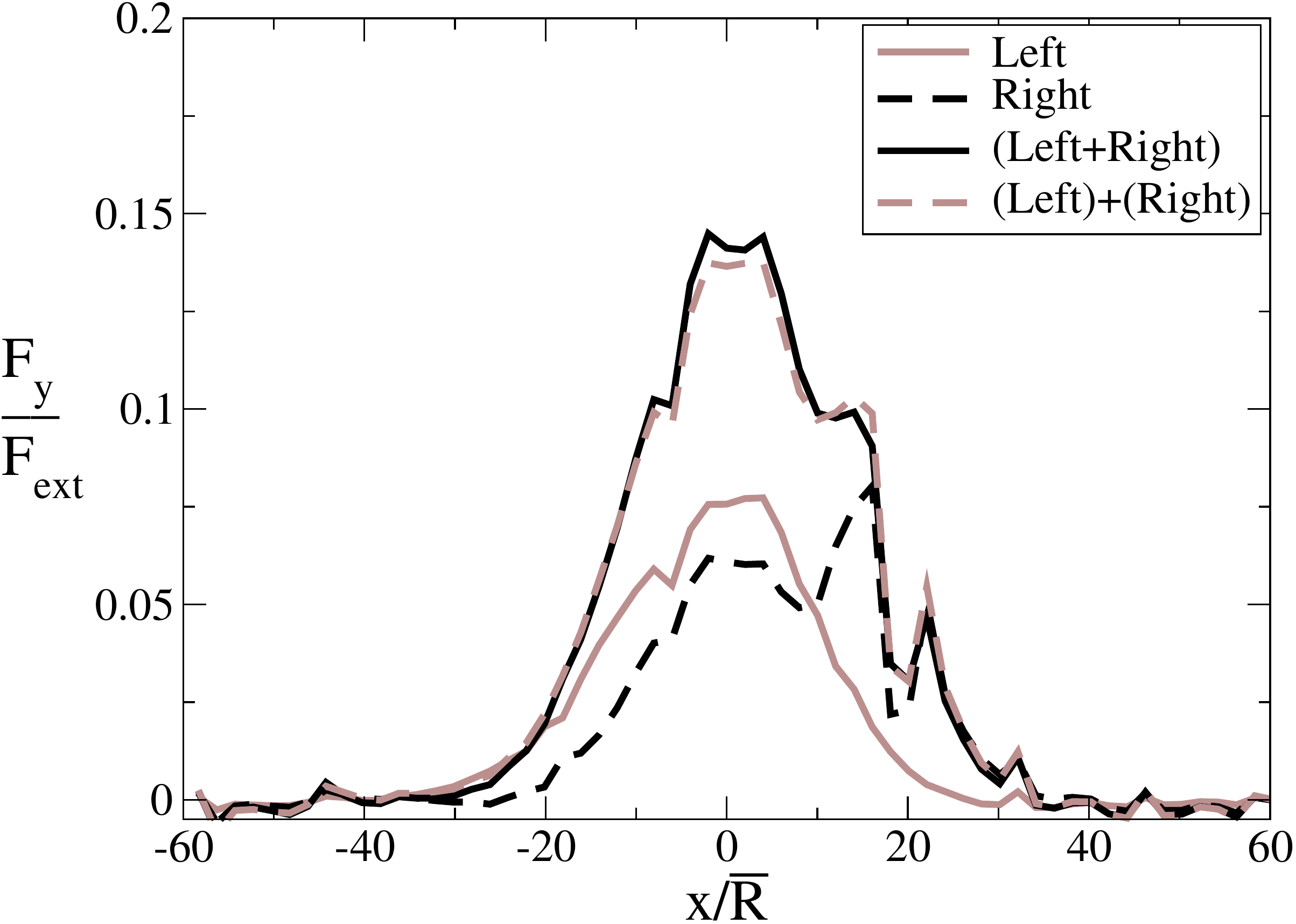}\label{fig:superp-poly1p-mu0.05_f1_dist8R}}&
    \subfigure[$d=16\bar{R}$.]{\includegraphics[clip,width=0.3\hsize]{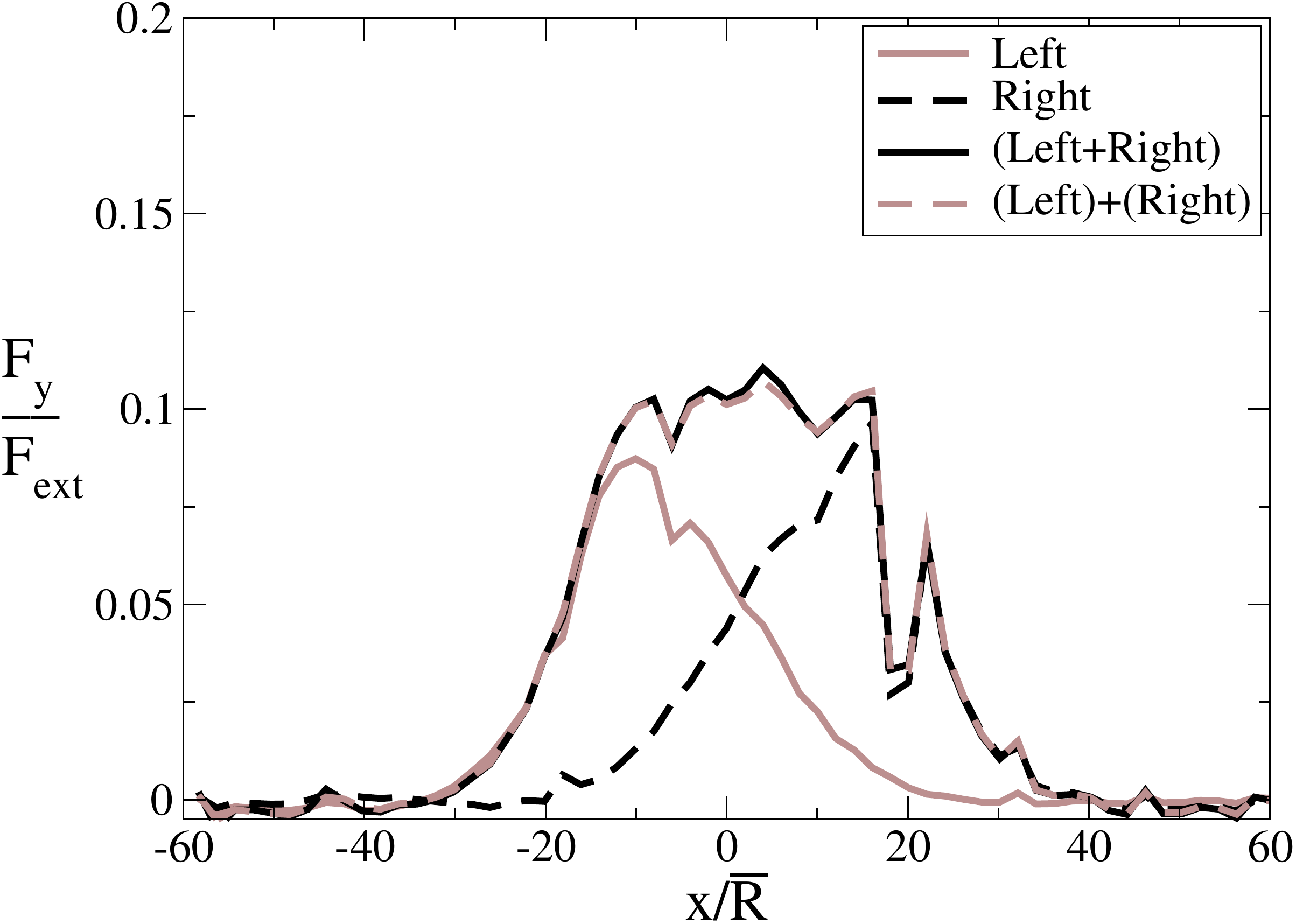}\label{fig:superp-poly1p-mu0.05_f1_dist16R}}&
    \subfigure[$d=24\bar{R}$.]{\includegraphics[clip,width=0.3\hsize]{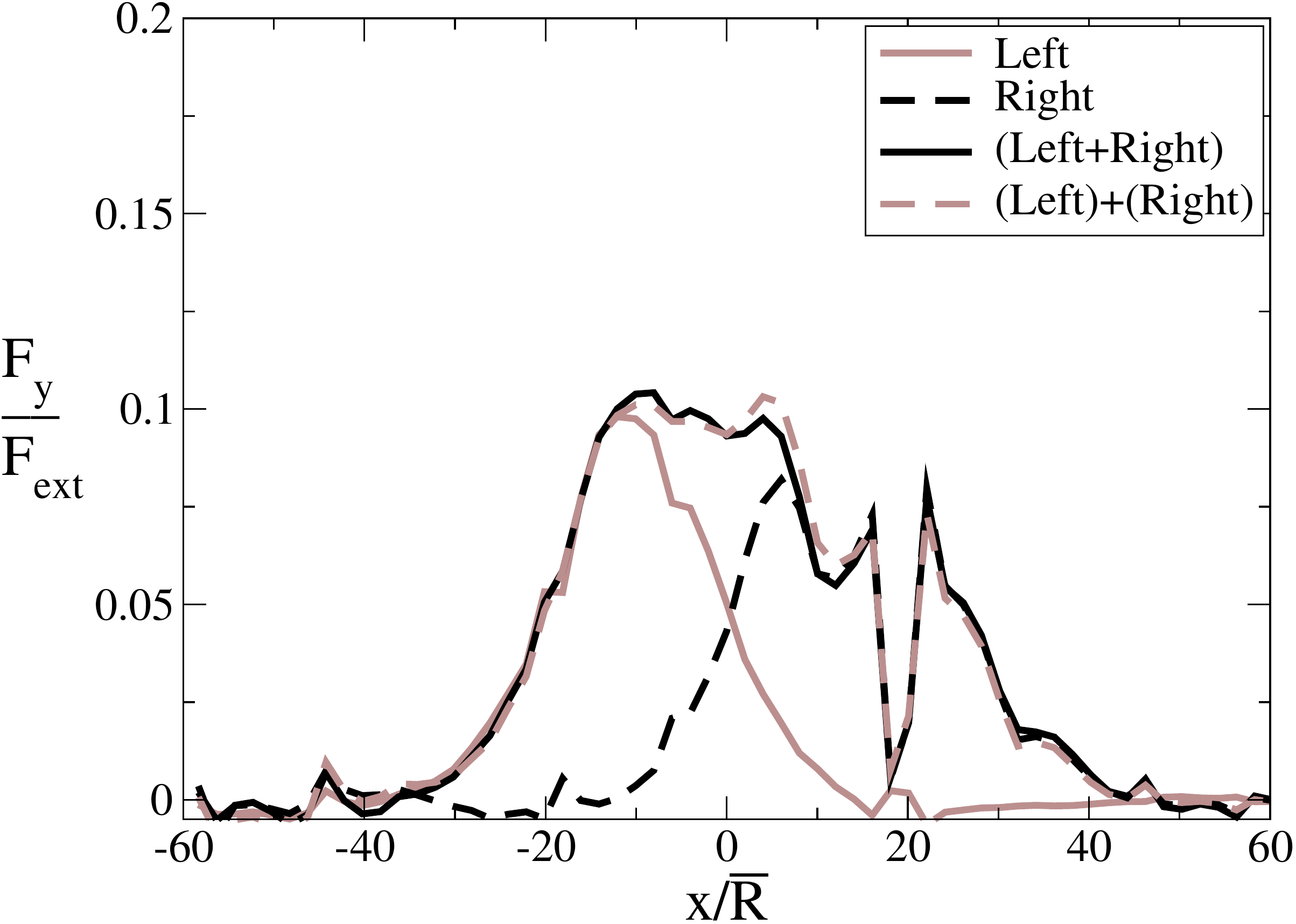}\label{fig:superp-poly1p-mu0.05_f1_dist24R}}
    \end{tabular}
    \caption{Superposition in a polydisperse system with $\delta=0.01$, 
      $\mu=0.05$,  and $F_{\rm ext}=\bar{m}g$ (same as
      Fig.~\ref{fig:superp-poly1p-mu0.05-f0.2} with a different applied force;
      the same realization of the disorder was used).
    \label{fig:superp-poly1p-mu0.05-f1}}
\end{figure*}

Interestingly, we find that superposition holds quite well even beyond the
linear regime, as shown in Fig.~\ref{fig:superp-poly1p-mu0.05-f1} (for $F_{\rm
  ext}=\bar{m}g$). A possible reason is that sufficiently far from where the
forces are applied, the rearrangements (changes in the contact network or
sliding) are less sensitive to the precise  positions of the applied forces. In more
general terms, it is possible that the response to a distribution of forces,
sufficiently far from the region where they are applied, is not very sensitive
to the details of the distribution.

This explanation  is similar to St.\ Venant's principle~(see, e.g.,~\cite{Mase70}),
which states (for a {\em linear elastic} system) that the difference in
stresses and strains in the interior of an elastic body due to two separate but
statically equivalent systems of surface tractions (same overall force and
torque), are negligible sufficiently far from the area where the  loads are
applied. In the case considered here, the system is certainly not strictly
elastic for $F_{\rm ext}\gtrsim 0.4\bar{m}g$ (due to the  effects of contact
network changes and sliding). However, the response to a localized force (in an
elastic medium) decays with the distance from the point of application as a
power law (e.g., the solutions for an infinite half plane or half space
[Boussinesq's problem], which decay like $1/z$ in 2D and $1/z^2$ in
3D~\cite{Landau86,Johnson85}).  Therefore, as the distance from the point of
application increases, the relative displacements of the particles become
smaller, and they are less likely to reach the nonlinear regime (the finite
region of open contacts discussed in Sec.~\ref{sec:friction} is an example of
such a nonlinear effect), so that far from the point of application of the
force, the system becomes ``more elastic'' and the response to a distributed load is the
same (or nearly the same) as the response to the resultant load.  This suggestion may apply to
recent experiments~\cite{Kolb04,Kolb06} on a 2D system subject to a
small cyclic displacement, which shows a $1/r$ decay of the displacement field
at large distances from the perturbation.

\section{Interpretation of Experimental Results}
\label{sec:interpretation}
The results described in the previous section help to interpret the
experimental results reviewed in Sec.~\ref{sec:response-experiments}.  To this
end, it is important to note that the different experiments described in
Sec.~\ref{sec:response-experiments} correspond to different regions of the
phase diagram [Fig.~\ref{fig:phase_diagram}] described in
Sec.~\ref{sec:disorder}.

As mentioned, experiments on 2D systems using photoelastic
particles~\cite{Geng01,Geng03} typically use a rather large applied force
($F_{\rm ext}\simeq 150mg$ in these experiments), required in order to obtain a
significant photoelastic effect.  In these experiments the interparticle
forces, rather than the macroscopic stress, were measured (the forces on the
floor have not been measured). These are reproduced quite well by our DEM
simulations with friction (see Sec.~\ref{sec:friction}).  It is yet unclear
whether these experiments are compatible with the phase diagram of
Sec.~\ref{sec:disorder}, since the macroscopic stress has not been measured
directly in these experiments.

The experiments reported in~\cite{Reydellet01,Serero01} used disordered systems
with a rather small applied force (a few times the particle weight), for which
a single peak was observed, in agreement with the phase diagram. Since these
experiments used sand, rather than spherical particles, the effective
coefficient of static friction may be much larger than $1$, so that sliding is
less likely to occur. Note that those experiments were performed on a 3D
system, while the simulations reported here are in 2D; however, we expect
similar qualitative features of the phase diagram in 3D systems.  The
difference in the responses of dense and loose sand~\cite{Serero01}, as well as
the deviations from the isotropic elastic prediction observed in both cases,
may be explained by a (small) anisotropy, induced by the packing construction
process (see also~\cite{Atman05}). We believe that the anisotropy induced by
the small applied force in these experiments should be negligible (note that
the systems studied in~\cite{Reydellet01,Serero01} are also deeper, in terms of
particle diameters, than the ones studied in~\cite{Geng01,Geng03}).

In contrast to the experiments reported in~\cite{Reydellet01,Serero01}, the
experiments reported in~\cite{Mueggenburg02} on ordered 3D packings used a very
large force of a few thousand times the particle weights. Several distinct
peaks were observed for shallow systems. These peaks became more diffuse for
deeper systems. In the absence of horizontal contacts, the coordination number
in the HCP and FCC lattice is reduced to $6$ (each particle is in contact with
$3$ particles in the layer above it and that below it).  This corresponds
to the isostatic limit (in the absence of friction). Apparently, the
experimentalists  attempted to avoid horizontal contacts by choosing the
appropriate wall spacing; even if some horizontal contacts were present, it is
likely that such a large force  opened the  horizontal contacts, at least in the
top layers. Since the force is large, it may also lead to sliding at most of
the contacts (at least in the top layers), so that for shallow systems, the
observed behavior is extremely anisotropic, and, as mentioned, near the
isostatic limit. However, in deeper systems, the opening of contacts and
sliding may be reduced (as we observed for $\mu=0.2$), reducing the anisotropy
(note that even the deepest layers in the described experiment were not very
deep: about 20 layers). In amorphous packings, a single peak was observed
(however, the method of application of the force was different: a force impulse
was used, which may correspond to a weaker force in our quasi-static
description). Thus, the results of the experiments reported
in~\cite{Mueggenburg02} are also consistent with the schematic phase diagram
presented in Sec.~\ref{sec:disorder}.

The phase diagram may also explain the striking difference observed
in~\cite{Moukarzel04} between the experimentally measured displacement response
in  packings of (frictional) disks subject to a localized displacement (which
exhibits a single peak), and the results of simulations of frictionless
isostatic packings (which exhibits two peaks). Note, however, that the mean
coordination number in the frictionless polydisperse systems studied here
(e.g., $z \sim 4.25$ for $\delta=0.01$) is well above the isostatic limit of
$z=4$. It is, however, much smaller than that of the ideal triangular lattice, $z=6$,
the ``missing'' contacts being predominantly horizontal. This suggests that the
anisotropic (``hyperbolic-like'') response of frictionless systems may arise,
at least for small polydispersity, from the anisotropic structure of these systems rather than
their (near)-isostaticity, as suggested in~\cite{Moukarzel04}.
\section{Concluding Remarks}
\label{sec:conclusions}
We performed simulations studying the response of a 2D granular slab to a
localized force, in particular the effects of the magnitude of the applied
force, the frictional parameters and the polydispersity. Our results indicate
that {\em anisotropy}, which is induced by changes in the contact network and
frictional sliding due to applied force, gives rise to a hyperbolic-like
response (as already suggested in~\cite{Savage98b,Cates99}).  However, we show
that this effect becomes less pronounced for large systems (which is more
likely to be the case in most engineering applications) and/or systems subject
to smaller applied forces for a given system size. This may explain why
small-scale experiments sometimes yield a hyperbolic-like response while
elasticity is typically used for describing small deformations in engineering
practice. In addition, {\em friction}, which is present in all real granular
systems, increases the range of linearity in the applied force and further
reduces the effect of stress-induced anisotropy, rendering the response even
closer to that predicted by isotropic elasticity.  The same behavior is
observed in polydisperse systems, which is the typical case in nature, although
in polydisperse systems the forces at which induced anisotropy becomes large
are smaller. The validity of superposition beyond the linear elastic regime
indicates that sufficiently far from the applied load, the system responds
(nearly) elastically even though it is not elastic everywhere.

The above considerations should also be relevant for the case of granular
piles, whose properties depend strongly on their construction
history~\cite{Vanel99b}. For instance, when the pile is grown from a point
source, it is shaped by successive avalanches, all of which are initiated in
the apex area (beneath the source). The nature of the granular flow (i.e., the
avalanches) in the case of formation by flow from an extended source is clearly
different from that corresponding to a point source. In particular, the degree
of anisotropy of piles constructed from a point source is larger than that of
piles built from an extended (and approximately uniform) source~\cite{Geng01b}.
In the former case, there is a dip in the pressure distribution on the floor,
whereas in the latter there is no dip. In~\cite{Vanel99b}, this effect was
described in terms of orientations of force chains, which may also reflect the
macroscopic anisotropy. However, we believe that once the (possibly
inhomogeneous) anisotropy in the pile is taken into account, the stress
distribution under the pile should be compatible with an elastic picture. The
question of how this anisotropy is created during construction will probably
require a more detailed description of the pile dynamics, possibly involving the
changes in the contact network (and sliding) which lead to macroscopic
anisotropy.

In general, the elastic description of granular solids is expected to be valid
for sufficiently large scales (and system sizes) and not-too-large applied
forces. Outside this range, such a description may need to be extended by using
a nonlinear, incrementally elastic continuum model, with stress-history
dependent elastic moduli (which reflect the induced anisotropy, and may exhibit
hyperbolic-like behavior). The nonlinear effects of broken contacts and
frictional sliding should be important in understanding the failure of granular
materials. In particular, the dominant effect of tangential forces in such
failure, e.g., in the transition to an asymmetric response described in
Sec.~\ref{sec:fric-largemu}, as well as the effects of applied torques
(Sec.~\ref{sec:apptorque}) and of rolling resistance, support the suggestion
that, at least on intermediate scales, continuum models of micropolar, or
Cosserat type~\cite{Kroner67,Jaunzemis67,Eringen68} may be required (see,
e.g.,~\cite{Borst91, Walsh04}).

We note, as a suggestion for a future experiment, that one should be able to
observe experimentally the crossover we obtain in the simulations with
increasing applied force, from a single peaked to a double peaked response; in
such an experiment, the setup should be sufficiently sensitive to measure the
response to a small force (in the linear regime) but also sufficiently robust
to enable the application of larger forces. In addition, the force would have
to be applied slowly, to avoid plastic flow of the material. Such an experiment
would be very useful in testing the applicability of the above physical
considerations to real granular materials.

\begin{acknowledgments}
  We thank A.~P.~F. Atman, R.~P. Behringer, P. Claudin, E.~Cl{\'e}ment, J.
  Geng, N. Mueggenburg, M. van~Hecke, W. van~Saarloos and T.~A. Witten for
  useful discussions. We gratefully acknowledge support from the Israel Science
  Foundation (ISF), grant no.\ 689/04, the German-Israeli Science Foundation
  (GIF), grant no.\ 795/2003 and the US-Israel Binational Science Foundation
  (BSF), grant no.\ 2004391. CG acknowledges financial support from a
  Chateaubriand Fellowship and from a European Community FP6 Marie Curie Action
  (MEIF-CT2006-024970).
\end{acknowledgments}

\end{document}